\documentclass[preprint]{aastex}
\begin{document}
\title{Flaring Behavior of the Quasar 3C~454.3 across the Electromagnetic Spectrum}
\author{Svetlana G. Jorstad\altaffilmark{1,2}, Alan P. Marscher\altaffilmark{1}, 
Valeri M. Larionov\altaffilmark{2,3}, Iv\'an Agudo \altaffilmark{4}, Paul S. Smith\altaffilmark{5},
Mark Gurwell\altaffilmark{6}, Anne L\"ahteenm\"aki\altaffilmark{7}, Merja Tornikoski\altaffilmark{7},
Alex Markowitz\altaffilmark{8}, Arkadi A. Arkharov\altaffilmark{9}, Dmitry A. Blinov\altaffilmark{2}, Ritaban Chatterjee\altaffilmark{1}, Francesca D. D'Arcangelo\altaffilmark{1}, Abe D. Falcone\altaffilmark{10}, Jos\'e L. G\'omez \altaffilmark{4},
Vladimir A. Hagen-Thorn\altaffilmark{2,3}, Brendan Jordan\altaffilmark{11}, 
Givi N. Kimeridze\altaffilmark{12}, Tatiana S. Konstantinova\altaffilmark{2}, 
Evgenia N. Kopatskaya\altaffilmark{2}, Omar Kurtanidze\altaffilmark{12},
Elena G. Larionova\altaffilmark{2}, Liudmilla V. Larionova\altaffilmark{2}, 
Ian M. McHardy\altaffilmark{13}, Daria A. Melnichuk\altaffilmark{2},
Mar Roca-Sogorb\altaffilmark{4}, Gary D. Schmidt \altaffilmark{5},
Brian Skiff\altaffilmark{14},  Brian Taylor\altaffilmark{1,14}, Clemens Thum\altaffilmark{15},
Ivan S. Troitsky\altaffilmark{2}, and Helmut Wiesemeyer\altaffilmark{16}}

\altaffiltext{1}{Institute for Astrophysical Research, Boston University, 725 Commonwealth Avenue, Boston, MA 02215}
\email{jorstad@bu.edu}
\altaffiltext{2}{Astronomical Institute, St. Petersburg State University, Universitetskij Pr. 28, Petrodvorets, 
198504 St. Petersburg, Russia}
\altaffiltext{3}{Isaac Newton Institute of Chile, St. Petersburg Branch, St. Petersburg, Russia}
\altaffiltext{4}{Instituto de Astrof\'{\i}sica de Andaluc\'{\i}a, CSIC, Apartado 3004, 18080,
Granada, Spain}
\altaffiltext{5}{Steward Observatory, University of Arizona, Tucson, AZ 85721-0065}
\altaffiltext{6}{Harvard-Smithsonian Center for Astrophysics, 60 Garden St., Cambridge, MA 02138}
\altaffiltext{7}{Mets\"ahovi Radio Observatory, Helsinki University of Technology TKK, Mets\"ahovintie 114, FIN-02540 Kylm\"al\"a, Finland}
\altaffiltext{8}{Center for Astrophysics and Space Sciences, University of California, San Diego, M.C. 0424, La Jolla, CA 92093-0424}
\altaffiltext{9}{Main (Pulkovo) Astronomical Observatory of RAS, Pulkovskoye shosse, 60, 196140,
St. Petersburg, Russia}
\altaffiltext{10}{Deptartment of Astronomy \& Astrophysics, Pennsylvania State University, 525 Davey Lab, University Park, PA 16802}
\altaffiltext{11}{School of Cosmic Physics, Dublin Institute of Advances Studies, Ireland}
\altaffiltext{12}{Abastumani Astrophysical Observatory, Mt. Kanobili, Abastumani, Georgia}
\altaffiltext{13}{Department of Physics and Astronomy, University of Southampton, Southampton, SO17 1BJ,
United Kingdom}
\altaffiltext{14}{Lowell Observatory, Flagstaff, AZ 86001}
\altaffiltext{15}{Institut de Radio Astronomie Millim\'etrique, 300 Rue de la Piscine, 38406 St. Martin d'H\`eres, France}
\altaffiltext{16}{Instituto de Radio Astronom\'{\i}a Milim\'etrica, Avenida Divina Pastora, 7, Local 20, E-18012 Granada, Spain}

\shorttitle{Flaring Behavior in 3C~454.3}
\shortauthors{Jorstad et al.}
\begin{abstract}
We analyze the behavior of the parsec-scale jet of the quasar 3C~454.3 
during pronounced flaring activity in 2005-2008. 
Three major disturbances propagated down the jet along different
trajectories with Lorentz factors $\Gamma>$10. The disturbances show a clear connection
with millimeter-wave outbursts, in 2005 May/June, 2007 July, and 2007 December. 
High-amplitude optical events in the $R$-band light curve precede peaks 
of the millimeter-wave outbursts by 15-50~days. 
Each optical outburst is accompanied by an increase in X-ray activity.
We associate the optical outbursts with propagation of the superluminal knots
and derive the location of sites of energy dissipation in the form of radiation.
The most prominent and long-lasting of these, in 2005 May, occurred closer to the 
black hole, while the outbursts with a shorter duration in 2005 Autumn and in 2007
might be connected with the passage of a disturbance 
through the millimeter-wave core of the jet. The optical outbursts, 
which coincide with the passage of superluminal radio knots through the core, 
are accompanied by systematic rotation of the position angle of optical linear 
polarization. Such rotation appears to be a common feature during the early 
stages of flares in blazars. We find correlations between optical variations and 
those at X-ray and $\gamma$-ray energies. We conclude that the emergence of 
a superluminal knot from the core yields a series of optical and high-energy 
outbursts, and that the mm-wave core lies at the end of the jet's acceleration 
and collimation zone.  We infer that the X-ray emission is produced 
via inverse Compton scattering by relativistic electrons of photons both
from within the jet (synchrotron self-Compton) and
external to the jet (external Compton, or EC); which one dominates depends
on the physical parameters of the jet. A broken power-law model of 
the $\gamma$-ray spectrum reflects a steepening of the synchrotron emission
spectrum from near-IR to soft UV wavelengths. We propose that the $\gamma$-ray 
emission is dominated by the EC mechanism, with
the sheath of the jet supplying seed photons for $\gamma$-ray 
events that occur near the mm-wave core. 

\end{abstract}

\section{Introduction}
During the past four years, the quasar 3C~454.3 (z=0.859) has displayed pronounced variability at all
wavelengths. In spring 2005 it returned to the night sky with unprecedented brightness, $R\sim$12.0~mag,
a level not seen at optical wavelengths over at least 50 years of observations \citep{VIL06}. An increase
in activity occurred at X-ray and radio wavelengths 
as well, with the 230 GHz radio variations having a delay of $\sim$2~months with respect to the optical
variability \citep{RAI08b}. This prominent outburst was followed by a more quiescent period 
at all wavebands from spring 2006 to spring 2007. During this interval, the optical spectrum possessed
characteristics typical of radio-quiet active galactic nuclei (AGN), such as a ``big blue bump'' and
``little blue bump,'' attributed to thermal emission from the accretion disk and broad emission lines
from surrounding clouds, respectively \citep{RAI07}. After the quiescent state, the quasar underwent a
new stage of high optical activity \citep{RAI08a} that continued to the end of 2008.  During this time span,
very bright $\gamma$-ray emission was detected \citep{VER08,TOSTI08}, with an excellent correlation
between the $\gamma$-ray and near-infrared/optical variations \citep{BON09}.

Models proposed to explain the observed variability and spectral energy distribution (SED) 
of 3C~454.3 across the electromagnetic spectrum involve processes
originating in the radio jet of the quasar. \citet{VIL07} suggest that the very high optical flux in
spring 2005 was connected with a disturbance (e.g., a shock) propagating along a curved
trajectory in the jet, with optical synchrotron photons emitted over a different volume than the
longer-wavelength radiation. As the emission zone of a given wavelength passes closest to the line of sight,
the flux peaks at that wavelength. This occurs first at optical and later at longer wavelengths. 
\citet{GHIS07} have found that the behavior of 3C~454.3 in 2005-2007 is consistent with the model
suggested by \citet{KG07}, in which the dissipation site of an
outburst depends on the bulk Lorentz factor and compactness of the  perturbation propagating
down the jet. Outbursts occurring closer to the black hole (BH) should have a more compact emitting region
with a lower bulk Lorentz factor, $\Gamma$, and a stronger magnetic field, $B$. Greater compactness of the
emission region intensifies the synchrotron flux as well as the high energy component 
produced via the synchrotron self-Compton (SSC)  mechanism,
while the external Compton (EC) high-energy component (inverse Compton radiation with seed photons from outside the jet) 
is suppressed owing to a weaker Doppler factor, $\delta$, resulting from the lower value of $\Gamma$. 
\citet{GHIS07} model the outburst
in 2005 as an event that occurred closer to the BH ($\Gamma\sim$8, $\delta\sim$13, $B\sim$15~G, and 
size of the emitting region $a\sim$5.5$\times$10$^{-3}$~pc) than the outburst in 2007 
($\Gamma\sim$16, $\delta\sim$16, $B\sim$9~G, and  $a\sim$8$\times$10$^{-3}$~pc). 
\citet{SMM08} argue that the optical, X-ray, and millimeter light curves during 
the outburst in 2005 require a release of a significant fraction of the jet energy when the
jet becomes transparent at millimeter wavelengths at the millimeter-wave
``photosphere.'' These authors conclude that this photosphere is located at $\sim$10~pc from the BH,
coinciding with the expected location of a torus of hot dust. \citet{SMM08} infer that the X-ray 
and $\gamma$-ray emission is most likely produced via the EC mechanism, with seed photons emitted by
the hot dust scattered by relativistic electrons in a plasma with bulk Lorentz factor $\Gamma\sim$20.
Interpretations of $\gamma$-ray observations with {\it AGILE} during autumn 2007 and the densely
sampled light curve provided by the {\it Fermi} Gamma-ray Space telescope starting in 2008 August,
combined with simultaneous observations at longer wavelengths, involve higher-energy 
electrons that emit synchrotron radiation at near-IR and optical wavelengths as well as scatter
external photons from the broad line region to energies up to $\sim$100~GeV \citep{VER09,BON09}.
   
\citet{J05} monitored the quasar 3C~454.3 at 43~GHz with the Very Long Baseline Array
(VLBA) bimonthly from March 1998 to April 2001, and determined parameters of the parsec-scale jet
during this period based on the apparent speed of superluminal knots (also referred to as ``components''
of the jet) and time scale of their flux variability. They found that the jet of 3C~454.3
during this time span had physical parameters as follows: 
$\Gamma$=15.6$\pm$2.2, $\Theta_\circ$=1.3$^\circ\pm$1.2$^\circ$,
$\delta$=24.6$\pm$4.5, and $\theta$=0.8$^\circ\pm$0.2$^\circ$, where $\Theta_\circ$ is the viewing angle
and $\theta$ is the opening half-angle of the jet. The Boston University group resumed VLBA monitoring
of the quasar in June 2005, and has continued to monitor the source within a program of roughly
monthly VLBA imaging
of bright $\gamma$-ray blazars at 43 GHz. In this paper, we analyze disturbances seen in the quasar jet during
the period of high optical activity in 2005-2008, and connect events 
in the jet with prominent variability at different wavebands. Observations cover the range
of frequencies from 10$^{10}$~GHz to 10$^{23}$~GHz, which provide significantly different
angular resolutions at which an event is observed  -
10-15~arcminutes at high energy frequencies, $\sim$5~arcseconds at millimeter and sub-millimeter wavelengths, $\sim$1~arcsecond in the optical and IR bands, and
$\sim$0.1~milliarcsecond (mas) with the VLBA at 43~GHz. This renders interpretation
of multifrequency behavior challenging, and makes analysis of variability at 
different wavelengths along with VLBI images the main tool for understanding 
processes and mechanisms involved in the physics of blazars. 
   
\section{Multifrequency Light Curves}
We have analyzed light curves of the quasar 3C~454.3 from $\gamma$-ray to radio wavelengths during the
period 2004-2009. We use relative Julian dates (RJD$\;\equiv\;JD-2450000$) to refer to the epochs of observation.

\subsection{Optical, Ultra-Violet, and near-Infrared Photometric data}

We use the optical light curve in $R$ band collected by the WEBT collaboration \citep{VIL06,VIL07,RAI08a,RAI08b}, which 
covers the period from 2003 December 28 to 2008 February 8 (RJD: 3001-4505). 
To this we add the data from the 2~m robotic
Liverpool telescope at the Observatorio del Roque de Los Muchachos (La Palma, Spain) and the 1.8~m 
Perkins telescope of Lowell Observatory (Flagstaff, AZ).
We have extended the light curve up to 2009 January 7 (RJD: 4839)
using the data obtained at the Perkins telescope, 70-cm telescope of the Crimean Astrophysical Observatory 
(Nauchnij, Ukraine), 40~cm telescope of St. Petersburg State University (St. Petersburg, Russia), 
2.2~m telescope of the Calar Alto Observatory (Spain), and 70~cm telescope of Abastumani Observatory
(Republic of Georgia). At some telescopes, the photometric data have been 
carried out not only in $R$ band but also in $B, V$, and $I$ filters, although less frequently than
in $R$ band. We use these data for broad-band spectral analysis. The data from 2008 June 23 
to December 10 (RJD: 4640-4810) are 
supplemented by measurements by the SMARTS consortium, posted
at their website\footnote{http://www.astro.yale.edu/smarts/glast/}. 

We have performed the $U$-band (3500 {\AA}) observations of the quasar with the 
Liverpool telescope from 2005 May 18 to December 10 (160 measurements) 
and reduced the data in a manner 
used in the WEBT collaboration \citep{RAI08a}. We have obtained archived UVOT Swift 
data from 2005 April 24 to 2009 January 27 in four filters, $U$ (3501 {\AA}),
$UW1$ (2634 {\AA}), $UM2$ (2231 {\AA}), and $UW2$ (2030 {\AA}) 
(86, 100, 89, and 95 
measurements, respectively). The UVOT data were reduced using the HEASOFT 6.5 package
and following the threads provided by the UVOT User's Guide and recommendations 
contained in the release notes with parameters similar to those used by \citet{RAI08a}.
We have adopted galactic extinction values derived by \citet{RAI08a}:
0.58 ($U$), 0.73 ($UW1$), 1.07 ($UM2$), and 1.02~mag ($UW2$).

We use infrared (IR) {\it JHK} data collected at the 1.1~m telescope of the Main (Pulkovo) Astronomical
Observatory of the Russian Academy of Sciences located at Campo Imperatore, Italy \citep{LAR08}. 
The values of Galactic extinction provided by the {\it NASA Extragalactic
Database} are adopted to correct the photometric estimates for Galactic absorption. We have applied the calibration 
of \citet{MEAD90} for all optical and IR measurements to transform magnitudes into flux densities. Figure~\ref{mainLC}
shows the optical light curve of the quasar in $R$ band, the UV  light curves
in $U$ and $UW$1 bands, and the near-IR  light curve in $K$ band from January 2004 to January 2009.

\subsection{Radio Light Curves}
The 230~GHz (1.3~mm) and 345~GHz (0.85~mm) light curves were obtained at the Submillimeter
Array (SMA), Mauna Kea, Hawaii from 2004 June 16 to 2009 January 31 (RJD: 3172-4862).
3C~454.3 is a bright quasar included in an ongoing monitoring program at the SMA to determine
the fluxes of compact extragalactic radio sources that can be used as calibrators at mm and sub-mm
wavelengths. Data from this program are 
updated regularly and are available at SMA website\footnote{http://sma1.sma.hawaii.edu/callist.html}.
Details of the observations and data reduction can be found in \citet{GUR07}.

The data at 230~GHz
are supplemented by measurements carried out at the 30~m telescope of Instituto de Radio Astronom\'{\i}a Milim\'etrica (IRAM, Granada, Spain). The IRAM 30~m Telescope observed simultaneously at 86.24~GHz (3.5~mm) 
and 228.93~GHz (1.3~mm) by making use of the A100/B100 and A230/B230 pairs of orthogonally linearly polarized heterodyne receivers, respectively. 
Every IRAM 30~m measurement was preceded by a cross-scan pointing of the telescope toward 3.5~mm and 1.3~mm calibration sources.
Such measurements consisted of a series of wobbler-switching on-offs with total integration times of 4~min to 8~min, depending on the total flux density of the source and atmospheric conditions.
Measurements of Mars and/or Uranus were obtained at least once per observing session in order 
to estimate and subtract residual instrumental polarization, and to calibrate the absolute total flux density scale. 
Whenever these planets were not visible, the compact H~II regions W3~OH, K3-50A, and NGC~7538, and/or the compact planetary nebula NGC~7027 were observed for calibration.  
The initial calibration of the amplitude was performed through the telescope's online data processing procedures within the MIRA-GILDAS software\footnote{http://www.iram.fr/IRAMFR/GILDAS/doc/html/mira$-$html}.
The remaining data reduction involved the removal of outliers, 
an elevation dependent calibration, and the calibration of the absolute flux density scale as described 
in \citet{Agu06}. The final flux density error of every measurement also included an 
additional 5\% non-systematic uncertainty added in quadrature. Finally, the resulting data were 
averaged for those observing epochs on which more than one measurement was obtained.

The 37~GHz (8 mm) observations were performed with the 13.7~m telescope at Mets\"ahovi Radio Observatory of 
Helsinki University of Technology, Finland. The flux density calibration is based on observations of DR~21,
with 3C~84 and 3C~274 used as secondary calibrators. A detailed description of the data reduction and 
analysis is given in \citet{FIN98}. The radio light curves are plotted in Figure~\ref{mainLC}. 

\subsection{High-Energy Light Curves}
We obtained 63 measurements of 3C~454.3 with the
{\it Rossi X-ray Timing Explorer} ({\it RXTE})
from 2005 May 11 to September 5 (RJD:3502-3619).
Each {\it RXTE} visit lasted $\sim$1 ksec.
We used data taken with {\it RXTE}'s Proportional Counter Array (PCA),
which is collimated to have a FWHM 1$\degr$ field of view.
However, there is a cataclysmic variable star, IM Peg, located about 0.72$\degr$
to the NW of 3C~454.3 \citep{Per97}.
To completely eliminate contributions to the observed
spectrum of 3C~454.3 from IM Peg, we chose an {\it RXTE}
on-axis pointing position 0$\fdg$52 to the SE of
3C~454.3; at this offset position, the PCA collimator efficiency
is 45$\%$ \citep{Jahoda06}; we corrected for this when
generating the PCA response matrices (see below).
Reduction of the PCA data followed standard extraction and
screening procedures, using HEASOFT version 6.0
software\footnote{http://heasarc.gsfc.nasa.gov/ftools/}.
PCA STANDARD-2 data were collected from Proportional
Counter Unit (PCU) 2 only. We use the ``L7-240''
background models, appropriate for faint sources.
Response matrices were generated for each spectrum
separately using the FTOOLS program PCARSP. We modeled the 3--15 keV
spectrum in XSPEC v.11 with a single power-law and Galactic
absorption corresponding to a hydrogen column density
$N_{\rm H}$ = 7.2$\times$10$^{20}$~cm$^{-2}$ \citep{Elvis89}, and
calculated the light curve of X-ray flux density at 4 keV.

For the period from 2005 May 10 to 2009 January 27 (RJD: 3501-4858) we used publicly available data from
the {\it Swift} satellite\footnote{http://www.swift.psu.edu/monitoring/} that were carried out with the
X-Ray Telescope at 0.3-10~keV (95 measurements) and processed through the XRT pipeline. We checked 
the data: the PC mode to WT mode transitions are well handeled, the rates are self-consistent, implying
that there are no problems with the point-spread-function correction factors; the uncertainties 
are $\lesssim$2\%.   

We obtained Photon and Spacecraft data from the Fermi Science Support Center (FSSC) 
for observations with the Large Area Telescope (LAT) from 2008 August 5 to 2009 February 10 
(RJD: 4684-4873) within a 15$^\circ$ radius centered on the quasar. 
We then calculated the $\gamma$-ray light curve of 3C~454.3  
at 0.1-300~GeV with daily binning using the software and following the Analysis Threads 
provided by the FSSC. The procedure included
selection of good data and time (programs: {\it gtselect, gtmktime}), construction of the exposure map for each day
({\it gtltcube, gtexpmap}), and modeling of the data by a single power law, optimized via a maximum-likelihood method ({\it gtlike}). We used the
response function generated by the FSSC, and created a  model file
that consisted of a power-law  model (prefactor and index) for 3C~454.3 and three other bright $\gamma$-ray sources in the field: PKS~2201+171,
CTA~102, and PKS~2325+093 \citep{Abdo09a}, and diffuse emission models for Galactic and extragalactic background provided by the FSSC. 
We consider the quasar to be detected if the test-statistic, $TS$,
calculated by the {\it gtlike} procedure exceeds 10, corresponding to at least a 3$\sigma$ detection level \citep{Abdo09a}. For such measurements we derived $\gamma$-ray flux estimates with 
uncertainties within 10-15\%. We have supplimented 
these $\gamma$-ray data by the measurements from RJD: 4649-4675 taken from Figure 1 of \citet{Abdo09b}, which are not provided by the LAT Data Server.  
Figure~\ref{OptHE} shows the high-energy light curves for three time intervals when
the X-ray and $\gamma$-ray measurements were available along with the
optical light curve. 
 
\section{Polarization Observations}

We have obtained linear polarization measurements of 3C~454.3 at optical and mm wavelengths.
All of these data were checked for consistency and corrected for statistical bias \citep{WK74}. 

\subsection{Optical Polarization Data} 

Our optical polarization monitoring in $R$ band of $\gamma$-ray blazars, including 3C~454.3, began
in 2005 May at the 70~cm telescope of the Crimean Astrophysical Observatory. The telescope is equipped 
with an ST-7 based photometer-polarimeter. The details of the observation and data reduction 
can be found in \citet{LAR08}. 

Since 2005 September, we have obtained polarimetric observations at the Perkins telescope of Lowell
Observatory (Flagstaff, AZ) with the PRISM camera\footnote{http://www.bu.edu/prism/} supplemented by
a polarimeter with a rotating half-wave plate. These involve a series of 5-7 
Stokes Q and U measurements for a given object. Each series
consists of four measurements at instrumental position angles 0, 45, 90, and 135$^\circ$
of the waveplate.
Since the camera has a wide field of view ($14^\prime\times14^\prime$), we use field stars
to perform both interstellar and instrumental polarization corrections. We use  
unpolarized calibration stars from \citet{SCHMIDT92} to check the instrumental polarization,
which is usually within 0.3\%-0.5\%, and polarized stars from the same paper to calibrate
the polarization position angle. 

In 2006 November the 40~cm telescope of St. Petersburg State University joined 
the program. The telescope is equipped with a nearly identical photometer-polarimeter
as that of the 70~cm telescope in Crimea, and the observations and data reduction are carried out 
in the
same manner \citep{LAR08}. However, the polarization observations are performed without a
filter, with central wavelength $\lambda_{\rm eff}\sim 670$~nm.

Starting in 2007 July, the MAPCAT (Monitoring AGN with Polarimetry at the Calar Alto
Telescopes)\footnote{http://www.iaa.es/$\sim$iagudo/research/MAPCAT/MAPCAT.html} program has made use of the Calar Alto 
Faint Object Spectrograph (CAFOS) in its imaging polarimetric mode on the 2.2~m Telescope at the Calar Alto Observatory (Almer\'ia, Spain).
Every $R$-band polarization measurement consisted of four imaging exposures of $\sim$50~s 
at 0$^{\circ}$, 22.5$^{\circ}$, 45$^{\circ}$,  and 67.5$^{\circ}$ of the $\lambda/2$ plate in 
CAFOS.
Polarization measurements of at least two polarized standard stars from \citet{SCHMIDT92} 
were performed to estimate the instrumental polarization, which is $\sim$0.2\%. 
The degree ($p$) and position angle ($\varphi$) of linear
polarization were obtained from their relation with the 8 flux measurements, as described in \citet{Zap05}.

We have obtained polarization data at Steward Observatory with the 1.54~m Kuiper 
during a 2005 campaign from October 25 to November 4 
\citep[details given in][]{FRANI09}, and within another, currently operating program at
that telescope and the 2.3~m Bok Telescope
to provide optical polarization measurements for bright $\gamma$-ray blazars 
from the {\it Fermi}
LAT-monitored blazar list\footnote{http://james.as.arizona.edu/$\sim$psmith/Fermi}.
The combined optical polarization data are presented in Figure \ref{Poldat}. 

\subsection{Millimeter-Wave Polarization Data}
The IRAM 30~m Telescope also performed linear and circular polarization at 86.24~GHz.
The A100/B100 pair of orthogonally linearly polarized heterodyne receivers is connected to the 
XPOL polarimeter \citep{Thu08}, which provides 500~MHz bandwidth per receiver. We have carried out
the initial calibration of the phase through the telescope's online data processing procedures 
within the MIRA-GILDAS software. The subsequent polarimetric data reduction follows that of 
\citet{Thu08} and \citet{Agu09}. These procedures involve the removal of outliers and 
of systematic residual instrumental polarization still present in the data, calibration of the absolute 
flux density scale, estimation of remaining non-systematic or statistical uncertainties still 
affecting the data, and quadratic addition of the latter to the final uncertainties of the measurements. 
Such final errors are $\sim0.5$\,\% in Stokes $Q$ and $U$ parameters, and $\sim0.3$\,\% in
Stokes $V$ parameter. The data are plotted in Figure \ref{Poldat}. 

\section{VLBA Observations}

We observed 3C~454.3 in the course of a program
of monthly monitoring of bright $\gamma$-ray blazars with the VLBA at 43~GHz
(7~mm)\footnote{http://www.bu.edu/blazars/VLBAproject.html}. 
During the period from June 2005 to May 2009, we obtained 44 total
and polarized intensity images of the quasar at a resolution of 
$\sim$0.3$\times$0.1~milliarcseconds (mas). Figure~\ref{VLBAyr} presents a sequence of 
total intensity images, with several images per year convolved with the same beam and with contours
based on the same global map peak.  
We performed the data reduction in the manner of \citet{J05} using the Astronomical Image Processing
System (AIPS) and Difmap \citep{DIF97}.
In addition to the standard resolution beam of uniform weighting of the {\it uv}-data,
we have convolved the images with  a
beam of FWHM diameter of 0.1~mas comparable to the resolution of the longest
baselines of the VLBA. Figure ~\ref{image7mm} shows the morphology of the parsec-scale 
jet of 3C~454.3, convolved with a standard beam along with an image with the smaller beam 
that features knots $K1$, $K2$, $K3$, and $C$ found in the jet (see \S 5). 
Although the synthesized beam of the VLBA observations along the jet is $\sim$0.1~mas, we have
convolved the core region of the jet with a circular beam with a FWHM diameter of 0.05~mas
for epochs when all antennas were in operation and the weather was favorable. Since this is only
30\% smaller than the FWHM angular resolution of the longest baseline, 0.07~mas, the array partially
resolves structures on 0.05~mas scales.
At the majority of suitable epochs, the resulting super-resolved map of the core region consists of
three components, as shown in Figure~\ref{hCore}. Such a structure is especially
apparent when the core is not disturbed by passage of a moving knot ({\it bottom panel}).
To support the conclusion that this fine structure
in the core is real, we have constructed similar super-resolved images of the quasar CTA102, 
obtained with the VLBA along with 3C~454.3. Since CTA102 is close to 3C~454.3 in the sky, the
observations have similar $uv$-coverage. The resulting images of CTA102 contain only single-component
cores. This suggests that the fine structure of the core  
of 3C~454.3 is not an artifact of the smaller beam.  

Since VLBA data yield only the relative orientation of the electric vector position angle
(EVPA) across a source, special attention was paid to the calibration of the absolute EVPA values.
We employed data from the NRAO VLA/VLBA
Polarization Calibration Page\footnote{http://www.vla.nrao.edu/astro/calib/polar/}, which
provides EVPA measurements with the VLA at 43~GHz, integrated over the entire source for several objects
from our sample observed with the VLBA (OJ287, 1156+295, 3C~273, 3C~279, BL~Lac,
and 3C~454.3). For other epochs without contemporaneous VLA observations,
we used the {\it D-terms} method discribed by \citet{Dterm}. The accuracy of the absolute EVPA 
calibration is within 4-$8^\circ$.
\citet{J07} provide an estimate of the Faraday rotation measure ($RM$) of $(-6.8\pm3.7)\times 10^3$~rad~m$^{-2}$
in the mm-VLBI core of 3C~454.3, corresponding to a correction for the 
EVPA at 43~GHz of $-19^\circ\pm 10^\circ$. Although the estimate of $RM$ has a large uncertainty
and is not simultaneous with the VLBA observations reported in this paper, we find that applying 
the $RM$ correction at 43~GHz improves the similarity between the behavior of EVPAs at
86~GHz and 43~GHz. Figure \ref{Poldat} shows the results of our measurements of degree 
($p_{\rm 43}$) and position angle ($\varphi_{\rm 43}$) of polarization in the core 
region of the quasar.

\subsection {Modeling of Images}

We have modeled our calibrated $uv$-data with point source brightness distributions
using the {\it Difmap} task MODELFIT. The number of point-like components required to fit the data was 
determined by the best agreement between the model and data according to $\chi^2$ values, 
with reduced $\chi^2$ ranging from 1.0 to 5. Since we have roughly monthly
observations with the VLBA, the model yielding the best agreement at a given
epoch was used as an input model for the following epoch in order to maximize consistency
across epochs given the complexity of the brightness distribution. We then performed 100 iterations
to adjust the input model to the $uv$-data of the second epoch, editing the model if deleting or
adding components was required to represent the image. We assumed that the core is a stationary
component at relative right ascension and declination (0,0), and then determined for each component 
the total and polarized flux intensity, $S_{\rm 43}$ and $S_{\rm 43}^{\rm p}$, respectively, 
distance from the core, $R$,  position angle relative to the core, $\Theta$,
degree of polarization, $p_{\rm 43}$, and position angle of polarization, $\varphi_{\rm 43}$.
We have identified components across the epochs in the manner described in \citet{J05}
and used their method, as well as the same cosmological parameters
($\Omega_{\rm m}=0.3$, $\Omega_\Lambda=0.7$, and $H_\circ$=70~km~s$^{-1}$~Mpc$^{-1}$),
to calculate the proper motions and apparent speeds of moving features.
Figure~\ref{mainLC} shows the light curve
of the VLBI core, while Figure~\ref{evol7} presents the positions of all components  
(except the core) brighter than 50~mJy found within 1~mas of the core during the period
2005 June to 2009 May (RJD: 3547-4982).

Our analysis of the VLBA images concentrates on the innermost
region of the jet. Figure~\ref{jetdir} shows the projected jet direction, $\Theta_{\rm jet}$, 
determined at each epoch according to the position angle of the brightest component 
within 0.1-0.3~mas of the core. The average jet direction is 
$\bigl<\Theta_{jet}\bigr>$=-95$^\circ\pm$8$^\circ$. Figure~\ref{jetdir} shows that
there are significant deviations of $\Theta_{\rm jet}$ from the average, which might reflect 
intrinsic changes of the jet direction or shifts in the brightness distribution across the
width of the jet.

\section{Disturbances in the Inner Jet}
Analysis of the images reveals three features moving with respect to the core
for which the apparent speed, $\beta_{\rm app}$,  and time of ejection (coincidence with the core), $T_\circ$,
can be estimated. We have designated the features as components $K1, K2,$ and $K3$ (Fig.~\ref{evol7}).
The majority of images show a brightness enhancement at 0.6-0.7~mas from the core,
which is most likely associated with component $C$ reported by \citet{J05} 
as a stationary knot at position $R\sim0.63$~mas. 
Knots $K1$ and $K2$ move ballistically only within 0.2~mas of
the core. Beyond 0.2~mas, $K1$ accelerates and fades, while $K2$ decelerates and brightens.
Figure~\ref{trajKs} shows light curves and trajectories of components.
Although the separation of $K3$ from the core is at the limit of our resolution, modeling requires the presence
of two components in the core region at all epochs starting from 2008 June, with  
a persistent increase in separation between the components at later epochs. Figures~\ref{imgK1},
\ref{imgK2}, and \ref{imgK3} show sequences of total and polarized intensity images
displaying the jet evolution that we attribute to the emergence
of components $K1$, $K2$, and $K3$, respectively. Knots $K1$ and $K2$ are distinct 
on the high-resolution total intensity maps and have better alignment of the position angle
of the polarization with the jet direction than does the polarization of the core. The brightest knot, $K3$,
fades dramatically while it is still very close to the core.

Since we are interested mainly in the time of ejection of the components and their speed in the vicinity of 
the core, we have used only those epochs when the components moved ballistically (the first 10 and 8 epochs
for $K1$ and $K2$, respectively, and all 11 epochs for $K3$) for deriving kinematic parameters. The parameters and their uncertainties were obtained in the manner 
described in \citet{J05}. 

We modeled components by point sources because
such an approach gives better consistency between epochs than, for example,
modeling by components with circular Gaussian brightness distributions. This suggests
that the knots are very compact, with angular size smaller than half the beam axis
along the jet, $\lesssim$0.07~mas. We have adopted a FWHM size of 0.05~mas for all
three components to estimate Doppler and Lorentz factors
and viewing angle independently by using the method developed for high-frequency
VLBI monitoring \citep{J05}. Table \ref{Kparm} lists for each knot the time of ejection, $T_\circ$, 
the highest flux 
density measured, $S^{\rm max}$, the epoch of $S^{\rm max}$, proper motion, $\mu$, 
apparent speed, $\beta_{\rm app}$, Lorentz, $\Gamma$, and Doppler, $\delta$, factors,
and viewing angle, $\Theta_\circ$.  There is a significant difference in the viewing angles
of components $K2$ and $K3$, and the Lorentz and Doppler factors
of $K3$ are twice those of $K1$ and $K2$, as well as twice those derived
from the parameters of several components detected in 3C~454.3 between 1998-2001 \citep{J05}.
Figure~\ref{jetdir} shows that the most dramatic
change in the projected jet direction occurred after ejection of component $K3$. If we assume that 
the average jet parameters derived by \citet{J05} give the most probable 
values of viewing and opening angles ($\Theta_\circ$=1.3$^\circ$ and $\theta$=0.8$^\circ$, respectively) 
then component $K3$ with $\Theta_\circ$=0.2$^\circ$ should occupy the side of the jet
closest to the line of sight, while component $K2$ with $\Theta_\circ$=2.5$^\circ$
should move along the far side of the jet. 
Figure~\ref{jetdir} supports such an interpretation, since,
after the ejection of component $K2$, the projected jet direction swings to the north, while 
after the ejection of $K3$, $\Theta_{\rm jet}$ turns to the south. Component $K1$ appears to follow
a path closer
to the jet axis. In addition, Figure~\ref{jetdir} shows a sharp swing of $\Theta_{\rm jet}$
near RJD$\sim$3750-3800, possibly indicating the emergence of a new knot
from the core that we cannot distinguish from $K1$, which is bright and close to the core
at this time (see Fig.~\ref{evol7}).

\section{Correlation Analysis between High Energy Variations and Light Curves at Optical and mm Wavelengths}

The X-ray observations were carried out less uniformly than observations at other 
wavelengths. 
There are three intervals when X-ray observations were obtained: I - RJD: 3502-3619,
II - RJD: 4292-4450, and III - RJD: 4613-4858 (see Fig.~\ref{OptHE}).
We have performed correlation analysis between the X-ray, optical, and 230~GHz light curves
separately for these intervals because (i) there are significant gaps in X-ray coverage
between the intervals,
(ii) X-ray measurements at different intervals are obtained with different instruments,
and (iii) correlation between X-ray and longer wavelength light curves can change with
time \citep{RITABAN08}. For interval III, we include the {\it Fermi} $\gamma$-ray light curve in
the correlation analysis. All light curves are grouped into 1-day bins.
We employ the code developed for correlation analysis \citep{RITABAN08}, which
calculates the discrete cross-correlation function, CCF \citep{EK88}, 
finds the peak of the CCF, $f_{max}$, and determines the position of the centroid near the peak
\citep{WP94}. This yields an estimate of the significance of the peak as well as the delay,
$\tau$, between variations at different wavelengths if the peak is significant at a
confidence level 0.05.
 
Interval I has the best X-ray data sampling, provided by
RXTE monitoring. We have correlated this X-ray light curve 
with the optical and 1~mm light curves over exactly the same period,
while performing the correlation between the R-band and 230~GHz light curves over
a longer period, RJD: 3499-3750, which has similar sampling at the two wavelengths.  
Figure \ref{ccf_p1}~({\it left panel}) gives the results of the CCF analysis, revealing very strong
correlation between variations at X-ray and longer wavelengths.
During interval II, there are only 21 measurements obtained 
with {\it Swift}, insufficient for correlation analysis.
We compute the CCF only between the optical and 230~GHz light curves for the period RJD: 4200-4505.  Figure \ref{ccf_p1}~({\it right panel}) shows a moderate correlation
between variations at the two wavelengths.
During interval III, the X-ray ({\it Swift}), optical, and 230~GHz light curves have good 
coverage, with 72, 137, and 115 points, respectively; daily $\gamma$-ray fluxes became 
available starting at RJD: 4684. We have added  $\gamma$-ray
measurements from RJD: 4649-4675 presented in Figure 1 of \citet{Abdo09b} to the
$\gamma$-ray light curve for correlation analysis between $\gamma$-ray and
optical variations. In the case of the X/$\gamma$-ray correlation, the $\gamma$-ray
data during RJD: 4649-4675 are not used due to the absence of X-ray measurements 
during this period (see Fig.~\ref{OptHE}). 
Figure \ref{ccf_p3}~({\it left panel}) presents the cross-correlation
functions between X-ray, optical, and 230~GHz light curves, while
the right panel shows the CCF between the $\gamma$-rays and lower frequencies. 
Table \ref{CCFparm} summarizes the results of the correlation
analysis for all intervals and wavelengths.

Table \ref{CCFparm} shows a strong correlation between optical
and 230~GHz variations with a delay of mm-wave emission with respect to the
optical variations. Time lags between the optical and 1~mm light curves  
decrease significantly, from $\sim$50~days to $\sim$10~days,
from interval I to III, respectively. \citet{RAI08a} have reported a change 
in delay between the optical and 230~GHz
light curves of the quasar during 2007-08 relative to 2005-07, 
consistent with the difference in $\tau_{\rm opt/230}$ 
that we find between intervals I and II. During interval III, the delay became even shorter. Analysis of the X-ray/optical correlation suggests that variations
at the two wavelengths are simultaneous within 1~day. However, a 
change from positive (optical variations lead) to negative (X-rays lead) delay might
have taken place from interval I to interval III, although we judge this to be of 
low significance given the uncertainties of the delays. 
Nevertheless, such a possibility is supported by the X-ray/230~GHz delays as well:
during interval I, the X-ray/230~GHz delay is shorter than the delay
between the optical and 230~GHz light curves; the opposite
situation ($|\tau_{\rm opt/230}|>|\tau_{\rm x/230}|$) is observed for interval III, 
but again the differences are within the uncertainties of $\tau$.

Interval III is especially interesting, since it contains the {\it Fermi} $\gamma$-ray light curve. 
The $\gamma$-ray variations show the best correlation with the optical light curve, with
no delay exceeding 1~day (Fig.~\ref{ccf_p3}, {\it right panel}), in agreement with the findings of \citet{BON09}.
However, these authors reported the absence of
a correlation between $\gamma$- and X-rays, while our analysis indicates the presence 
of such a correlation, although the peak of the CCF is less prominent than that
between the $\gamma$-ray and optical light curves. 
From this analysis, we conclude that the $\gamma$/X-ray variations are
simultaneous within 2 days. Our X- and $\gamma$-ray light curves 
include additional data after RJD: 4750, which probably explains the discrepancy
between our correlation result and that obtained by \citet{BON09}.

We have performed correlation analysis between the 230~GHz and 37~GHz light curves
using all data from RJD: 3000 to RJD: 4850 (see Fig.~\ref{mainLC}). The CCF 
between the two wavelengths presented in Figure \ref{ccf_rad} shows a global
maximum with a delay  $\tau_{\rm 230/37}$=$-$215$\pm$30~days and two local maxima 
at $\tau_{\rm 230/37}$=$-$73$\pm$30~days and $\tau_{\rm 230/37}$=0$\pm$10~days. 
This is connected with a change of the delay between the variations at the
two wavelengths with time. Indeed, the CCF for the period from RJD: 4000 to 
RJD: 4850, which excludes the largest mm-wave outburst, 
gives a well defined peak at $\tau_{\rm 230/37}$=0$\pm$7~days
(Fig. \ref{ccf_rad}, {\it dotted line}). The change of the delay between 
the 230~GHz and 37~GHz light curves from $\sim$200~days to 0~days 
reflects the transition of mm-wave emission of the source from optically thick 
to optically thin (see \S 8), which also leads to a decrease of the delay between
optical and mm-wave variations.

\section{Comparison of Timing of Flux Variability and Ejection of Superluminal Components}

We decompose the light curves at optical and radio wavelengths into individual flares with exponential rise
and decay by applying the method of \citet{val99} \citep[see also][]{RITABAN08}. The $R$-band light
curve is smoothed with 1-day binning, while the mm-wave curves are smoothed with 5-day binning.
The method gives the epoch of peak flux, 
$T_\nu$, output power, $E_\nu$ (area under the curve fitting a flare), and width, 
$w_\nu$ (average of the rise and decay times). The decomposition yields 10 outbursts in
the $R$-band light curve and five outbursts in the mm-wave curves, each with $w_\nu>$10~days (Table \ref{Outparm}). 
We consider these to be the most reliable flares, based on the sampling of the data. 
We have cross-identified outbursts at different wavebands by assuming that events
associated with each other should have a difference in times of the peaks corresponding
to the delay found from the correlation analysis of light curves at the two wavelenths, i.e. 
$T_{\nu_1}-T_{\nu_2}\approx\tau_{\nu_1,\nu_2}$ (see \S 6). 
We also analyzed relative values of the parameters $E_\nu$ and $w_\nu$ to find 
additional support for the cross-identification. This implies,
for example, comparison of $E_{\rm opt}/E_{\rm opt}^{\rm max}$ with $E_{\rm 230}/E_{\rm 230}^{\rm max}$, where $E_{\rm opt}^{\rm max}$ and $E_{\rm 230}^{\rm max}$ are
the maximum output powers derived among optical and mm-wave outbursts, respectively.
The results of cross-identification are shown in Table \ref{Outparm}. We have 
associated the optical outbursts peaked at RJD: 3501, RJD: 4301, and RJD: 4437 with 
the 230~GHz outbursts with maxima at RJD: 3534, RJD: 4330, and RJD: 4454, respectively,
because differencies in the peaks ($-$33, $-$29, and $-$17~days) agree very well with
the delays found between the two wavelengths (Table \ref{CCFparm}).
The rest of the optical flares do not have counterparts at mm-wavelengths. All
230~GHz outbursts have corresponding 37~GHz outbursts, based on
the results of the correlation analysis.   

We have also related the mm-wave outbursts to the ejection of superluminal knots,
based on the assumption that an outburst and an ejection are connected if 
the latter occurs within the width of the outburst. Table~\ref{Delparm} gives the delays between 
the time of the ejection and the peak of the corresponding outburst at different wavelengths,
$\Delta T_\nu$ = $T_\nu-T_\circ$. It shows that each ejection occurred within 20$-$30~days
of the peak of a 230~GHz outburst. We denote the 230~GHz outbursts as $M_{K1}$, $M_{K2}$, 
and $M_{K3}$ in accordance with the designation of related superluminal component (Fig.~\ref{mainLC}).
We then associate a newly appearing superluminal knot with an optical outburst 
if the latter is related to a mm-wave outburst. As a result, the three major 
optical outbursts with the highest power are associated with newly emerging 
superluminal components. We denote these optical outbursts as $O_{K1}$, $O_{K2}$, and $O_{K3}$ in a similar manner as the mm-wave outbursts connected
with the ejections (see Fig.~\ref{mainLC} and Table~\ref{Outparm}).

Analysis of Tables~\ref{Outparm} and \ref{Delparm} reveals that (i) all ejections trigger
an increase in optical and mm-wave activity;  (ii) the connection
between optical outbursts and ejections of superluminal components is not one to one, since
either one ejection can trigger multiple optical oubursts or there are optical outbursts 
that are not connected with a new knot; and 
(iii) all ejections occur during the rising branch of a 37~GHz outburst, as
previously reported by \citet{SAV02}. 
In the case of the $K1$ event there is a pattern that the strongest optical outburst 
is followed by the 230 GHz outburst and by the ejection of a superluminal knot from 
the 43~GHz core. The delays between the peaks of the outbursts and time of ejection
of the knot can be interpreted in terms of spatial separation of sites of the outbursts
with respect to the core.  In the case 
of events $K2$ and $K3$, such an interpretation is possible as well, taking 
into consideration uncertainties in $\Delta T_\nu$. However, another interpretation,
which better corresponds to the derived delays and reflects the
differences in properties of $O_{\rm K2}$ and $O_{\rm K3}$ with respect to
$O_{\rm K1}$,  suggests that  
the sites of both the optical and 230~GHz outbursts are located in the mm-wave core
and that the outbursts occurred during the ejection of a superluminal knot.  
A delay between the peaks of associated optical and 230~GHz outbursts can be caused 
by the difference in time when the maximum number of relativistic electrons radiating 
at optical and mm-waves was achieved: the time of the maximum number of 
relativistic electrons accelerated to optical-emitting energies
coincides with the ejection time, while acceleration of electrons to mm-wave-emitting 
energies continues for 20-30~days longer. 

The extensive optical data collected by the WEBT collaboration \citep{RAI08a} 
contain a number of
epochs with several measurements during a single date. These authors have noted strong
intraday variability in 3C~454.3 in the optical $R$ band on a time scale of 
$t_{\rm var}\sim$1.5~hr in December 2007. We have calculated the means 
and standard deviations for $R$ 
magnitudes for each Julian date with 7 or more measurements. Figure~\ref{Rsigma} 
presents the standard deviations, $\Sigma R$, vs.\ epoch, along with the average
uncertainty for individual measurements on a given date, $\sigma R$ . 
There is possible intraday variability during the ejection of component $K1$, and pronounced
intraday variability during the emergence of knot $K3$. According to our estimations,
component $K3$ has the highest Doppler factor that we observed in the jet of 3C~454.3 from
2005 to 2009. This supports the idea that the knot is responsible 
for the short time scale of the variability seen at optical wavelengths.
The angular size of the region that can prodice such a variability,
$a_{\rm var}\lesssim c\Delta t_{\rm var}(1+z)/(\delta D_{\rm A})\approx 10^{-9}$~arcseconds,
where $D_{\rm A}$ is the angular distance to 3C~454.3. The size of 
$K3$ that we adopt from the VLBA imaging is $\sim$10$^4$ larger 
than $a_{\rm var}$. However, analysis of polarization properties in quasars
\citep{J07,FRANI07} shows that the  polarized optical emission, although partially 
co-spatial with the mm-wave core, occupies a volume that is thousands of times less than 
does the polarized emission at 43~GHz. Therefore, it is possible
that the site of an optical flare lies within a superluminal knot passing through the core,
and is confined to such a small region in the knot that it can vary on extremely short timescales. 

\section{Spectral Behavior} 

\begin{figure}
\epsscale{1.0}
\plotone{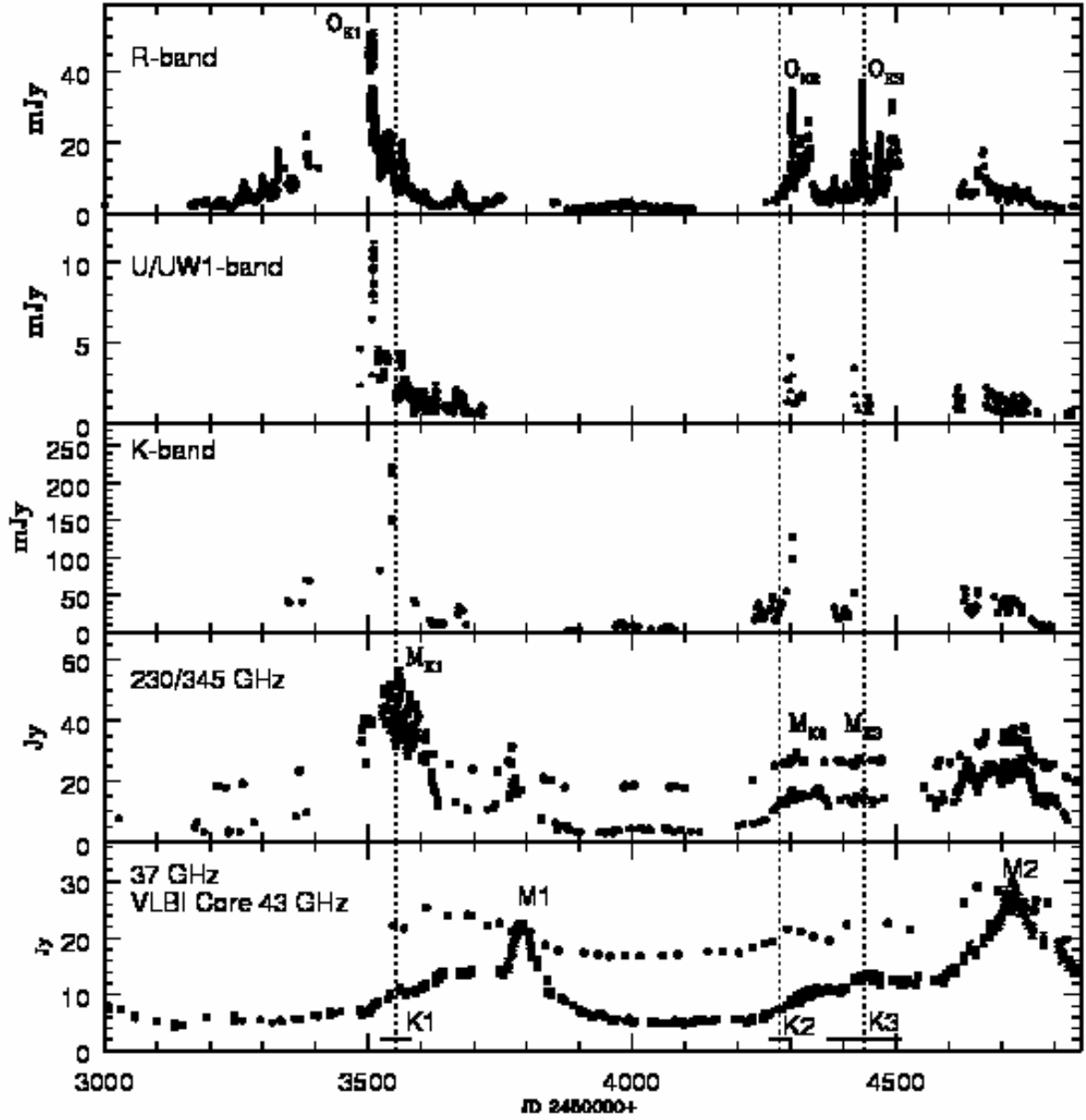}
\caption{Light curves of the quasar 3C~454.3 at different frequencies.
Dotted lines show times of ejections (times of coincidence with the position of the VLBI core at 43~GHz)
of superluminal components $K1$, $K2$, and $K3$
and solid line segments show uncertainties in the ejection times; {\it second panel:}
open circles -- Liverpool telescope data in $U$-band, open triangles --Swift UVOT data
in $U$-band, and filled triangles --  Swift UVOT data in $UW1$-band; 
345~GHz (open circles, {\it fourth panel}) and 43~GHz measurements 
(open circles, {\it fifth panel}) are shifted by +15~Jy for clearity.
Symbols designate the most prominent outbursts at different wavelengths.} \label{mainLC}
\end{figure}

We use 340~GHz, 230~GHz, 86~GHz, and 37~GHz data to calculate mm-wave spectral indices, $\alpha_{\rm mm}$
($S_\nu\propto\nu^{-\alpha}$). The spectral index for a given Julian date was calculated
if there were measurements at a minimum of three wavelengths within 2 days of
the date. When there were several measurements over a short time period, the observations nearest to each 
other were used. Figure~\ref{RadInd} reveals that the mm-wave spectral index exhibited significant variability.  
The uncertainty in $\alpha_{\rm mm}$ reflects the scatter about the best-fit power law.
During a quiescent state $\alpha_{\rm mm}$ is typical of a compact,
flat-spectrum radio source, $\alpha_{\rm mm}^q=$0.18$\pm$0.04, where $\alpha_{\rm mm}^q$ denotes a quiescent state 
chosen during the period RJD: 3900-4200 (Fig.~\ref{mainLC}) and denoted by $QS$.
The spectrum became strongly inverted during outburst $O_{K1}$ and slightly inverted during outburst $O_{K2}$.
The spectral index reaches a local minimum at epochs of superluminal ejections. After an ejection, $\alpha_{\rm mm}$ changes
sharply, rising to the value of $\alpha_{\rm mm}^{\rm q}$, and then continuing to increase 
up to a value $\alpha_{\rm mm} = 0.5$-0.6, signifying that the emission is close to being optically
thin. The epochs of the two global maxima of $\alpha_{\rm mm}$ coincide with the peaks of the sharp mm-wave
outbursts at RJD: 3781 ($M1$) and RJD: 4705 ($M2$), which are identified  neither  
with optical outbursts nor with ejections of superluminal components (Table~\ref{Outparm}, Fig~\ref{mainLC}).
The peaks of $M1$ and $M2$ do, however, coincide with the epochs of
maximum 43~GHz flux of knots $K1$ and $K3$, respectively (Table~\ref{Kparm}).
 
\citet{RAI07,RAI08a,RAI08b} plot the spectral energy distributions (SEDs) of the quasar at different
brightness levels. These authors interpret the behavior of the optical-UV part of the spectrum
as a result of superposition of beamed jet emission, blended line emission
from the broad line region (little blue bump), and emission from the accretion disk (big blue bump),
which agrees with previous findings by \citet{SMITH88}.
The compound SED explains the flattening of the spectrum with a decrease in brightness,
since at a low emission state the contribution of the blue bumps becomes more pronounced.
We use the method suggested by \citet{HT97} to determine the
spectral characteristics of a component of the source responsible for flux variability
observed on timescales of days and weeks. Such a component  most likely originates
via the synchrotron mechanism and has a power-law spectral energy distribution
from IR to UV wavelengths. The method allows one 
to determine spectral indices of the synchrotron emission separately from the
thermal emission, which varies on much longer timescales. The detailed description of the method and its application to multifrequency optical and near-IR data for 
the event $O_{\rm K1}$ are given in \citet{HT09}. We have performed a similar analysis 
for other events. Table~\ref{SpecInd}
gives the near-IR and optical spectral indices of synchrotron components during outbursts $O_{K1}$, 
$O_{K2}$ + $O_{K3}$, and $M2$, as well as the quiescent period $QS$.
Outbursts $O_{K2}$ and $O_{K3}$ are combined owing to the small number of
multifrequency optical and IR observations obtained separately for each outburst.

We have derived spectral indices of synchrotron components in the
UV-region, applying the same method to multifrequency Swift UVOT measurements.
We have constructed relations between
fluxes in $U$, $UM2$, and $UW2$ bands relative to fluxes in $UW1$ band,
with the largest number of measurements, during two periods RJD: 4292-4450
($O_{K2}$ + $O_{K3}$) and RJD: 4613-4858 ($M2$), when multifrequency UV-data
were available (see Fig.~\ref{mainLC}). Figure \ref{UVff} shows 
the flux-flux dependences, which can be fit by straight lines, 
$S_{\rm i}=A_{\rm i}+B_{\rm i}S_{\rm UW1}$. The slopes of the regressions,
$B_{\rm i}$, are used to construct the relative SED of the synchrotron component
responsible for the variability. Figure~\ref{Specsyn}
shows the relative SEDs of synchrotron components at the UV bands along
with the those obtained for IR and optical parts of the electromagnetic spectrum
(Table~\ref{SpecInd}). The relative UV SEDs are very similar during both events.
However, they cannot be represented by a single power-law 
owing to significant flattening of the spectrum at wavelengths below 2630~{\AA}. 
We have determined the spectral
indices for the soft UV (2634-3500~{\AA}), $\alpha_{\rm UV1}^{\rm syn}$, and
for the hard UV (2030-2634~{\AA}), $\alpha_{\rm UV2}^{\rm syn}$, emission separately (Table~\ref{SpecInd}).
We stress that this hardening of the UV spectrum cannot be connected to the contribution
of the blue bumps to the SED owing the short timescale of variability of the emission. 

Figure~\ref{Specsyn} and Table~\ref{SpecInd} show that
the optical spectra of the variable components are similar at the flaring states and  
flatter than $\alpha_{\rm opt}^{\rm q}$. This is a characteristic of optical variability
often observed in blazars (e.g., \citealt{LAR08}): variable sources are bluer at 
brighter flux levels. In the quiescent state $\alpha_{\rm IR}^{\rm q}$ is flatter 
than $\alpha_{\rm opt}^{\rm q}$ by $\sim$0.5, as expected for a synchrotron source 
with relativistic electron energy distribution  $N(E)=N_\circ E^{-3.8}$ when 
optically emitting electrons suffer strong radiative losses.
The significantly steeper $\alpha_{\rm IR}^{\rm syn}$ with respect 
to $\alpha_{\rm opt}^{\rm syn}$
during event $O_{K1}$  is unusual. This effect can be caused by 
undersampling the IR-data during the brightest stage of optical emission
(see Fig~\ref{mainLC}).
However, if it is real, a hardening of the synchrotron spectrum can occur 
at electron energies 
where Klein-Nishina (KN) effects become important \citep{DA02,MOD05},
if inverse Compton losses dominate over synchrotron losses, i.e.,
$u_{phot}\gg B^2/(8\pi)$, where $u_{\rm phot}$ is energy density of the
photon field and $B$ is strength of the magnetic field. Such a condition
might occur during the dramatic outburst $O_{\rm K1}$ that yields
diminishing inverse Compton losses at frequencies 
$\nu\ge 2.6\times 10^{15}\delta B/(1+z)(\nu'/10^{15})^{-2}$~Hz,
where $\nu'\approx\Gamma\nu_{\rm seed}$ and  $\nu_{\rm seed}$ is
the frequency of the external photon field. 
If we assume that $\delta$ and $\Gamma$
correspond to knot $K1$ and $\nu_{\rm seed}\approx 5\times10^{13}$~Hz, 
a hardening of the synchrotron spectrum at optical wavelengths, 
$\nu\ge 4\times 10^{14}$~Hz, requires the magnetic field $B\approx$4.5~mG, while
for $B\approx$1~G the external photon field should peak in U-band, 
$\nu_{\rm seed}\approx 7.5\times10^{14}$~Hz. 

The soft UV spectrum is steep, with $\alpha_{\rm UV}^{\rm syn}\approx$2.3,
an even steeper synchrotron spectral index at 2500~{\AA}, $\sim$ 2.7,   
was derived by \citet{SMITH88} from multifrequency polarization data of 3C~454.3.
The hardening of the synchrotron UV spectrum beyond 2500~{\AA} that we observe 
during events $O_{\rm K2}$+$O_{\rm K3}$ and $M2$  can be caused by inverse Compton
losses in the Klein-Nishina regime, as described above. According to the SED presented 
by \citet{Abdo09b}, the inverse Compton peak exceeds the synchrotron peak
during $M2$ event by at least a factor of $\sim$5, which implies that the 
conditions for inverse Compton losses should be achieved \citep{MOD05}. For events
$O_{\rm K2}$+$O_{\rm K3}$ and $M2$  
the jet parameters are uncertain. We adopt the average jet parameters
$\Gamma$=16 and $\delta=25$ \citep{J05} that result in a hardening of the synchrotron
spectrum at $\nu\ge 2\times 10^{15}$ if 
$\nu_{\rm seed}\approx 3\times10^{15}$~Hz and $B\approx$50~mG or 
if $B\approx$1~G and $\nu_{\rm seed}\approx 2\times10^{14}$~Hz.

X-ray spectral indices at 2.4-10~keV were calculated in the process of the RXTE data 
reduction. We use the X-ray spectral index from the epoch
closest to the time of the peak of $O_{K1}$ to characterize the outburst.
For outbursts $O_{K2}$, $O_{K3}$, and $M2$ we have derived $\alpha_{\rm x}$ at 2.4-10~keV 
from {\it Swift} data obtained at epochs closest to the dates of the peaks of the outbursts using
the same spectral model as for the {\it RXTE} data. The X-ray spectral index 
for the quiescent state is from \citet{RAI07}, derived from XMM-Newton measurements on 
2006 December 18/19. Table~\ref{SpecInd} summarizes the X-ray spectral indices and  
shows that during
outbursts $O_{\rm K2}$ and $O_{\rm K3}$  $\alpha_{\rm x}$ is similar to that
in the quiescent state, while 
during the most powerful optical outburst, $O_{K1}$,  $\alpha_{\rm x}$ became
steeper than  $\alpha_{\rm x}^{\rm q}$.

We have determined $\gamma$-ray spectral indices, $\alpha_{0.1-300Gev}$, using a single power-law model. Figure~\ref{Gindex} shows variations of the $\gamma$-ray spectral index  during RJD: 4684-4873. 
The average value $\bigl<\alpha_{0.1-300Gev}\bigr>$=1.46$\pm$0.29 
corresponds to the average of the optical and near-IR spectral indices of 
the synchrotron component during outburst $M2$.
\citet{Abdo09b} have found that the $\gamma$-ray spectrum of 3C~454.3 during $M2$ event
is described by a broken power-law model with $\alpha_\gamma^{\rm low}$=1.27$\pm$0.12
and $\alpha_\gamma^{\rm high}$=2.50$\pm$0.35 and a break at 2.4$\pm$0.6~GeV.
This implies a close correspondence between $\alpha_\gamma^{\rm low}$ and 
$\alpha_{\rm IR}^{\rm syn}$ and between $\alpha_\gamma^{\rm high}$ and the
spectral index of soft synchrotron UV emission, $\alpha_{\rm UV1}^{\rm syn}$
(see Table~\ref{SpecInd}).

\section{Polarization Behavior}

Figure~\ref{Poldat} shows the parameters of the polarization from the whole source at
both optical wavelengths and 86~GHz, and at 43~GHz in the VLBI core region.
The data reveal a very wide range of variability of the degree of optical
polarization, from $\sim$0\% to 30\%.
The range and timescale of the variability decrease with wavelength, in agreement with
the findings of \citet{J07}. The average values of the degree of polarization are
$\bigl<{\rm p}_{\rm opt}\bigr>$=7.4$\pm$5.3\%, $\bigl<{\rm p}_{\rm 86}\bigr>$=2.3$\pm$1.4\%, and 
$\bigl<{\rm p}_{\rm 43}\bigr>$=1.6$\pm$0.8\%. At all three wavelengths the position angle of polarization covers the entire range from $-$180$^\circ$ to 0$^\circ$, although $\varphi_{\rm opt}$
varies  more rapidly than that of the mm emission. We have collected 253 simultaneous measurements 
of the optical flux and degree of polarization, which produce a statistically 
significant correlation between these parameters (linear coefficient of correlation, $\rho$=0.32).
The connection between the flux and degree of polarization is weaker in the 43~GHz core ($\rho$=0.21) 
and absent at 86~GHz  ($\rho$=0.05). 
 
We have constructed distributions of alignment of EVPAs at optical 
and 86~GHz wavelengths from the whole source, and at 43~GHz from the VLBI core region with respect to
the jet direction.
The distributions are presented in Figure~\ref{hjet} for the all measurements at each 
frequency, as well as for cases when the degree of polarization is higher than the average
at a given frequency. The optical and 43~GHz core EVPAs extend across
all possible directions with respect to the jet axis, although $\varphi_{\rm opt}$ 
tends to be perpendicular to the jet, while $\varphi_{\rm 43}$ aligns more often with the
jet direction. These tendencies are more prominent when the degree of polarization is high. 
This might be partially caused by the larger uncertainties in EVPA when $p$ is low.
The EVPA at 86~GHz maintains good alignment with the jet direction independent of degree
of polarization. However, this evidence of better alignment may be biased by the poorer
time coverage at 86~GHz since, in general, Figure~\ref{Poldat} shows  good agreement 
between the EVPA measurements at 86 and 43~GHz.

Table \ref{O7evpa} presents the optical polarization position angle and EVPA in the 43~GHz core
for essentially simultaneous observations (within a week, less than the time scale of significant
variability at 43~GHz). The optical and
86~GHz polarization position angles measured within 2~days of each other are listed in Table \ref{O3evpa}.
There are 19 simultaneous pairs of optical and 43~GHz measurements, with
8 pairs (42\%) showing alignment between $\varphi_{\rm opt}$ and $\varphi_{\rm 43}$ within 
the 1$\sigma$ uncertainty of the measurements. The optical and 86~GHz EVPAs agree in only 2 out of
12 pairs (17\%). Alignment between $\varphi_{\rm opt}$ and 
$\varphi_{\rm 43}$ is observed independently of direction of polarization
with respect to the jet axis, while
two cases of agreement between $\varphi_{\rm opt}$ and  $\varphi_{\rm 86}$ occur 
when the EVPAs are close to the jet direction. If optical and 43~GHz EVPAs are independent,
Bayes theorem yields a probability of $<$3$\times$10$^{-8}$ that the number of observed
alignments between the optical and 43~GHz polarization position angles occur by chance.
This implies that the polarized emission at different wavelengths is governed  by 
common properties or processes rather than by chance, in agreement with previous
findings \citep{GRSS06,J07,FRANI07}.

\subsection{Rotation of Position Angle of Optical Polarization}
 Analysis of the optical polarization behavior during
optical outbursts shows that each ejection 
coincides with a rotation of the optical polarization position angle (Fig.~\ref{PolRot}). 
The rotations have similar rates for events $K1$ and $K2$, $v_{\rm rot}^{\rm K1}$=8.7$\pm$1.1  and 
$v_{\rm rot}^{\rm K2}$=7.7$\pm$0.8 degrees per day, while the rotation during the ejection
of $K3$ is slower, $v_{\rm rot}^{\rm K3}$=5.2$\pm$0.7 degrees per day. We interpret this
as evidence that the rotation of $\varphi_{\rm opt}$ originates in the acceleration and collimation zone (ACZ).
We can rule out a stochastic rotation of the position angle due to plasma turbulence \citep{MAR08,MAR10}, 
given that all three events have the same sense of rotation and coincide with 
optical flares. 
In the ACZ, Poynting flux is converted to bulk kinetic energy
through the Lorentz force, which rotates the MHD outflows and forms helical trajectories
of disturbances propagating down the jet \citep{VK04,VLAH06,kom07}. The Lorentz factor
should increase linearly with cross-sectional radius $r$, while the speed of rotation should decrease
as $r^{-1}$ to conserve angular momentum \citep{VLAH06}. 
If the polarized optical emission is produced in the ACZ by an emission feature that follows
streamlines that execute a spiral path about the axis, then the speed of 
rotation of different disturbances should be inversely proportional to the final bulk 
Lorentz factor. The jet flow must travel farther from the BH to attain a higher Lorentz factor,
hence the size scales of the site of EVPA rotation should be larger and the position of the core should shift downstream relative to states with a lower terminal Lorentz factor.
In support of these predictions, the lowest rate of rotation is observed during event $K3$,
which corresponds to the highest value of $\Gamma$,  
while $v_{\rm rot}^{\rm K1}\sim v_{\rm rot}^{\rm K2}$, in agreement with 
$\Gamma_{\rm K1}\sim\Gamma_{\rm K2}$ (see Table~\ref{Kparm}). Furthermore, there is a
discontinuity in the motion of knot $K2$ near the time of ejection of $K3$ (see Fig.~\ref{evol7}),
which could be explained if the 43~GHz core shifted downstream at this time.
At the end of the optical EVPA rotation in events $K2$ and $K3$, the EVPA became chaotic,
with average value perpendicular to the jet (see Fig.~\ref{PolRot}).    
This is consistent with an optical outburst arising from an interaction between a disturbance propagating
through a turbulent section in the jet and a standing
shock system in the mm-wave core \citep{FRANI07}. 

Rotation of the optical EVPA associated with a disturbance propagating down the jet has been observed 
previously in BL~Lacertae \citep{MAR08} and in the quasars 3C~279 and PKS~1510$-$089
\citep{LAR08,MAR10}.
There are several other properties of such rotations that are similar in BL~Lac, 3C~279,
PKS~1510$-$089, and 3C~454.3:
(i) at the beginning of the rotation, the degree of polarization is low ($\sim$1-2\%);
(ii) maximum fractional polarization is achieved when the E-vector aligns with the jet direction 
(for event $K1$, $P_{\rm max}$=22\%, $\varphi_{\rm max}$=$-$77$^\circ$; for $K2$, $P_{\rm max}$=26\%, 
$\varphi_{\rm max}$=$-$87$^\circ$; event $K3$ is undersampled); and
(iii) at the end of the rotation, the degree of polarization decreases. 
We infer that a rotation of the optical polarization position angle during a
prominent optical outburst is a common feature of blazars.
In addition,
in 3C~279 both the optical and 43~GHz core EVPAs rotate as a superluminal knot emerges from the core
\citep{LAR08}. A few measurements of the EVPA in the core of 3C~454.3, 
simultaneous with optical polarization observations, agree with the optical EVPAs 
(Fig.~\ref{PolRot}). However, the agreement occurs when $\varphi_{\rm opt}$ and $\varphi_{\rm 43}$
are closely aligned with the jet direction (Table~\ref{O7evpa}), so that it is possible that
the core EVPA remains near that preferred angle instead of rotating with $\varphi_{\rm opt}$.   

Despite agreement between $\varphi_{\rm opt}$ and $\varphi_{\rm 43}$ for a number of 
essentially simultaneous observations (see Table~\ref{O7evpa}), the majority of the high 
optical polarization measurements occur when the optical EVPA is perpendicular 
to the jet, while at 86~GHz  and in the 43~GHz core, the majority of the E-vectors at high degrees of polarization
lie along the jet direction. There are multiple possible reasons for this discrepancy: (i) high
mm-wave polarization does not exeed 7\%, while the optical degree of polarization $\ge$20\% quite often; 
this is qualitatively consistent with the mm-wave polarization being affected by synchrotron
self-absorption, which lowers the degree of polarization and rotates the EVPA by 90$^\circ$;
(ii) the magnetic field in the core region has a helical structure, with 
the optical and mm-wave emission regions located in different parts of the helix; and (iii)
the jet has a spine-and-sheath structure such that, in a quiescent state, the optical polarization comes 
from a transition zone between the fast spine stream and slower sheath flow (such emission would be 
highly polarized perpendicular to the outflow; \citealt{LAING80,FRANI09}), while the mm-wave emission region 
is dominated by turbulence and has low polarization. In an active state, a disturbance in the jet 
defines the optical and mm-wave polarization properties: at millimeter wavelengths the disturbance
orders the magnetic field perpendicular to the jet flow, while $\varphi_{\rm opt}$ depends on
whether the disturbance or shear layer dominates the polarized flux. 
 
\section{Discussion}

We have found a connection between mm-wave and optical outbursts and disturbances propagating down the radio jet
in 3C~454.3. At least three major optical outbursts, $O_{K1}, O_{K2}$, and $O_{K3}$,
observed in 2005-2008 are related to superluminal knots identified in the VLBA images.
\citet{J07} have found that in blazars the polarization properties of superluminal knots 
at 43~GHz are consistent with weak shocks and the optical polarized emission originates 
in shocks, most likely situated between the 86~GHz and 43~GHz VLBI cores. It
appears that shocks and their interaction with underlying jet structure might be
generally responsible for optical and high energy outbursts.  
 
\subsection{Location of the Optical Emission in the Jet}

Comparison of the timing of optical outbursts and epochs of ejection of superluminal knots from the
core can be used to determine the location of the sites where kinetic and internal energy is
dissipated in the form of radiation. The delay between the peak 
of an outburst and the time of the ejection of the associated superluminal component,
along with the apparent speed of the knot, determines the location of 
the optical outburst in the jet with respect to the core, 
$\Delta r_{\rm opt} = \beta_{\rm app}c\Delta T_{\rm opt}/\sin\Theta_\circ$. 
According to Tables~\ref{Kparm} and \ref{Delparm}, the peaks of outbursts $O_{\rm K1}, O_{\rm K2}$, and $O_{\rm K3}$ occurred 12.6$\pm$5.6~pc upstream, 3.5$\pm$3.7~pc downstream, and 2$\pm$18~pc upstream of the core, respectively. The average location for 
$O_{\rm K2}$ and $O_{\rm K3}$ is 2.8$\pm$3.4~pc downtream of the core.
In the case of a conical structure of the jet, \citet{J05}
have estimated the half opening angle of the jet in 3C~454.3
to be $\theta=0.8^\circ\pm0.2^\circ$. 
We have obtained an average size of the core during our period of observations
using modelling by circular Gaussians, $a_{\rm core}$=0.068$\pm$0.011~mas. 
This value of $a_{\rm core}$ gives an upper limit for the transverse 
size of the core, which, along with the opening angle of the jet, provides an estimate of 
the distance of the mm-wave core from the central engine, $R_{\rm BH}\le$18$\pm$3~pc. 
Therefore, if the perturbation that created $K1$ is reponsible for $O_{K1}$, then
the maximum of the outburst took place when $K1$ was located at a distance 
5.4$\pm$5.2~pc from the central engine, 
while the maxima of $O_{\rm K2}$ and $O_{\rm K3}$ outbursts occurred at a larger distance
from the BH, 21$\pm$4~pc, perhaps when knots $K2$ and $K3$ passed through 
the mm-wave core. This implies that
the dissipation site for $O_{\rm K1}$ was closer to the central engine (BH) than
that for outbursts $O_{\rm K2}$ and $O_{\rm K3}$. These results are consistent with the recently advanced ideas
that the dissipation sites of optical and high energy outbursts can occur at different
locations and involve different sources of seed photons \citep{GHIS07,MAR08,MAR10,RITABAN08}. 

\subsection{The Structure of the mm-Wave Core}
The most powerful optical 
outburst, $O_{K1}$, preceded the passage of $K1$ through the 43~GHz core (ejection)
by $\sim$50~days. In addition, there are optical
flares (RJD: 3537 and RJD: 3561, see Table~\ref{Outparm}) that occurred within 1$\sigma$ uncertainty
of the time of ejection of $K1$.
Outbursts $O_{K2}$ and $O_{K3}$ were simultaneous (within 1$\sigma$ uncertainty)
with the passage of $K2$ and $K3$, respectively, through the core. 
We isolate segments
of the optical light curve within the time interval $-\sigma T_\circ$ and +2$\sigma T_\circ$
of each of knots $K1$, $K2$, and $K3$ (Fig.~\ref{multi}, the values of $\sigma T_\circ$ are given 
in Table~\ref{Kparm}).
Each ejection is accompanied by a series of flares (denoted in Fig.~\ref{multi} as 1, 2, and 3)
that starts within 1$\sigma$ of $T_\circ$. The duration of the series (time between the first
and third peaks) is comparable, 50, 50, and 65~days, for events $K1$, $K2$, and $K3$, respectively.

The profile of event $O_{\rm K1}$ is very different from the profiles of $O_{\rm K2}$
and $O_{\rm K3}$. The average of the rise and decay times for $O_{\rm K1}$ has a width of $\sim$47~days (Table~\ref{Outparm}),
with a rise time of 68~days  and a decay of only 26~days. Outbursts $O_{K2}$ and $O_{K3}$, 
as well as the flares connected
with the passage of $K1$ through the core, have much shorter durations and similar rise and 
decay times (see Fig.~\ref{multi}). 
A possible interpretation is that, in the case of $O_{\rm K1}$, located deeply within the ACZ, the energization of
relativistic electrons is gradual and continuous, while for outbursts occurring in the VLBI core,
relativistic electrons are energized abruptly owing to interaction between the 
disturbance and the core. The latter implies that the three-flare structure of optical outbursts
seen in Figure~\ref{multi} is related to the physical structure of the mm-wave core.
Indeed, the super-resolved images of 3C~454.3 (Fig.~\ref{hCore}) support the idea 
that the core at 43~GHz might have a three-component structure.

The core might be a system of alternating conical
shocks and rarefactions, as suggested by \citet{GOM97}, \citet{MAR06}, and \citet{CAW06}.
In this case, the distance between constrictions, $z_{\rm max}$, can be calculated for an ultra-relativistic
equation of state according to \citet{DM88}: $z_{\rm max}\approx 3.3\times\Gamma~a_{\rm core}^{\rm t}/\eta$,
where $a_{\rm core}^{\rm t}$ is the transverse radius of the core in mas and
$\eta=p_{\rm ext}/p_{\circ,{\rm jet}}$ is the ratio of the external pressure to initial pressure
(i.e., at the upstream boundary of the core) in the jet.
\citet{J05} have found that the relation between the viewing angle and Lorentz factor in blazars 
is consistent with $\eta\sim$3. If the core of 3C~454.3 is a system of three conical shocks, 
then the longitudinal size of the core in projection on the sky can be determined as
$a_{\rm core}^{\ell,{\rm theor}}=3~z_{\rm max}\sin{\Theta_\circ}$. On the other hand, if the
three-flare structure of the optical outbursts observed in the vicinity 
of the ejection is caused by a disturbance moving through this system of conical shocks, then
$a_{\rm core}^{\ell,{\rm obs}}=\Delta T_{\rm s}\mu$, where $\Delta T_{\rm s}$ is the duration of the
optical outbursts and $\mu$ is the proper motion of the knot. Table~\ref{Zmax} lists values 
of $a_{\rm core}^{\ell,{\rm theor}}$ and $a_{\rm core}^{\ell,{\rm obs}}$ for events $K1$, $K2$, and $K3$,
which are consistent with the proposed association between a propagating disturbance and with the multi-component
structure of the VLBI core providing a viable explanation for the three-flare pattern observed in the optical 
light curve. From standard gas dynamics, the presence of multiple standing shocks
in three dimensions requires a high level of azimuthal symmetry of the pressure at the boundary of the jet in order
for the structure to remain intact.

\subsection{Location of Millimeter-Wave Outbursts in the Jet}

We have derived the locations of 230~GHz outbursts 
$M_{\rm K1}$, $M_{\rm K2}$, and $M_{\rm K3}$ 
in the same manner as for optical outbursts $O_{\rm K1}$, $O_{\rm K2}$, and $O_{\rm K3}$
owing their association with superluminal knots.
The maxima of $M_{\rm K1}$, $M_{\rm K2}$, and $M_{\rm K3}$ occurred when a disturbance
was located 4.6$\pm$6.1~pc upstream, 8.2$\pm$4.8~pc downstream, and 15$\pm$37~pc downstream
of the core, respectively, with the average value 3.8$\pm$4.5~pc downstream of the
43~GHz core. This implies that the passage of a disturbance through the 230~GHz core
should occur during the rising branch of a mm-wave outburst. 

The sharp mm-wave outbursts $M1$ and $M2$ (see Fig.~\ref{mainLC})
seem unrelated to the appearance of new knots although we cannot 
reject the possibility that new knots were ejected 
near the time of $M1$ and $M2$ but could not be resolved on the images because they were
weaker than knots $K1$ and $K3$, which were ejected before outbursts $M1$ and $M2$, respectively. 
However, the VLBA observations support the idea  that the outbursts are caused 
by curvature in the inner jet that decreases the angle between a disturbance
($K1$  or $K3$) and the line of sight so that the flux 
peaks when the disturbance passes through the minimum viewing angle, as
suggested by \citet{VIL07}. Indeed, Figure~\ref{trajKs} shows curvature in 
the trajectories of $K1$ and $K3$ at a distance $\sim$0.05-0.07~mas from the core. 
This is the location of knots $K1$ and $K3$ when outbursts 
$M1$ and $M2$ peaked. Comparison of Tables~\ref{Kparm} and \ref{Outparm} indicates
that $M1$ peaked at a time close to the maximum flux of knot $K1$, and that the maximum
of $M2$ occurred when knot $K3$ had the highest flux. The mm-wave
spectral index during outbursts $M1$ and $M2$ corresponds to optically thin emission 
(see \S 8), as expected for radiation coming from downstream of the core. 
In addition, Table \ref{Outparm} shows that the timescale of $M1$ is $\sim$4-5
times shorter than $w_{\rm mm}$ of $M2$, consistent with $K1$ moving closer
to the jet axis (a smaller radius of curvature) and $K3$ moving along periphery
of the jet (a larger radius of curvature), which agrees with the viewing angles
of the knots (Table \ref{Kparm}). We use the delay between the times of ejection
of $K1$ and peak of $M1$, along with the kinematic parameters of $K1$, to derive the location of the curvature of the jet, which is 54$\pm$16~pc downstream the core. 

\subsection{Location of the High Energy Emission in the Jet}

The analysis of spectral characteristics of outbursts in 3C~454.3 reveals that neither during
outbursts nor in quiescent states is the X-ray emission a continuation of the optical synchrotron spectrum.
The most likely mechanism of X-ray production is inverse Compton scattering of low-energy 
photons by relativistic electrons, as suggested by \citet{SMM08} and \citet{RAI08b}.

\begin{figure}
\epsscale{1.0}
\plotone{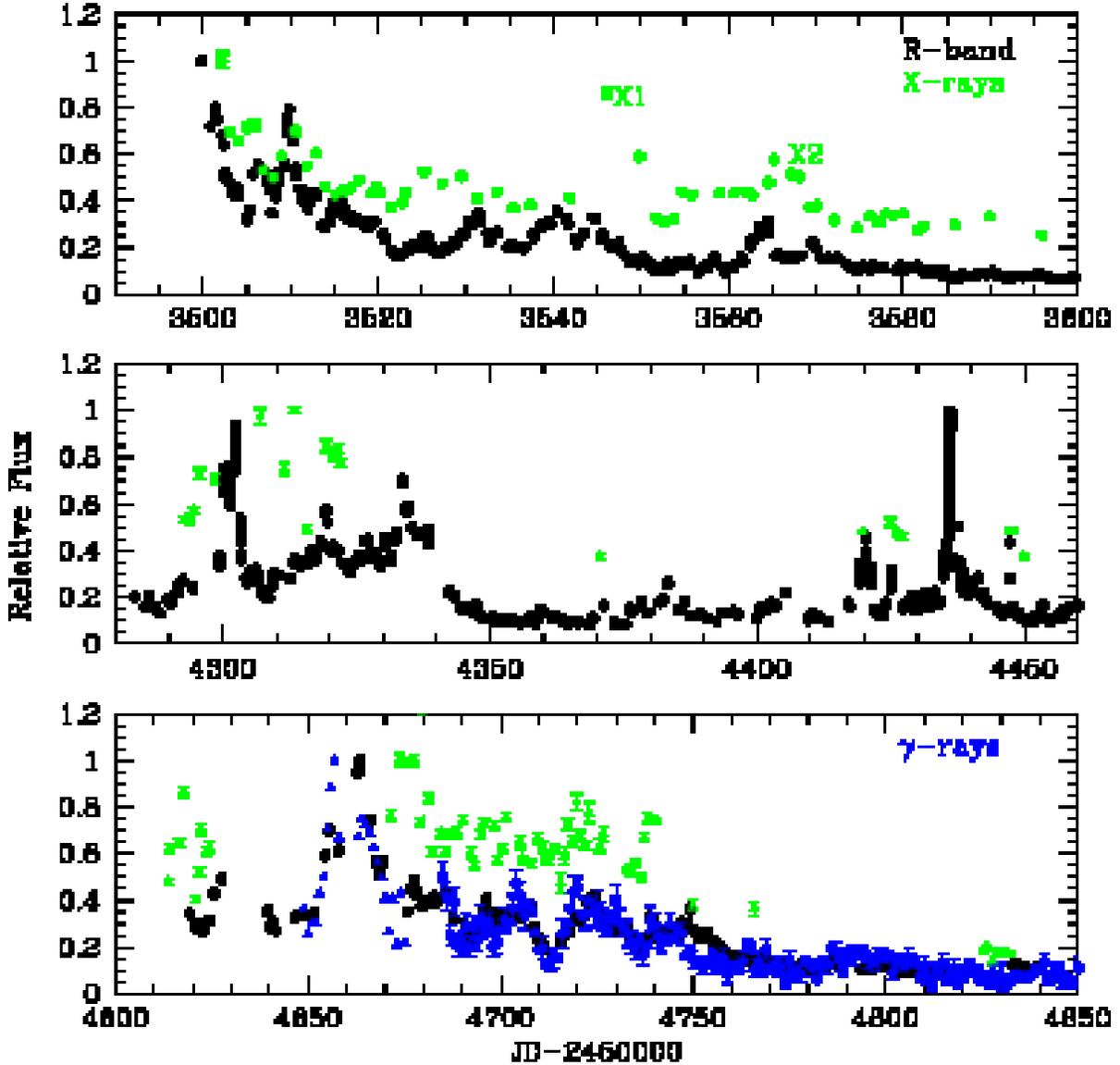}
\caption{Normalized $R$-band (black), X-ray (green, {\it RXTE} measurement - circles and {\it Swift} measurements - triangles), and $\gamma$-ray (blue) light curves for three 
periods. The maximum flux within each panel is adopted as the normalization factor. 
{\it Top panel:} $S_{\rm R}^{\rm max}$=63.6~mJy, $S_{\rm X}^{\rm max}$=10.13~$\mu$Jy (X-ray flux is
at 4~keV); 
{\it middle panel:} $S_{\rm R}^{\rm max}$=37.17~mJy, $S_{\rm X}^{\rm max}$=2.138~cts/s
(X-ray flux is at 0.3-10~keV); 
{\it bottom panel:} $S_{\rm R}^{\rm max}$=17.82~mJy, $S_{\rm X}^{\rm max}$=1.614~cts/s,
$S_{\gamma}^{\rm max}$=1.2$\times$10$^{-5}$~phot/cm$^2$/s (X-ray flux is at 0.3-10~keV, $\gamma$-ray
flux is at 0.1-300~GeV). Blue triangles show the $\gamma$-ray measurements taken from
Fig. 1 of \citet{Abdo09b}. Error bars for the optical observations are less than the size 
of symbols.
}\label{OptHE}
\end{figure}

\begin{figure}
\epsscale{1.0}
\plotone{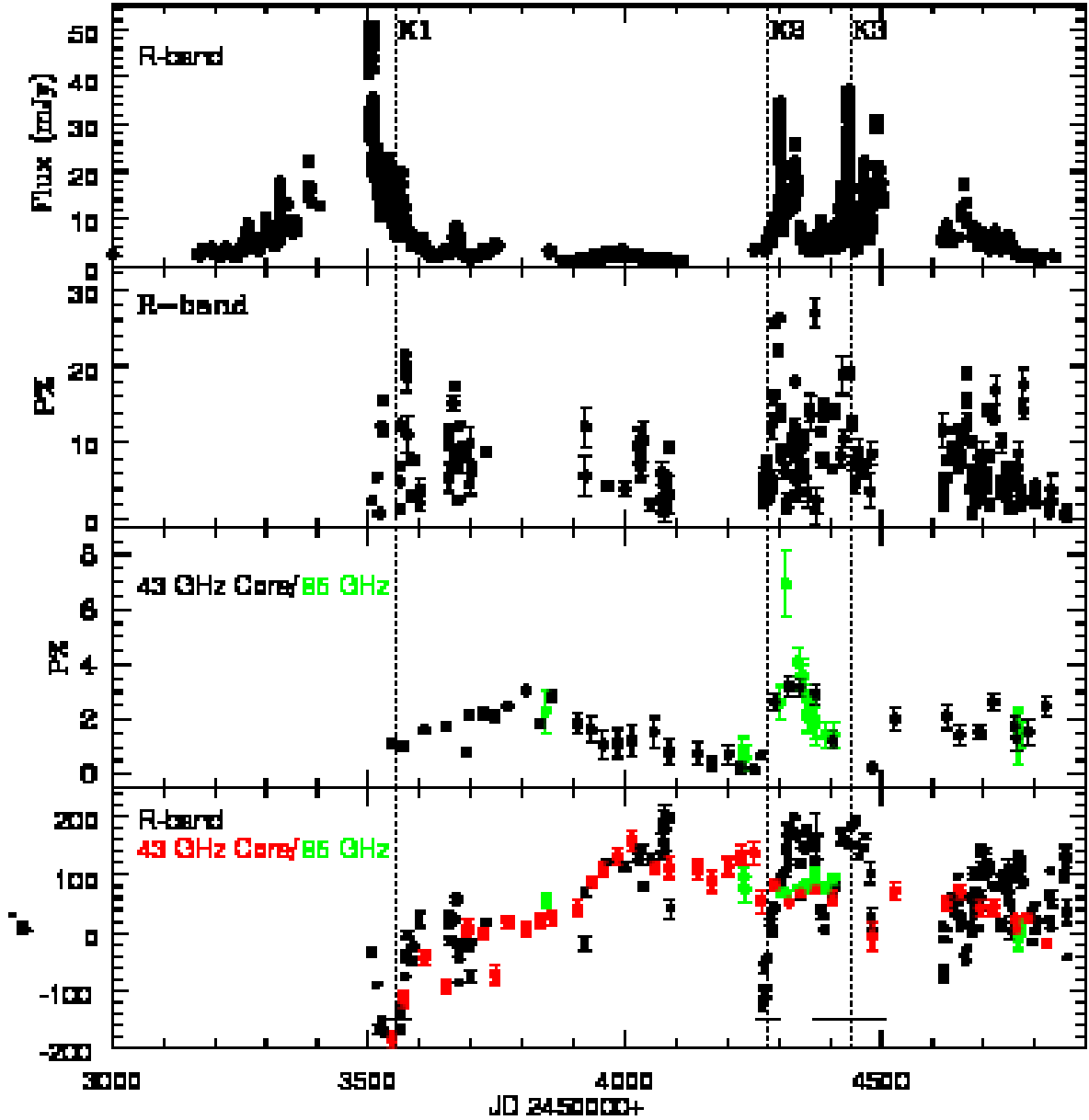}
\caption{Flux and polarization curves of the quasar 3C~454.3 at different frequencies: 
{\it top panel} - total flux density in R-band; {\it second panel} - 
degree of polarization at optical wavelengths; 
{\it third panel} - degree of polarization at 86~GHz (green circles) and in the VLBI core
at 43~GHz (black circles);  and  {\it bottom panel} - position angle of polarization at
optical wavelengths (black circles), at 86~GHz (green circles), and  in 43~GHz VLBI core 
(red circles); dotted lines show times of ejections of superluminal components
and solid line segments show uncertainties in the ejection times. Values of $\varphi_{\rm 7mm}$ 
are corrected
for $RM$ according to \citet{J07}. } \label{Poldat}
\end{figure}

\begin{figure}
\epsscale{0.35}
\plotone{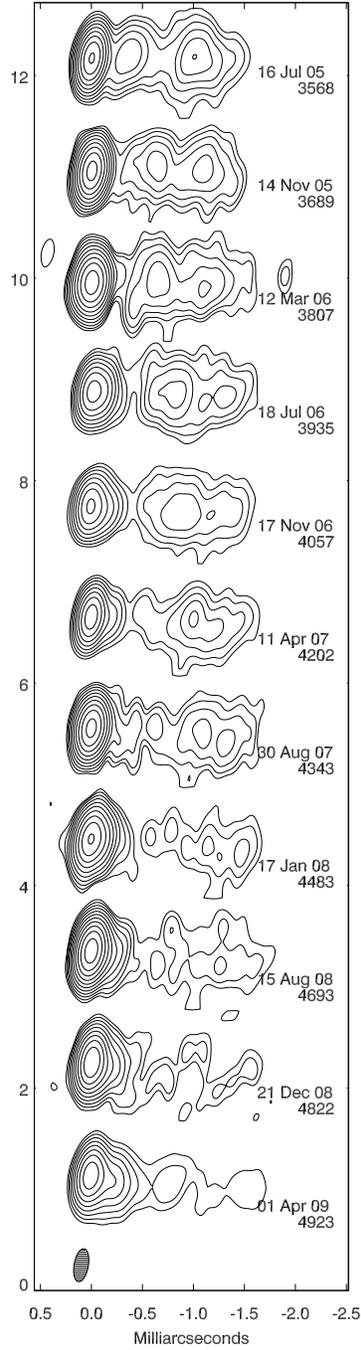}
\caption{43 GHz total intensity images of the quasar 3C~454.3
with $S_{\rm peak}$=19.43~Jy/beam and beam=0.14$\times$0.30~mas$^2$ at $PA$=$-10^\circ$.
Contours represent 0.0625, 0.125, 0.25,...64\% intensity of the peak.} \label{VLBAyr}
\end{figure}

\begin{figure}
\epsscale{1}
\plottwo{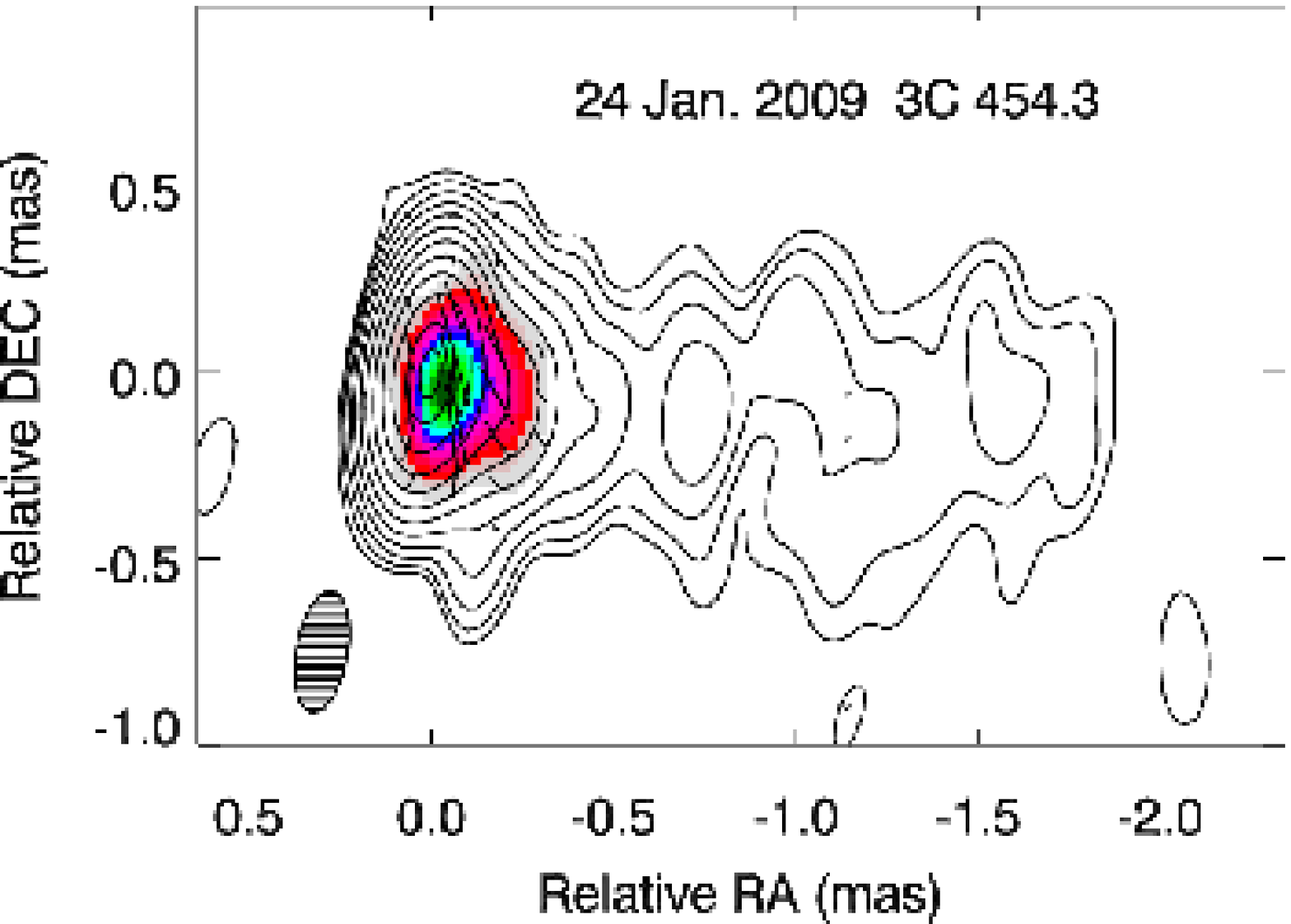}{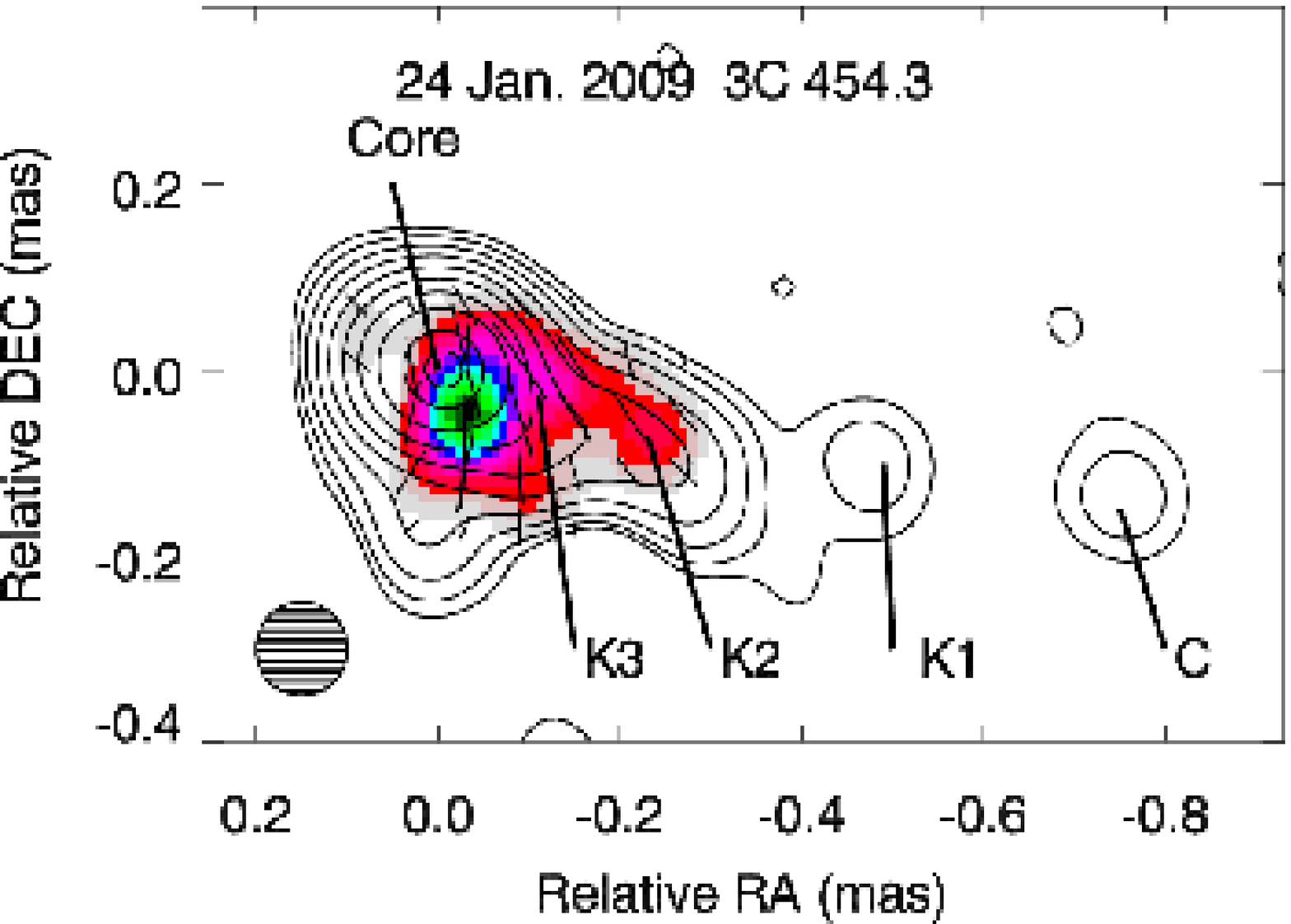}
\caption{43 GHz total (contours) and linearly polarized (color scale) intensity images of 3C~454.3 with
different convolving beams; {\it top :} 0.33$\times$0.14~mas$^2$ at PA=-10$^\circ$, $S_{\rm peak}$=6.20~Jy,
$S_{\rm peak}^{\rm p}$=0.076~Jy; {\it bottom:} 0.10$\times$0.10~mas$^2$,  $S_{\rm peak}$=5.66~Jy,
$S_{\rm peak}^{\rm p}$=0.067~Jy. Line segments show the plane of polarization and letters
identify components in the jet.} \label{image7mm}
\end{figure}

\begin{figure}
\epsscale{2.0}
\plottwo{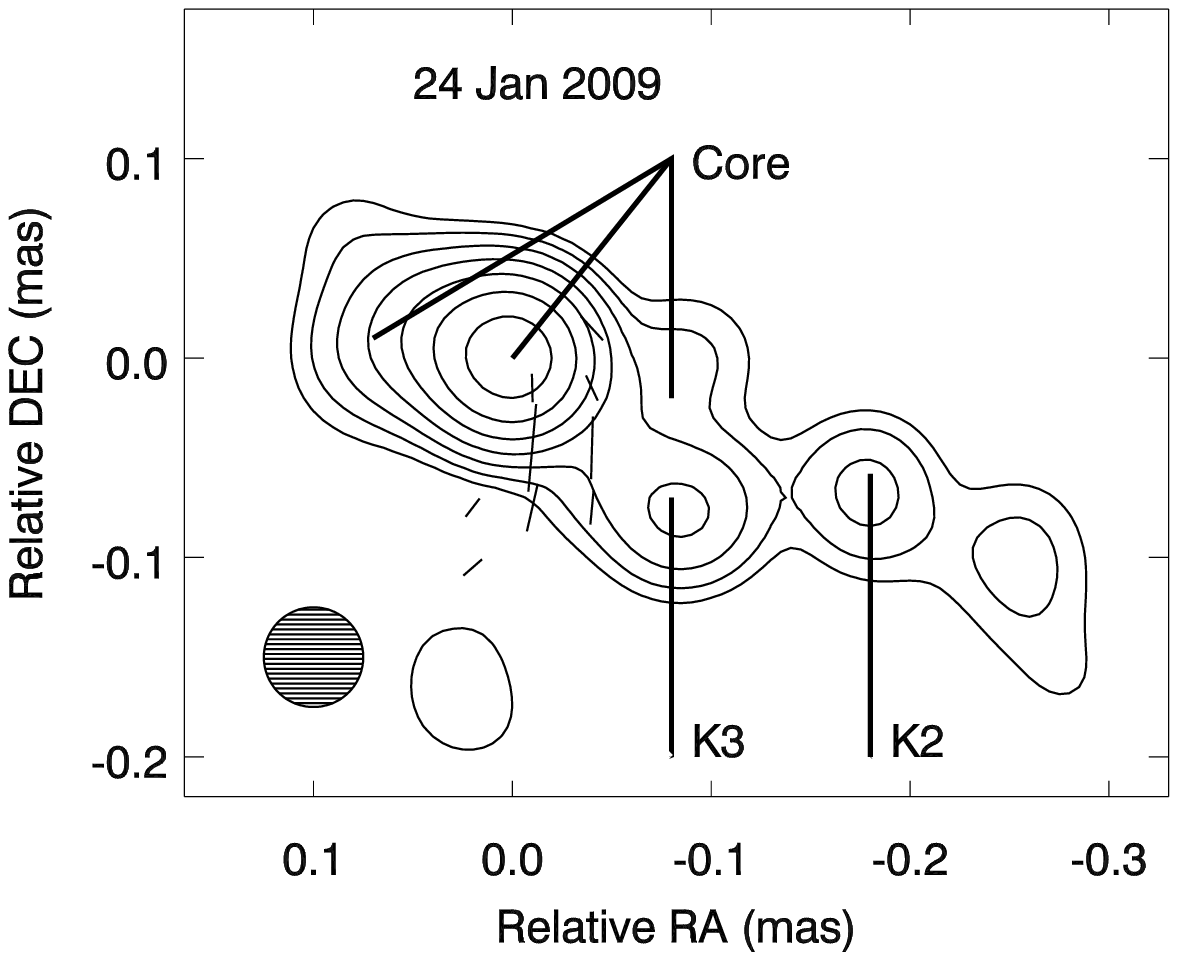}{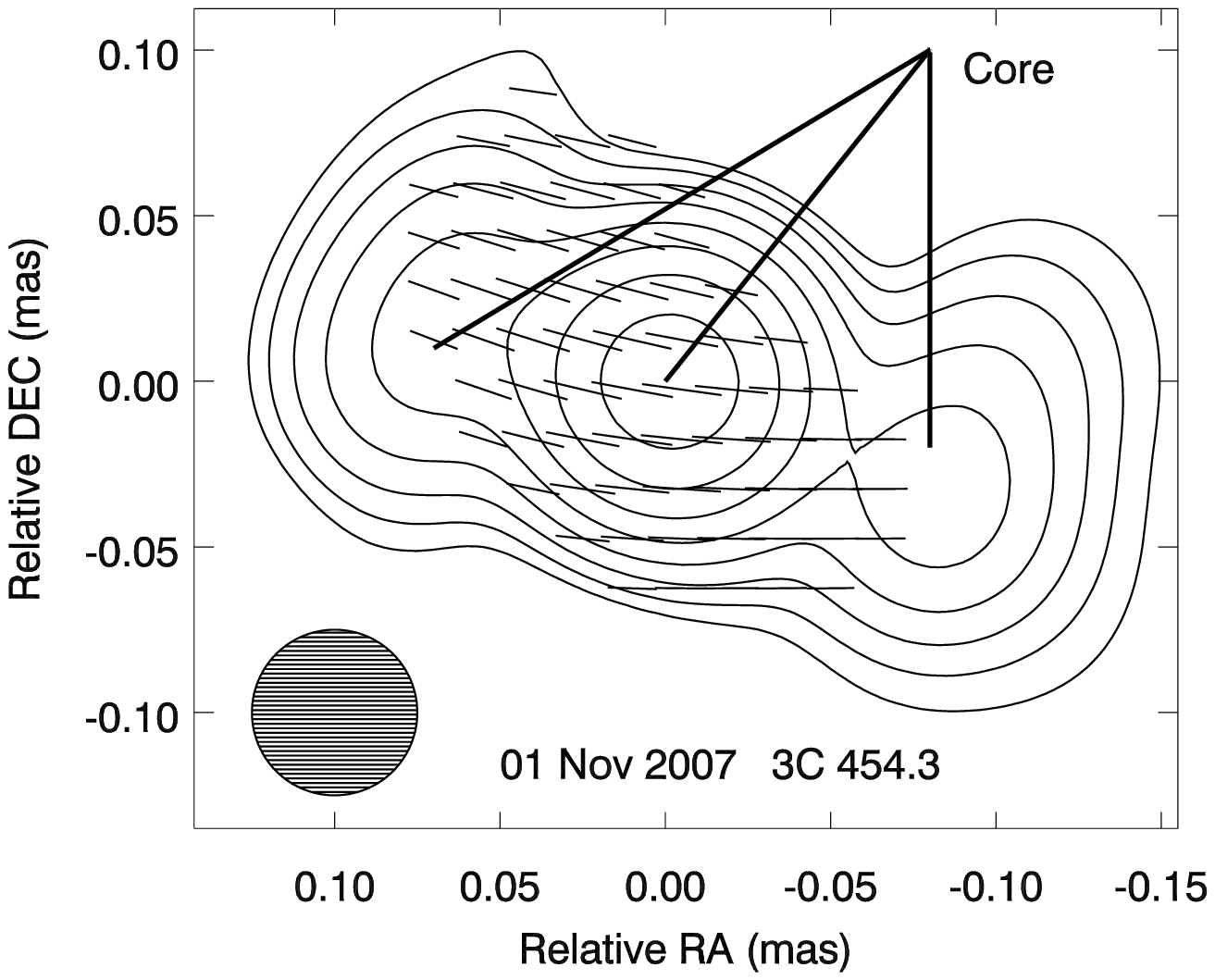}
\caption{43 GHz total (contours) intensity images of 3C~454.3 with
convolving beam 0.05$\times$0.05~mas$^2$ at different epochs; {\it top :} 2009 January 24 (RJD:4856) $S_{\rm peak}$=5.10~Jy,
$S_{\rm peak}^{\rm p}$=0.067~Jy; {\it bottom:} 2007 November 1 (RJD:4406) $S_{\rm peak}$=5.84~Jy,
$S_{\rm peak}^{\rm p}$=0.050~Jy. Line segments show the plane of polarization and letters
identify components in the jet. The tripod of lines indicates a three-component
structure of the core.} \label{hCore}
\end{figure}

\begin{figure}
\epsscale{1.0}
\plotone{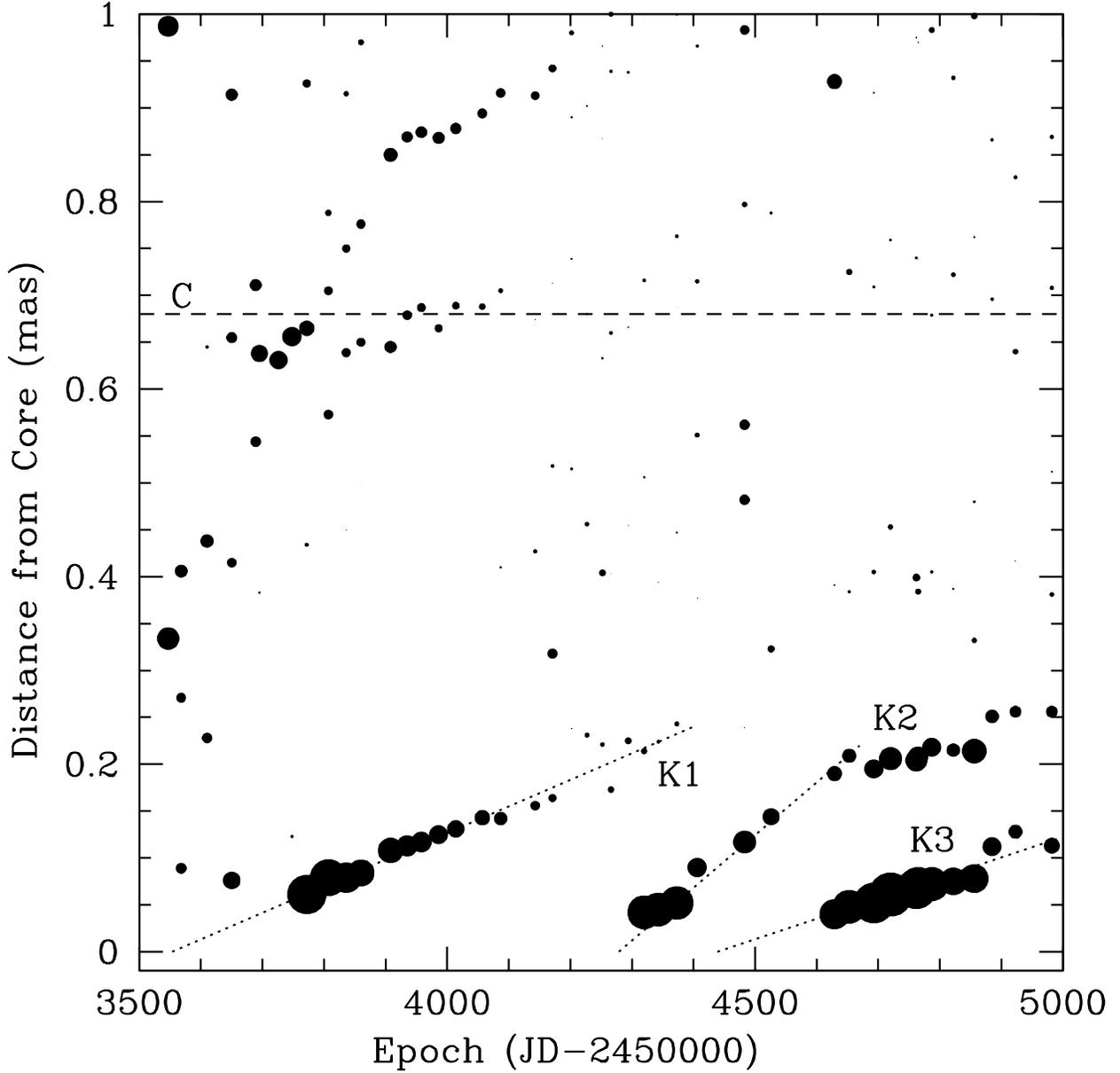}
\caption{Distance of components from the 43~GHz core as a function of time. The size of each symbol 
is proportional to the logarithm of the flux density of the knot. Dotted lines 
indicate an approximation for the presumably ballistic motion of components $K1$, $K2$, and $K3$
near the core. The dashed line shows the position of stationary knot $C$ reported by \citet{J05}.} \label{evol7}
\end{figure}

\begin{figure}
\epsscale{1.0}
\plotone{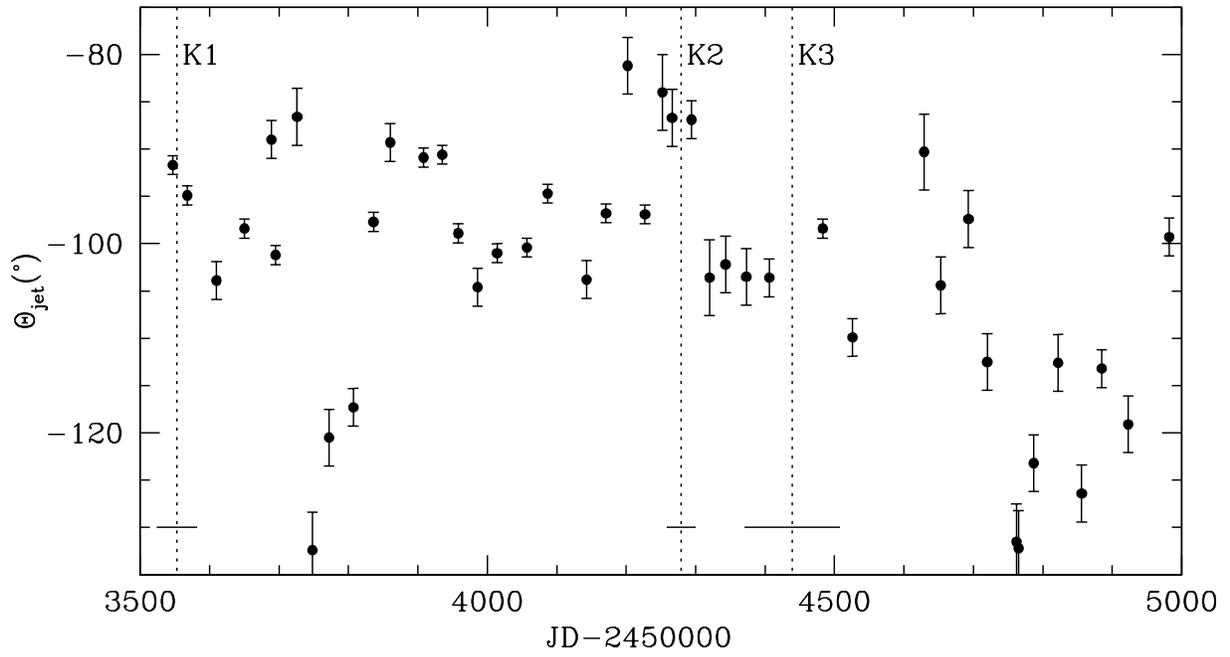}
\caption{Direction of the jet within 0.3~mas from the core as a function of time. Dotted lines
show the times of ejections of superluminal components $K1$, $K2$, and $K3$ from the 43~GHz core
and solid line segments show uncertainties in the ejection times.} \label{jetdir}
\end{figure}

\begin{figure}
\epsscale{1.0}
\plotone{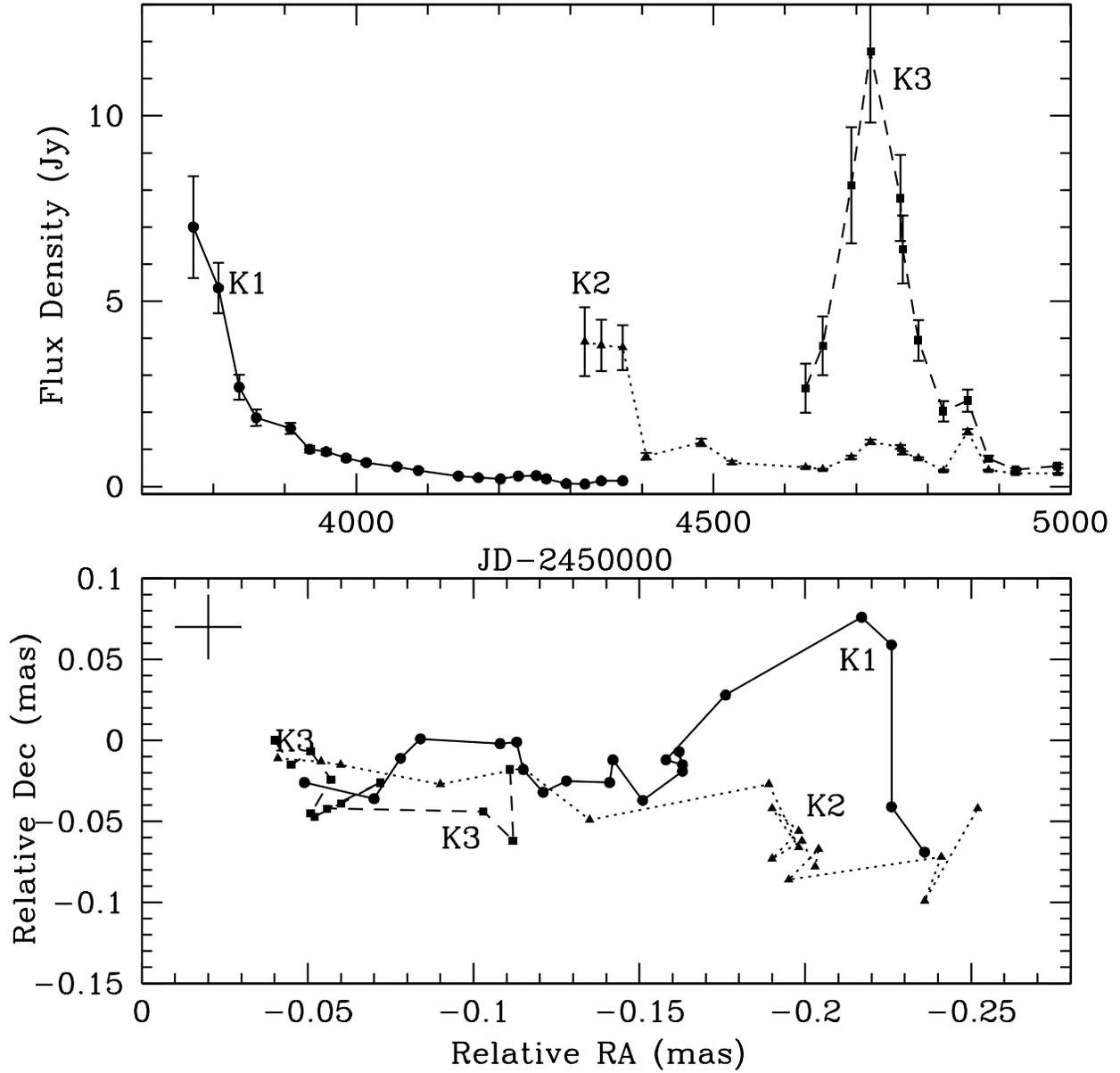}
\caption{Light curves ({\it top}) and trajectories ({\it bottom}) 
of moving knots; the cross indicates typical uncertainties in position of components ({\it upper left coner of the bottom panel})}. \label{trajKs}
\end{figure}

\begin{figure}
\epsscale{0.8}
\plottwo{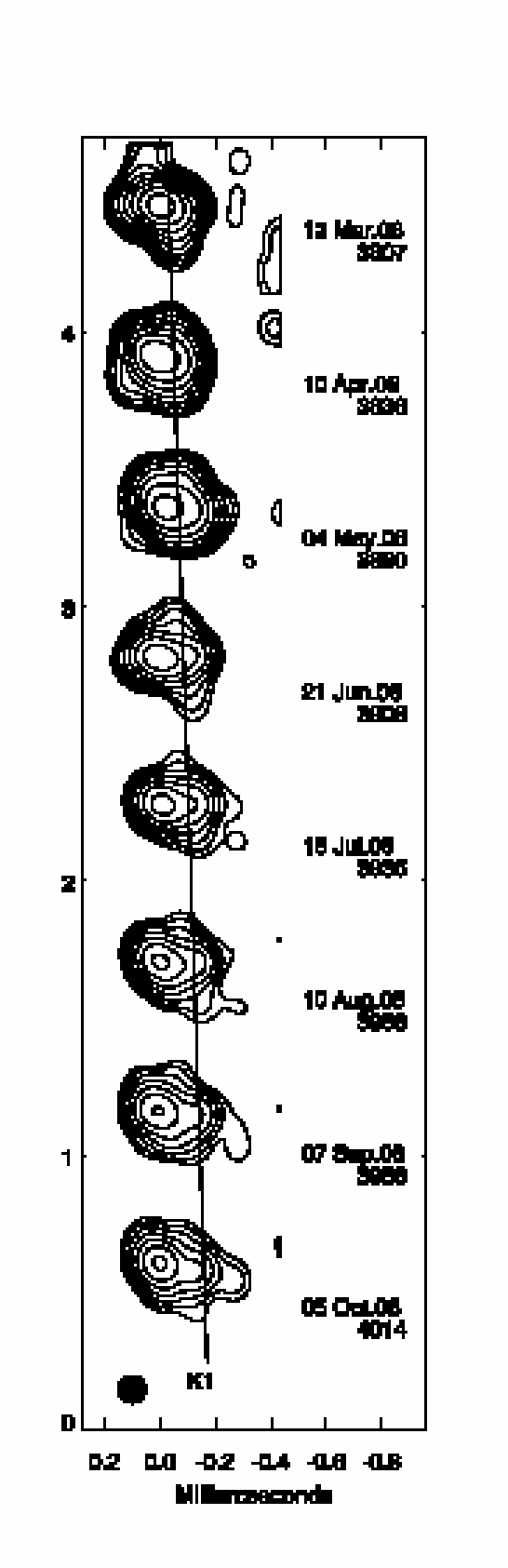}{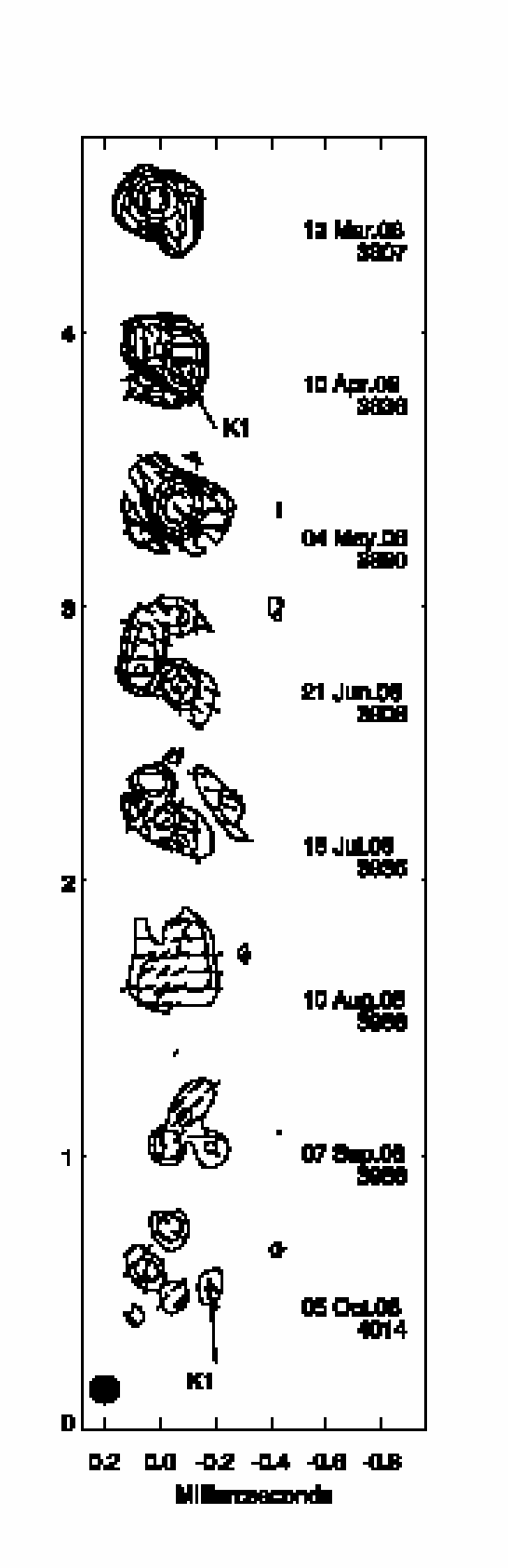}
\caption{43 GHz total ({\it left}) and polarized ({\it right}) intensity images of the quasar 3C~454.3
during period when $K1$ emerged from the core,
$S_{\rm peak}$=8.25~Jy/beam, $S_{\rm peak}^{\rm p}$=0.241~Jy/beam, beam=0.1$\times$0.1~mas$^2$. Total intensity
contours at 0.25,0.5,...64 \% of the peak and polarized intensity contours at 2,4,...64 \%
of the peak. Sticks over
the polarized intensity contours indicate the plane of polarization.} \label{imgK1}
\end{figure}

\begin{figure}
\epsscale{0.8}
\plottwo{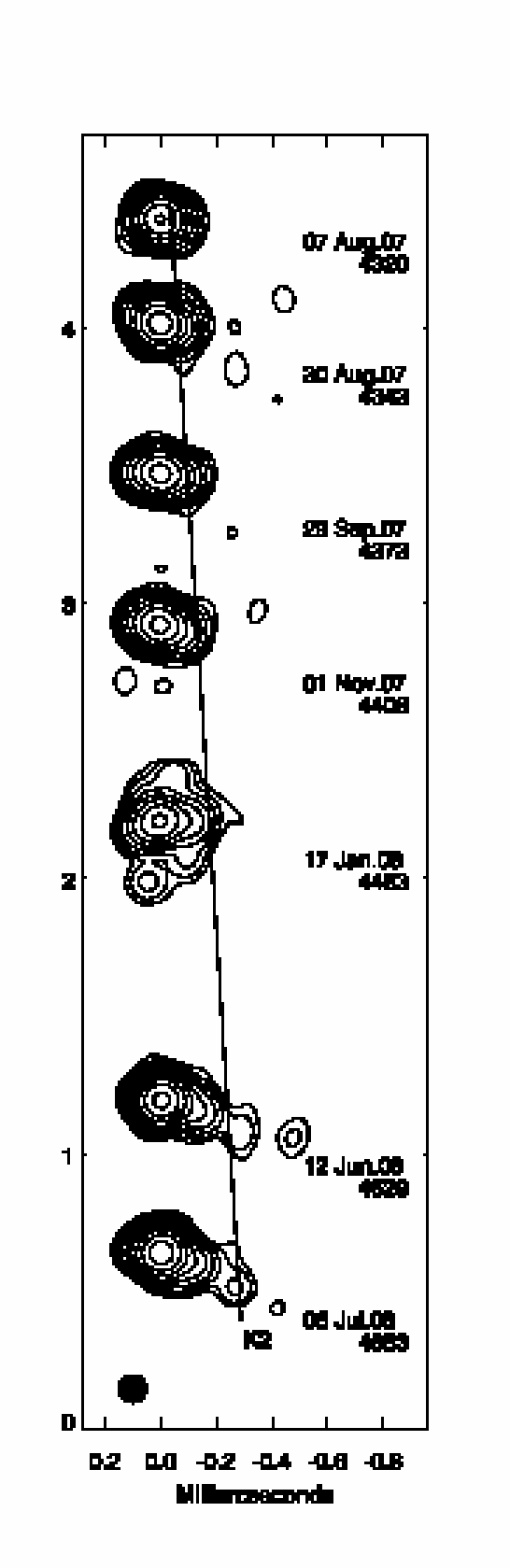}{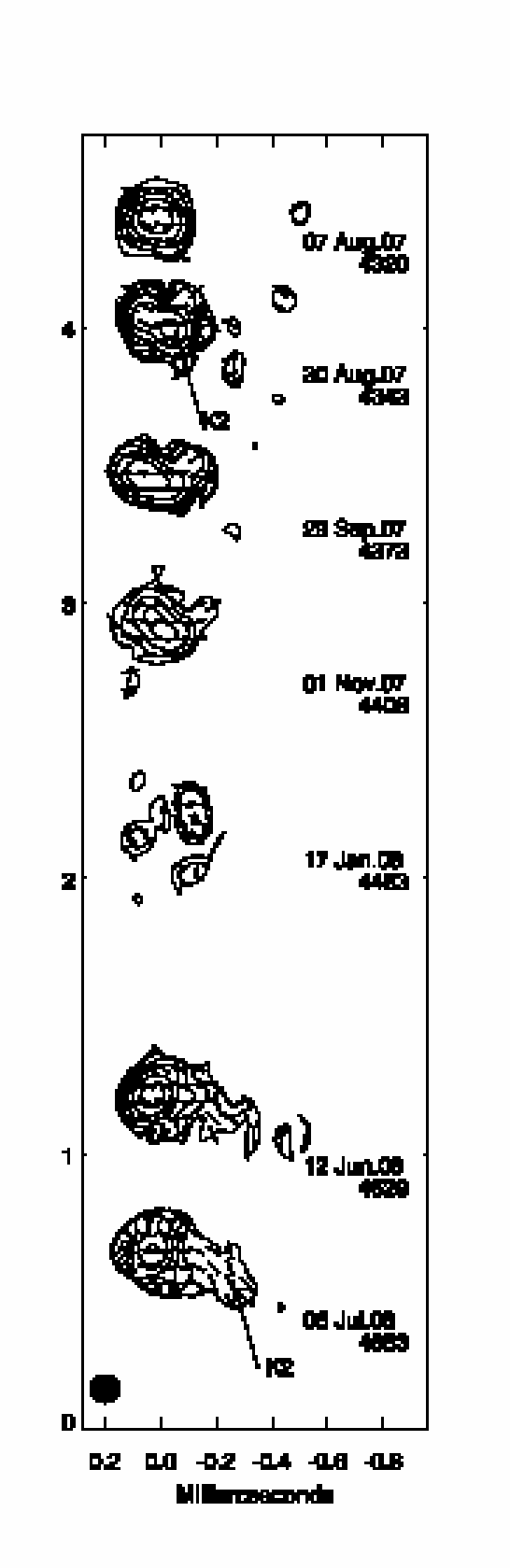}
\caption{43 GHz total ({\it left}) and polarized ({\it right}) intensity images of the quasar 3C~454.3
during period when $K2$ emerged from the core,
$S_{\rm peak}$=15.69~Jy/beam, $S_{\rm peak}^{\rm p}$=0.268~Jy/beam.  Total intensity
contours at 0.25,0.5,...64 \% of the peak and polarized intensity contours at 
2,4,...64 \% of the peak. Sticks over
the polarized intensity contours indicate the plane of polarization.} \label{imgK2}
\end{figure}

\begin{figure}
\epsscale{0.8}
\plottwo{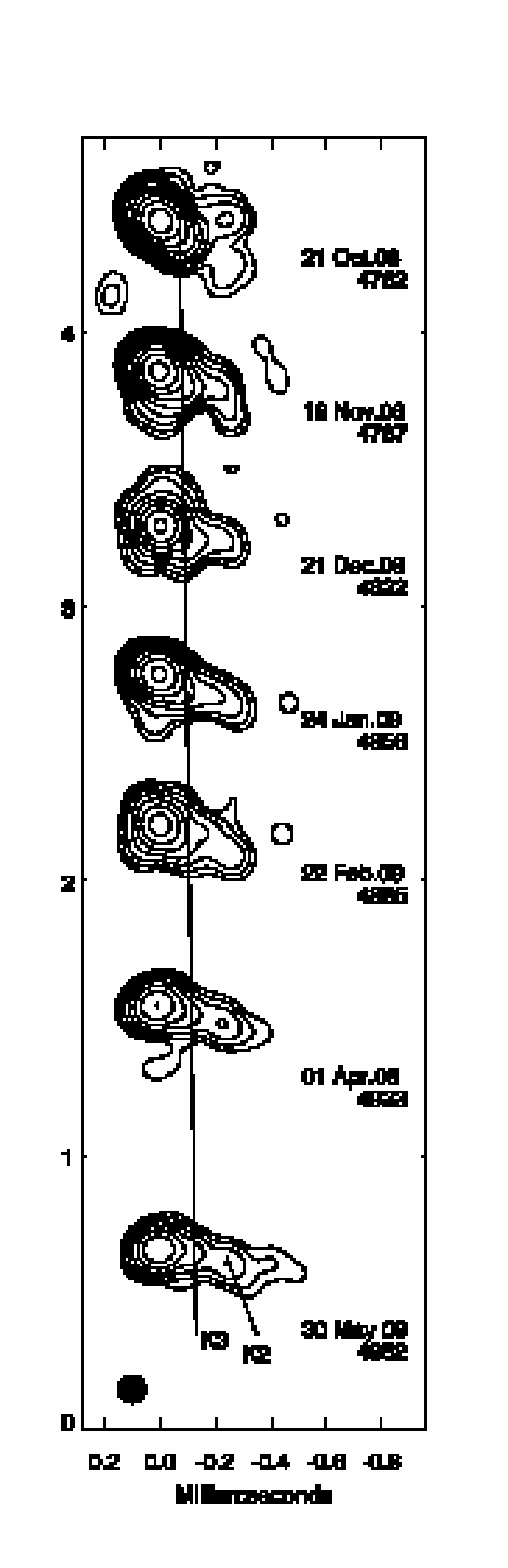}{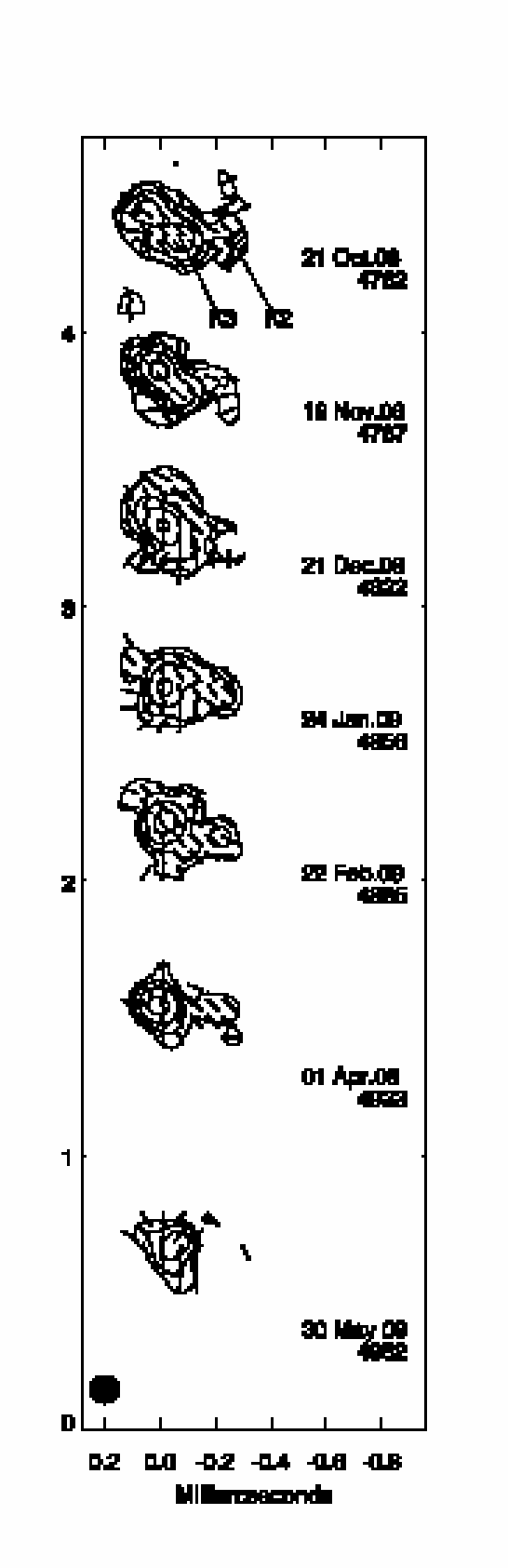}
\caption{43 GHz total ({\it left}) and polarized ({\it right}) intensity images of the quasar 3C~454.3
during period of $K3$ emerging from the core,
$S_{\rm peak}$=14.28~Jy/beam, $S_{\rm peak}^{\rm p}$=0.142~Jy/beam, beam=0.1$\times$0.1~mas$^2$. 
Total intensity contours at 0.25,0.5,...64 \% of the peak and polarized intensity contours at 2,4,...64 \%
of the peak. Sticks over
the polarized intensity contours indicate the plane of polarization.} \label{imgK3}
\end{figure}

\begin{figure}
\epsscale{1.0}
\plottwo{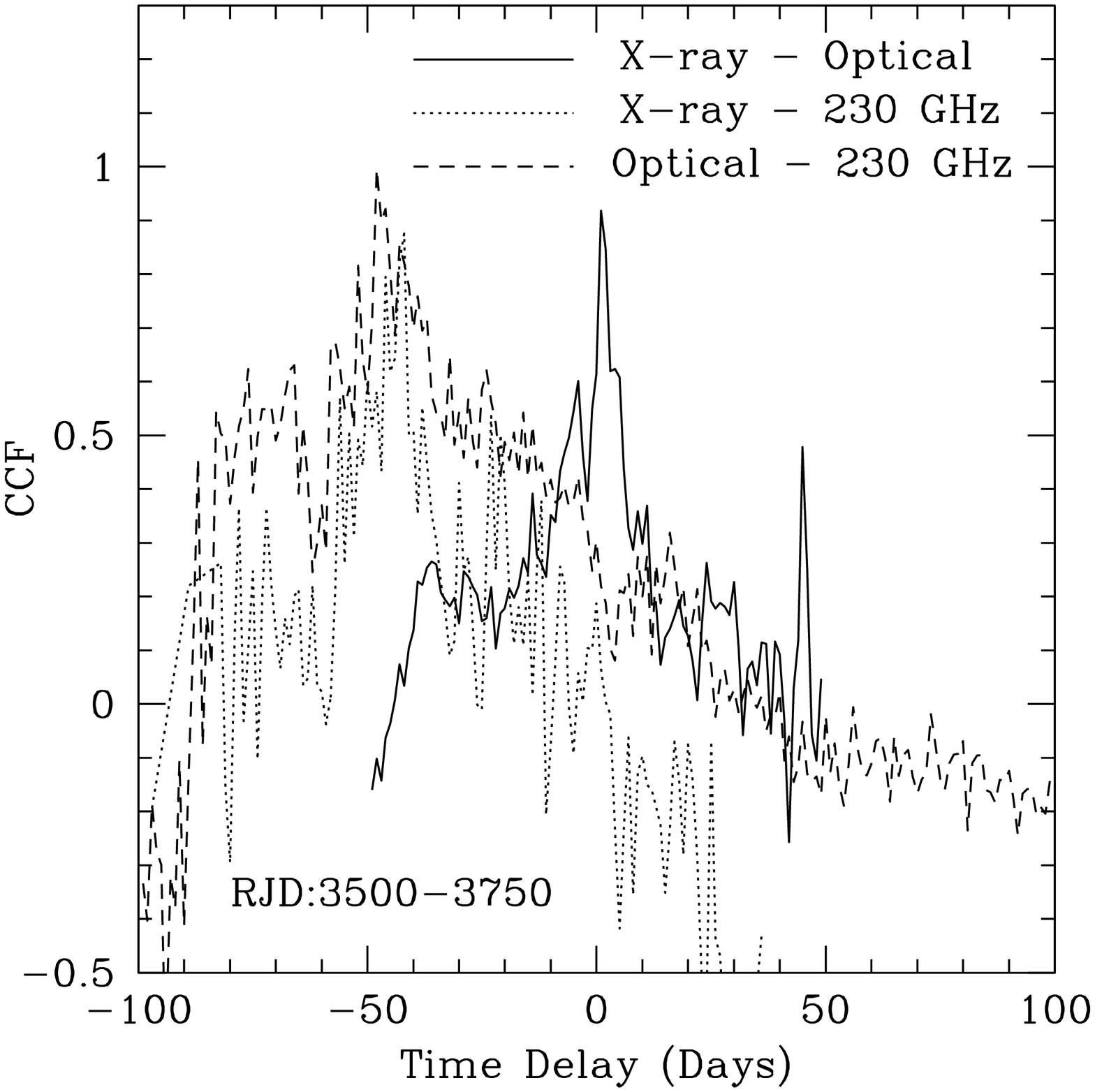}{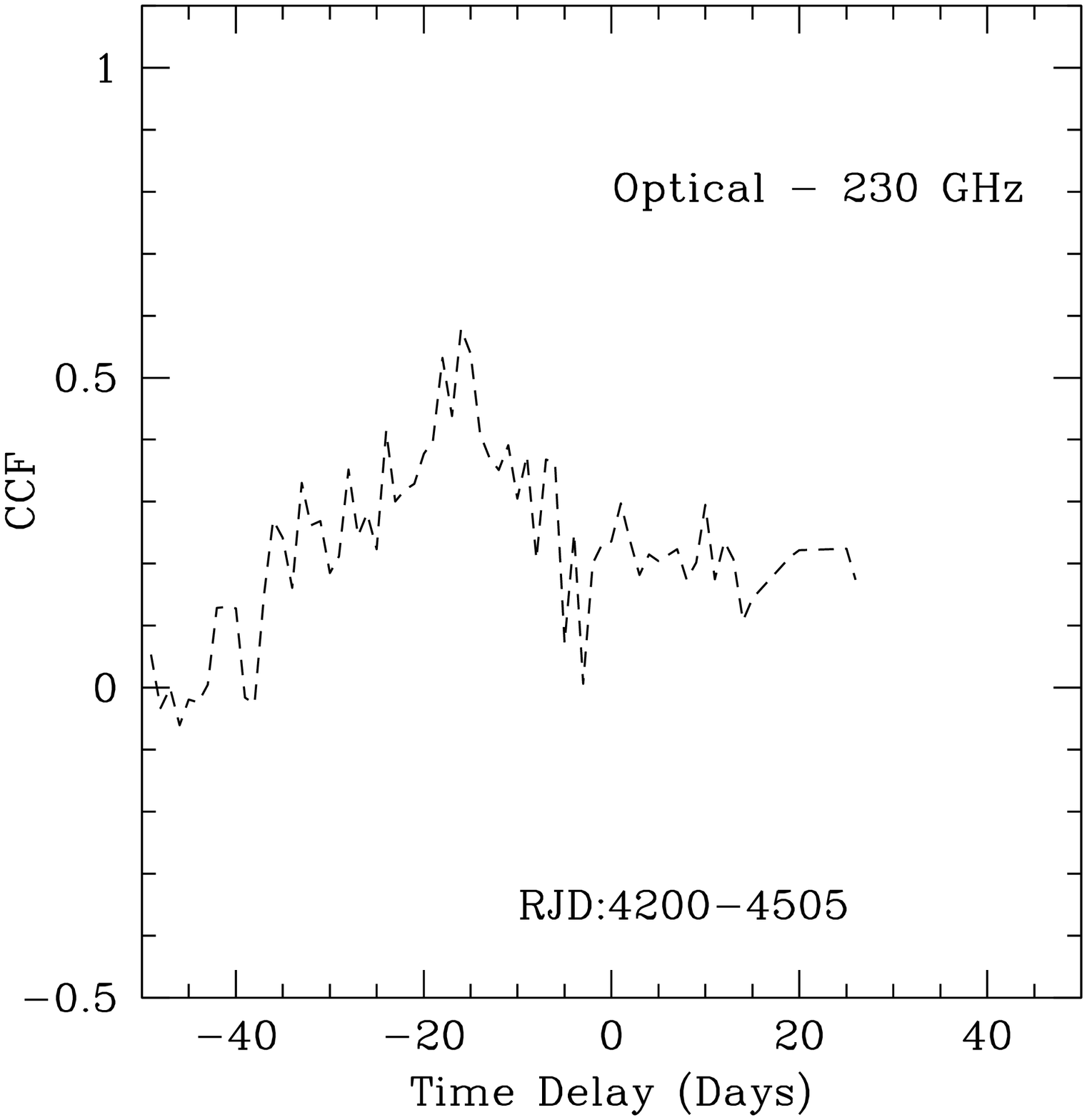}
\caption{ Cross-correlation function between X-ray, optical, and 230~GHz light curves for the 
indicated time intervals. A negative 
delay corresponds to higher frequency variations leading. 
} \label{ccf_p1}
\end{figure}

\begin{figure}
\epsscale{1.0}
\plottwo{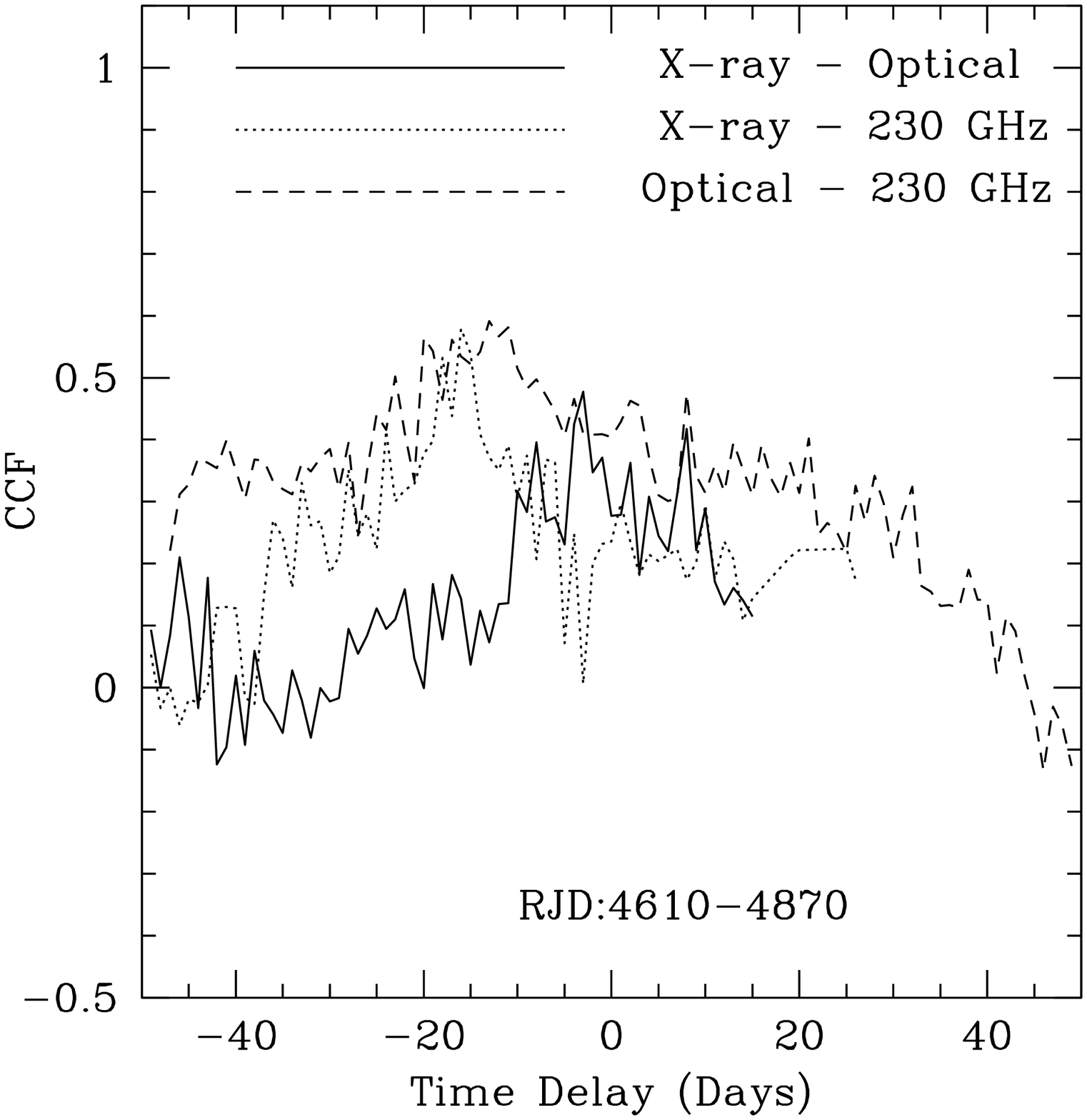}{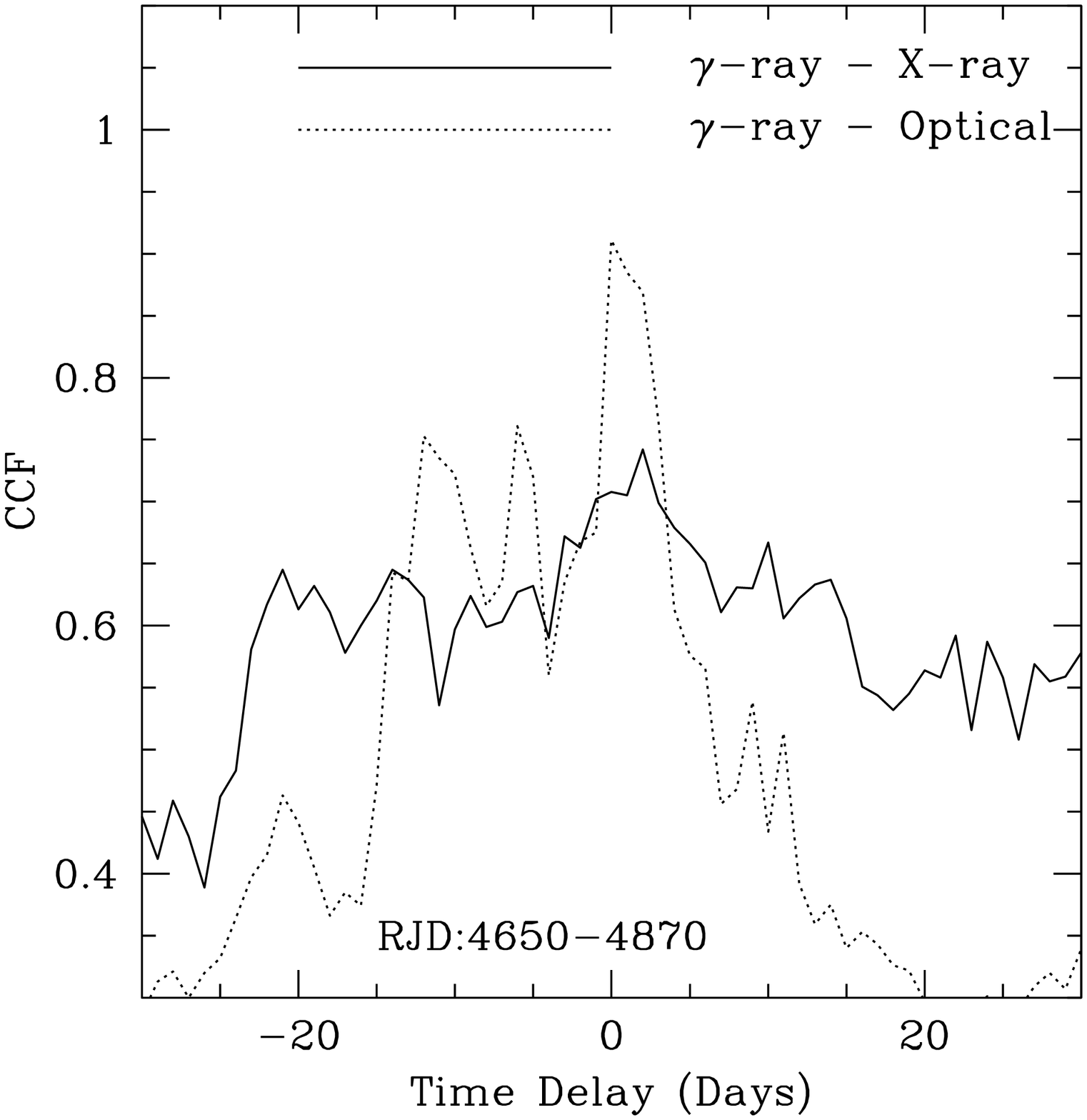}
\caption{{\it Left:} Cross-correlation function between X-ray, optical, and 230~GHz light curves
for the time interval RJD: 4610-4860.
{\it Right:} Cross-correlation function between $\gamma$/X-ray variations for the time interval RJD: 4689-4870, and between $\gamma$-ray and optical light curves 
for the time interval RJD: 4650-4870. A negative 
delay corresponds to higher frequency variations leading.
} \label{ccf_p3}
\end{figure}

\begin{figure}
\epsscale{1.0}
\plotone{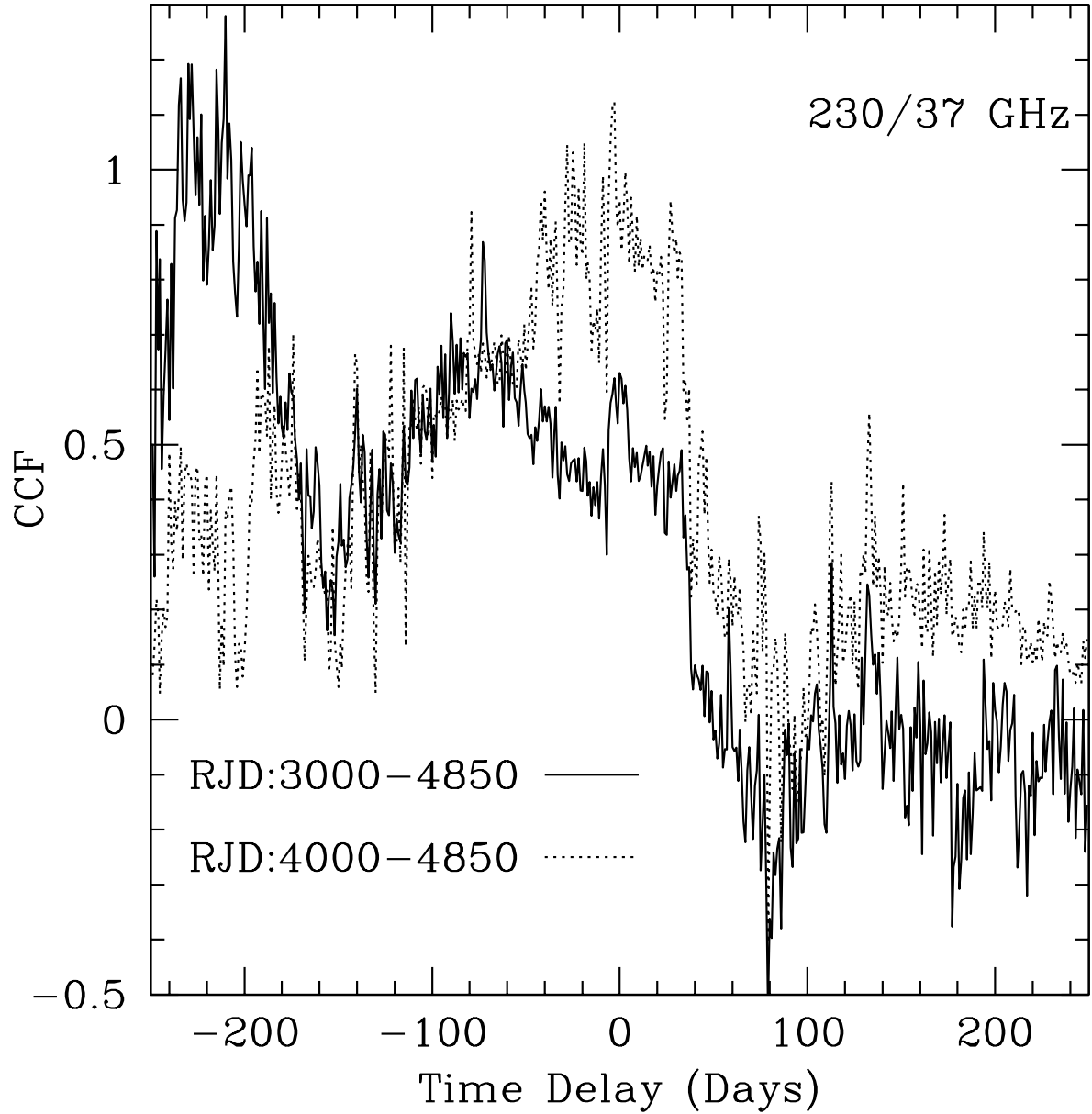}
\caption{Cross-correlation function between the 230~GHz and 37~GHz light curves
for periods RJD:3000-4850 (solid line) and RJD:4000-4850 (dotted line).
} \label{ccf_rad}
\end{figure}

\begin{figure}
\epsscale{1.0}
\plotone {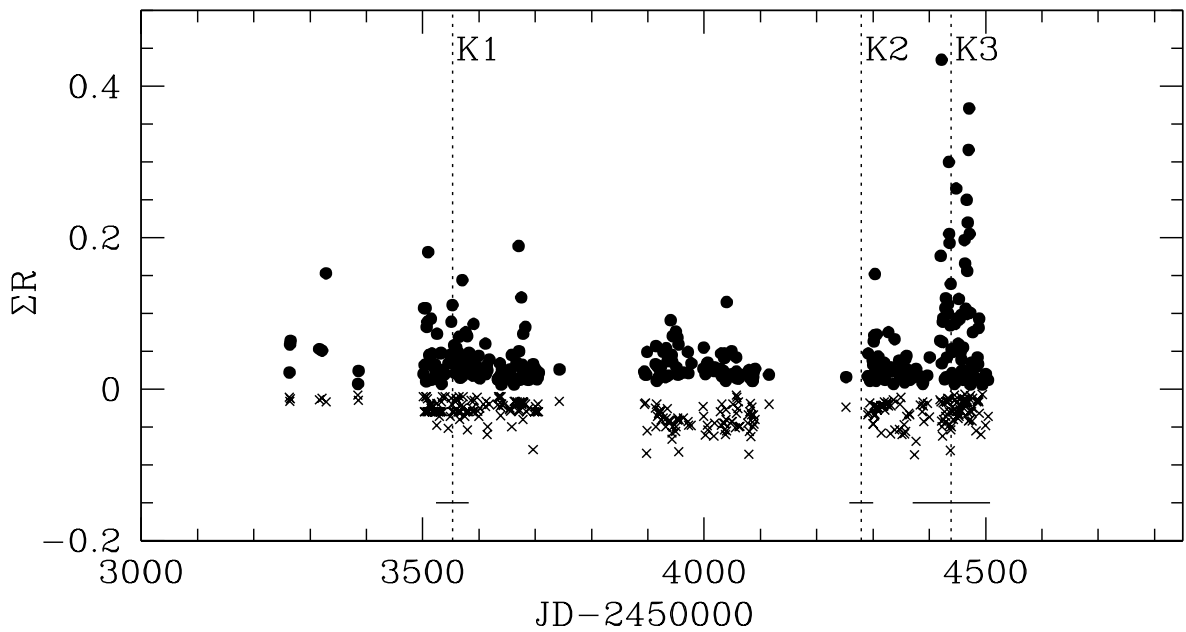}
\caption{Values of standard deviation of R-band magnitudes averaged within a day (filled circles).
The average uncertainties of individual measurements within a day are shown by crosses and are
given negative values to distinguish them from the standard deviations. Dotted lines
show times of ejections of superluminal components.} \label{Rsigma}
\end{figure}

\begin{figure}
\epsscale{1.0}
\plotone{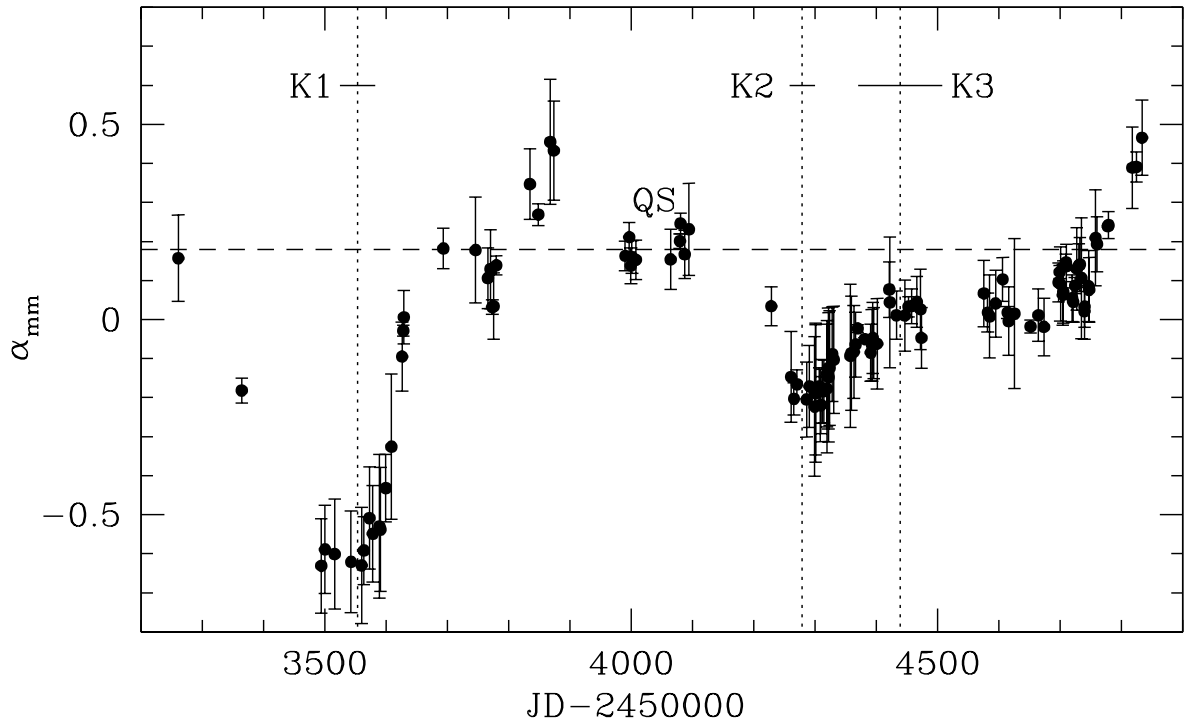}
\caption{Spectral index at 230-37~GHz (1-8~mm). Dashed line shows $\alpha_{mm}$ during 
a quiescent state, $QS$. Dotted lines show times of ejections of superluminal components
and solid line segments show uncertainties in the ejection times.} \label{RadInd}
\end{figure}

\begin{figure}
\epsscale{1.0}
\plotone{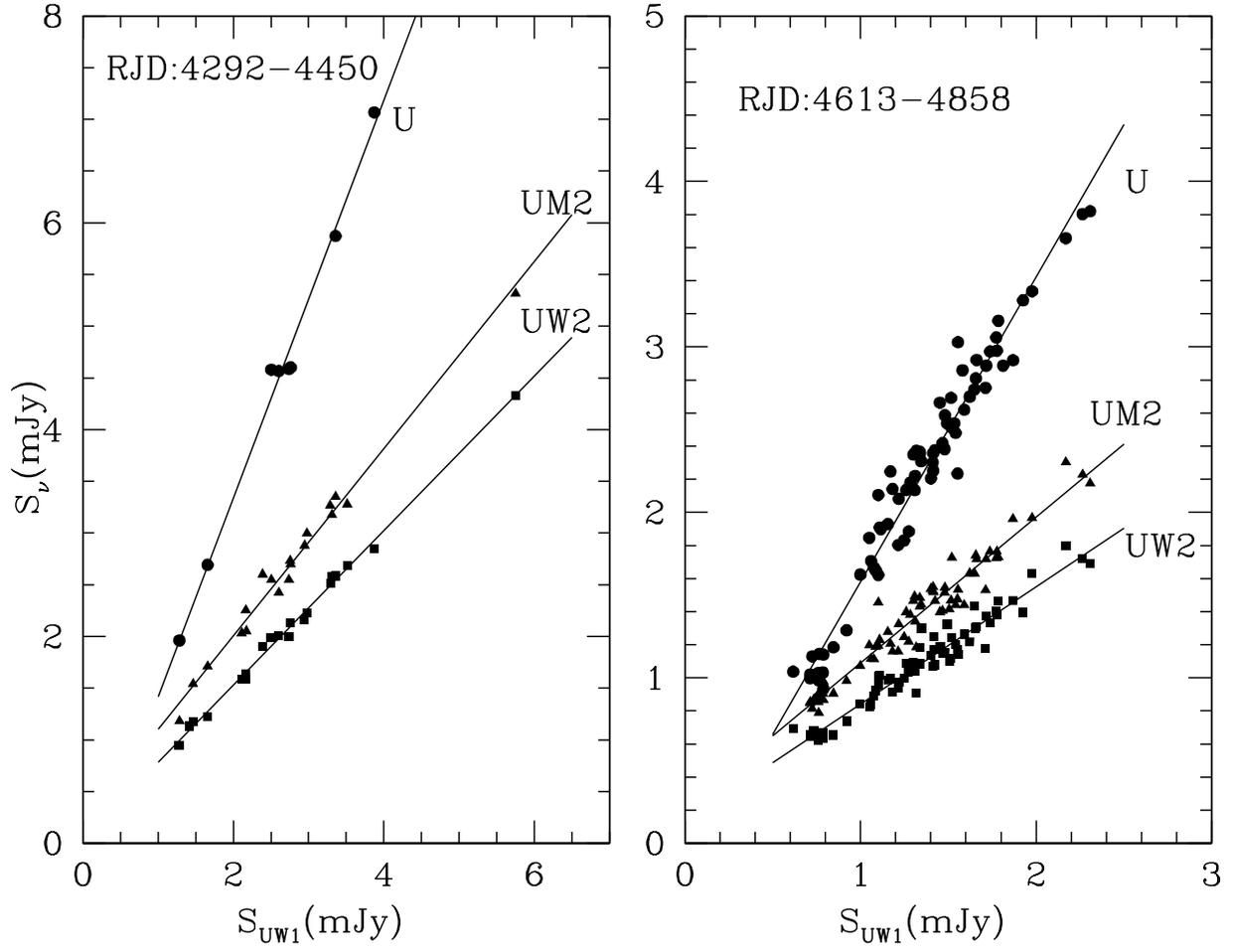}
\caption{Flux-Flux dependences for UVOT bands for periods RJD: 4292-4450 ({\it left})
and JD: 4613-4858 ({\it right}). } \label{UVff}
\end{figure}

\begin{figure}
\epsscale{1.0}
\plotone{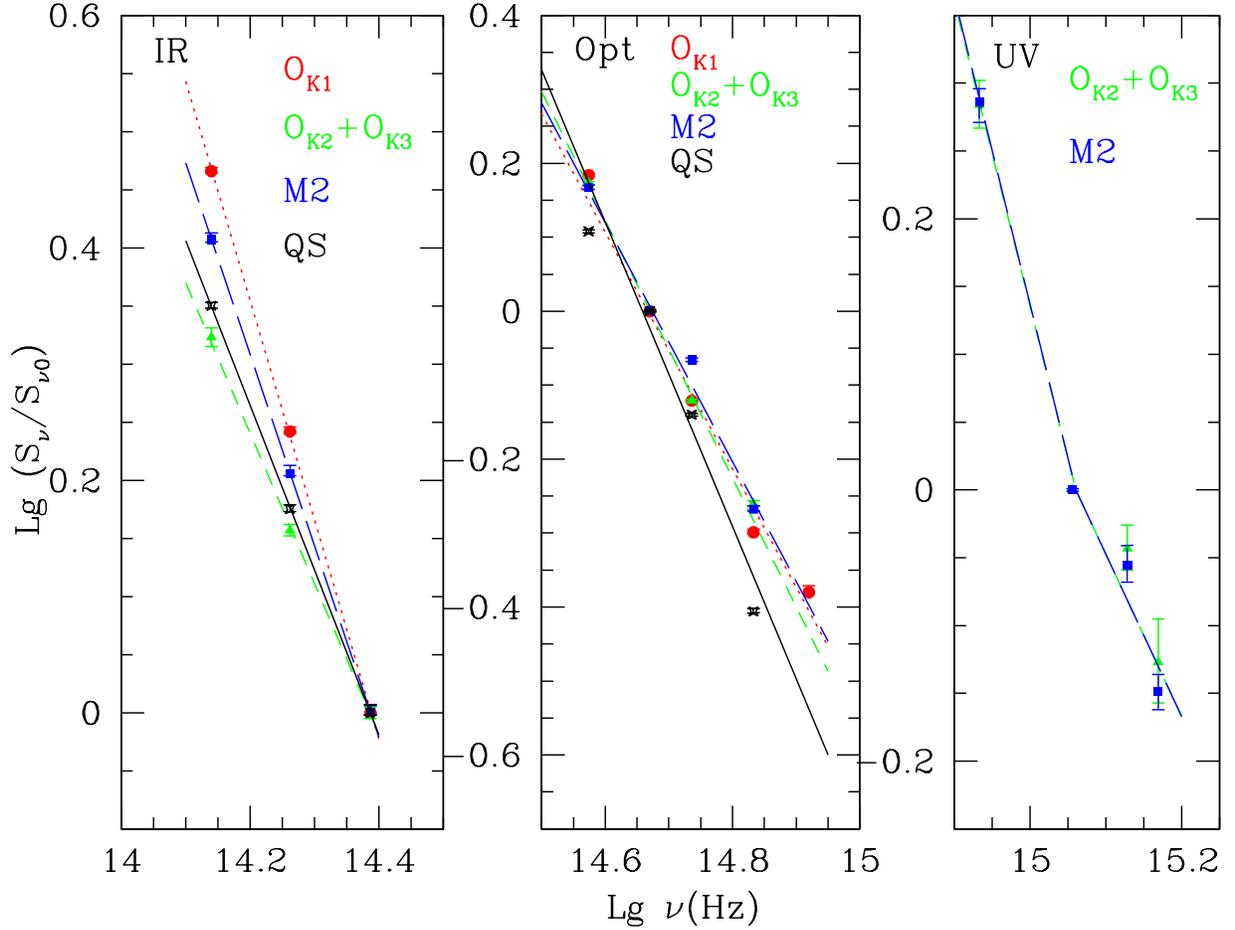}
\caption{Near-IR ({\it left}), optical ({\it middle}), and UV ({\it right}) relative spectral energy
distributions of synchrotron emission from 3C~454.3, normalized to K-band (Lg($\nu_\circ$)=14.140),
R-band (Lg($\nu_\circ$)=14.670), and UW1-band (Lg($\nu_\circ$)=15.056), respectively, during
different events.} \label{Specsyn}
\end{figure}

\begin{figure}
\epsscale{1.0}
\plotone{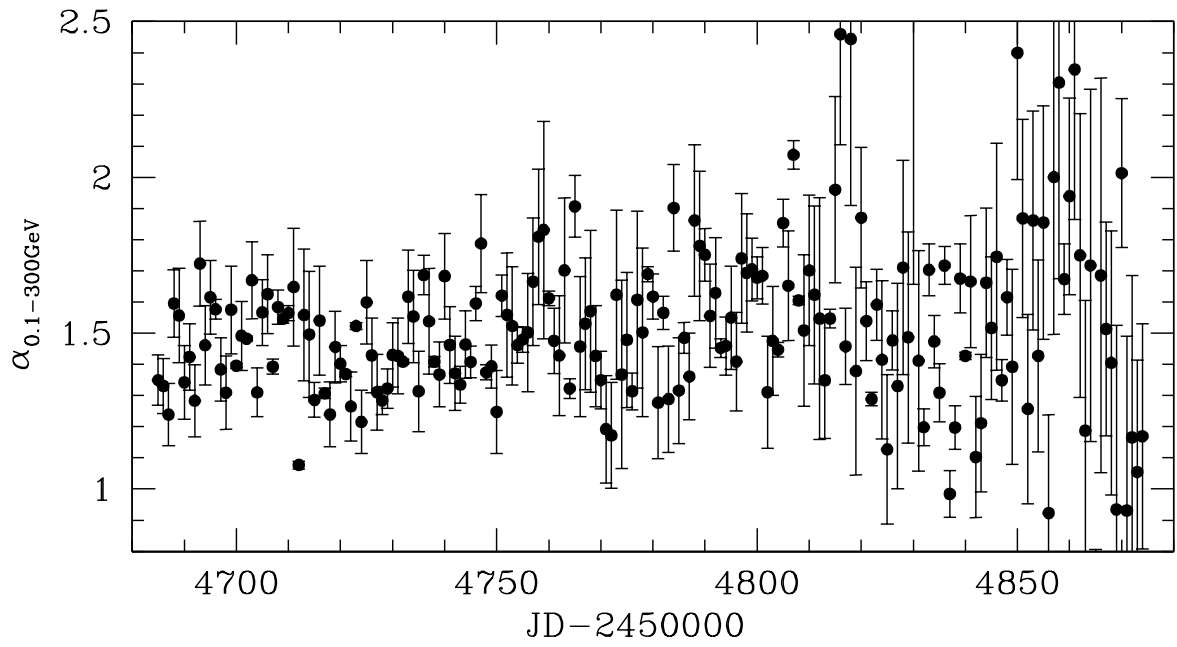}
\caption{Spectral index at 0.1-300~GeV.} \label{Gindex}
\end{figure}

\begin{figure}
\epsscale{1.0}
\plotone {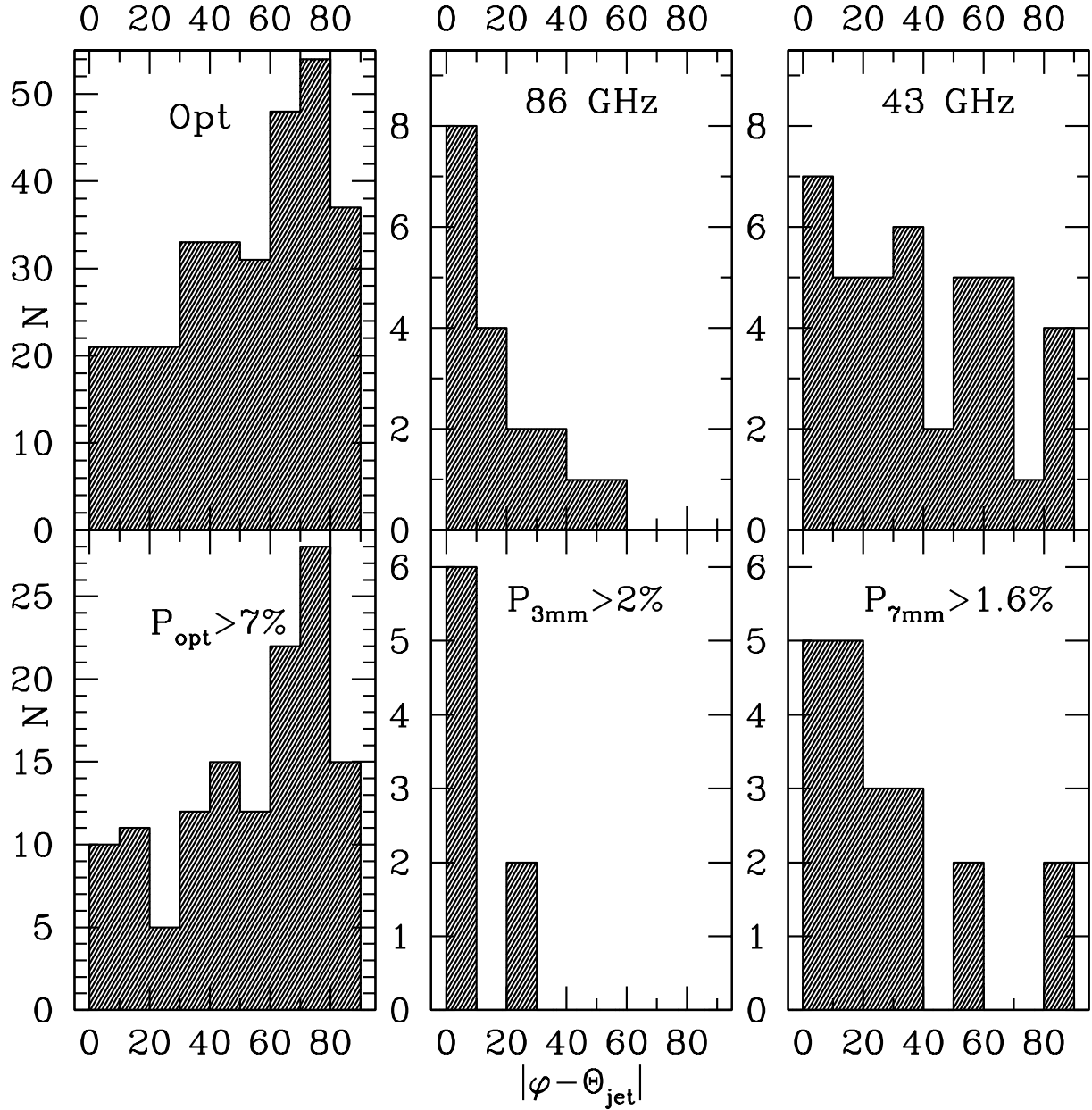}
\caption{Distributions of position angle of polarization with respect to the jet
direction at different wavelengths for all ({\it top panels}) and 
high polarization ({\it bottom panels}) measurements; values of $\varphi_{\rm 7mm}$ are corrected
for $RM$ according to \citet{J07}.}\label{hjet}
\end{figure}

\begin{figure}
\epsscale{1.0}
\plotone {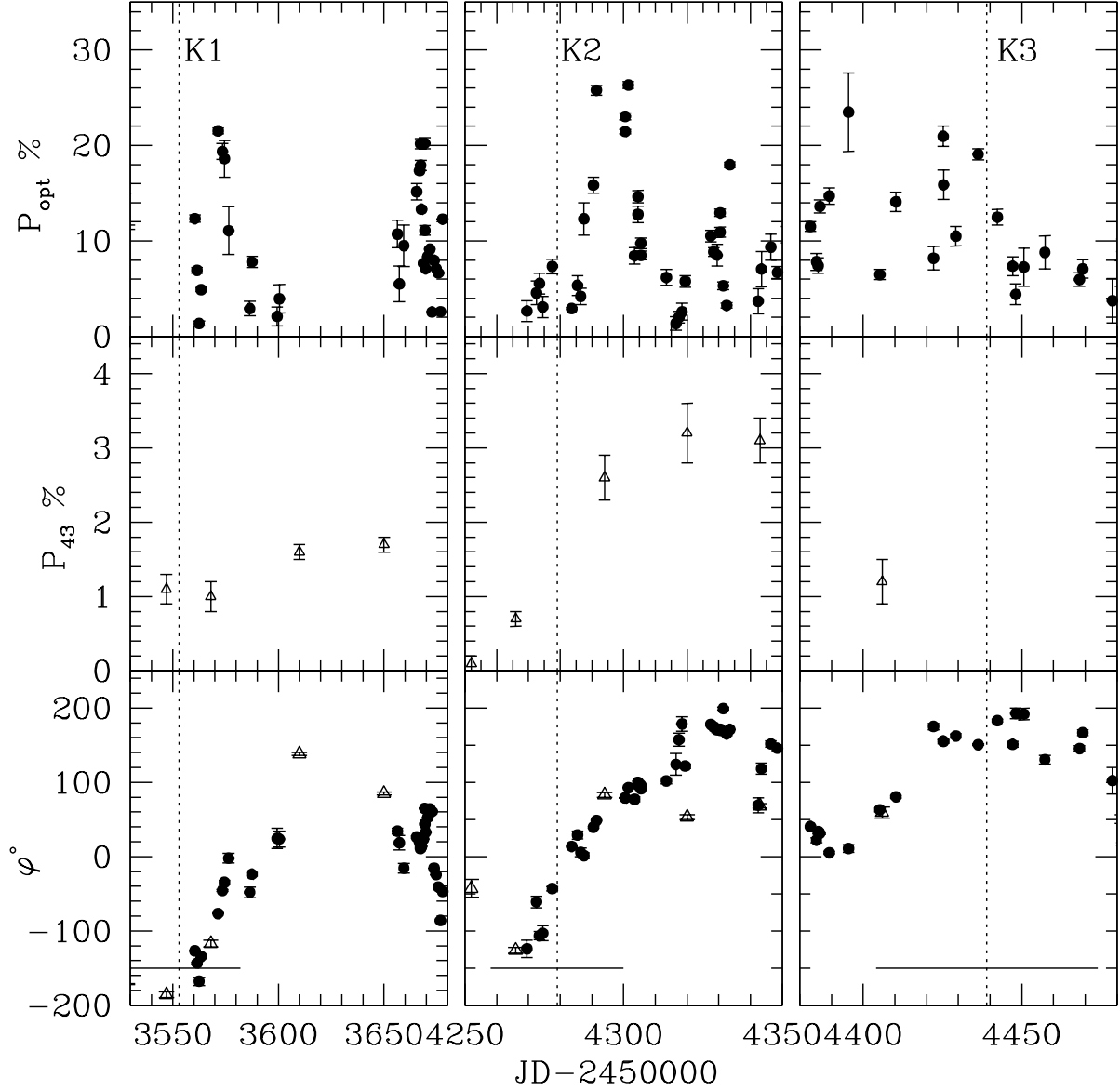}
\caption{Polarimetric behavior at optical wavelengths (filled circles) and in 43~GHz core (open
triangles) near the time of ejection of superluminal components (dotted lines), solid line segments show uncertainties in the ejection times.} 
\label{PolRot}
\end{figure}

\begin{figure}
\epsscale{1.0}
\plotone {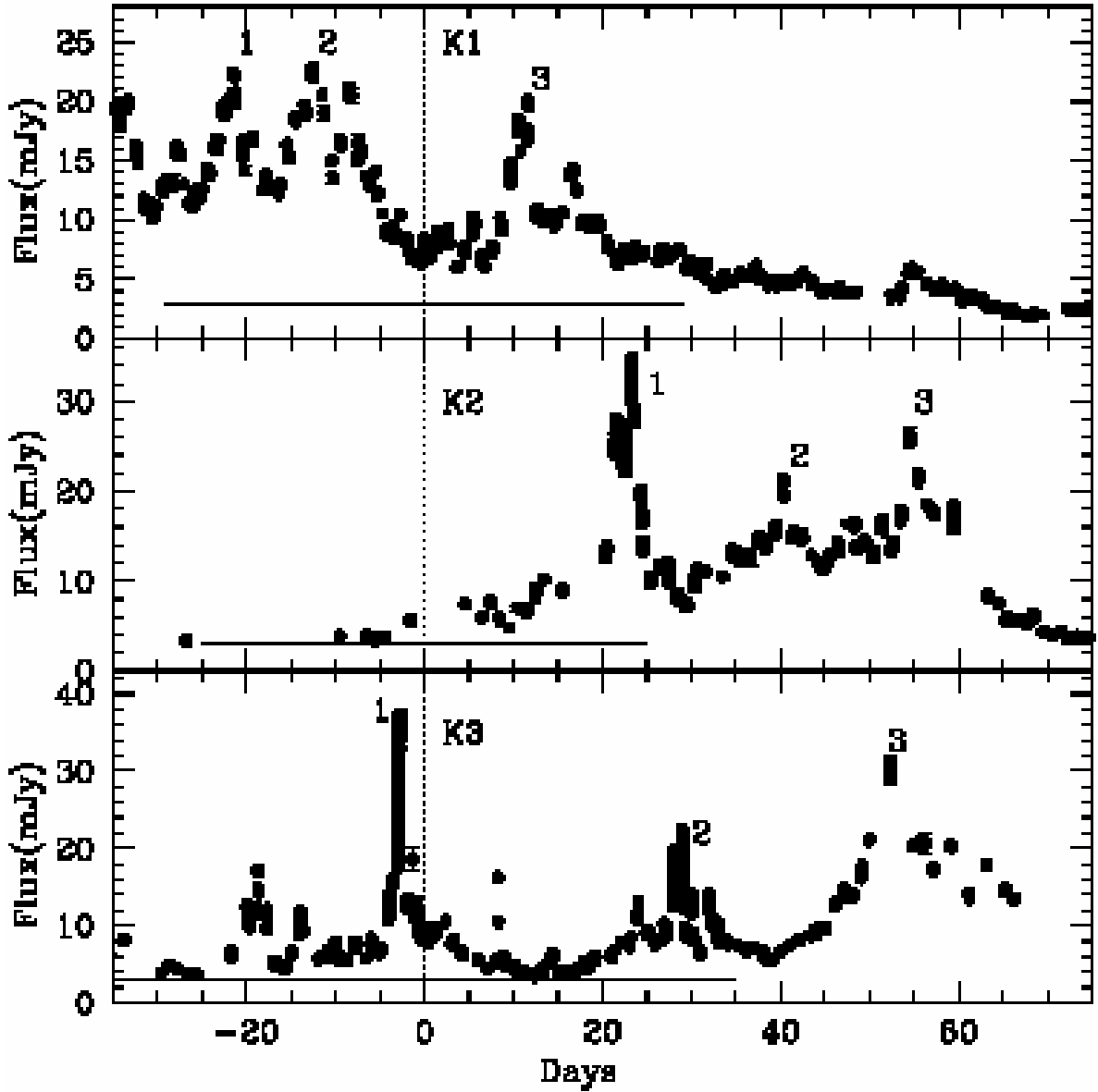}
\caption{Optical photometric variability  within $-$1$\sigma T_\circ$ to $+$2$\sigma T_\circ$ 
time interval relative to the time of the ejection, $T_\circ$ of each 
superluminal component (dotted line); values of $T_\circ$ and their uncertainties 
are given in Table~\ref{Kparm}.} 
\label{multi}
\end{figure}

Figure~\ref{OptHE} displays normalized high-energy light curves, with
the optical light curve superposed, during outburst $O_{\rm K1}$, the period covering 
$O_{\rm K2}$ plus $O_{\rm K3}$, and $M2$. Table~\ref{CCFparm} indicates that the
strongest correlation between optical and X-ray variations was during $O_{\rm K1}$,
with a delay $\lesssim$1~day. Inspection of Figure~\ref{OptHE} reveals that this correlation was
dominated by the first $\sim$15 days of the decaying branch of the outburst.
(Unfortunately, the rising branch was not observed at X-rays and has a seasonal gap at 
optical frequencies; see Fig.~\ref{mainLC}).
According to the discussion above, the optical outburst took place upstream of the 43~GHz core.
The X-ray outburst could have been produced via the EC mechanism, with the seed photons coming from the dusty torus,
as argued by \citet{SMM08}. Similarity in the decay at optical and X-ray 
frequencies implies that the primary cause of the outburst was an increase in the number
of relativistic electrons. However, the EC model predicts that the X-ray spectrum should be 
rather flat \citep{GHIS07}, whereas $\alpha_{\rm X}$ is steepest during $O_{\rm K1}$ (Table~\ref{SpecInd}). Although this is comfortably less than $\alpha_{\rm opt}$,
it is striking that the X-ray spectrum was flatter during the quiescent state. 
This can be understood if the X-ray emission during the quiescent state is dominated by the
EC process while the SSC mechanism contributes significantly in the X-ray 
production during $O_{\rm K1}$. A possible delay of X-ray variations by $\sim$1~day relative to
optical variations (Table~\ref{CCFparm}) supports such a hypotheses \citep{McH99}.

We have suggested that the optical flares observed within 1$\sigma$ uncertainty of the time of ejection
of $K1$ (Fig.~\ref{multi}) originated in the VLBI core as the result of compression of the
disturbance by standing shocks in the core. Figure~\ref{OptHE} shows that
two X-ray flares (designated as $X1$ and $X2$) occurred during the same time span, which
implies that these X-ray flares originated in the mm-wave core as well, probably
via the SSC mechanism. The SSC model is supported by tentative delays of several days  
relative to the optical peaks for flares $X1$ and $X2$, as can be inferred
from Figure~\ref{OptHE}. The higher
amplitude of the X-ray flares relative to the synchrotron flares (cf.\ Fig.~\ref{OptHE}) is also
expected in the SSC case if an increase in the number of relativistic electrons
causes the events \citep{RITABAN08}. 

The measurements during the $O_{\rm K2}$+$O_{\rm K3}$
period (middle panel of Fig.~\ref{OptHE}) show an increase in X-ray activity during 
$O_{\rm K2}$ and possibly $O_{\rm K3}$. 
\citet{VER09} have found a good correlation between $\gamma$-ray and optical variations
at the beginning of $O_{\rm K3}$, and argue that the 
$\gamma$-ray emission is dominated by EC scattering of photons from the broad-line region
by relativistic electrons in the jet. Although we agree that EC  
is probably the dominant mechanism for $\gamma$-ray production, we argue that the optical
outburst $O_{K3}$ appears related to the 43~GHz core located $\sim$20~pc from the
broad-line region. We propose that seed photons for the $\gamma$-ray production
are local to the mm-VLBI core, arising from a slower sheath surrounding the spine,
although synchrotron radiation from the disturbance itself 
plus the mm-wave core should contribute to the production of $\gamma$-rays as well
\citep{MAR10}.
The presence of such a sheath is supported by the wide range of apparent speeds observed
in the jet, from 0.12 to 0.53~mas~yr$^{-1}$\citep{J05}.  
 
The prominent mm-wave outburst $M2$ (see Fig.~\ref{mainLC}) featured moderate $\gamma$-ray, X-ray, and optical
flares. This high-frequency activity was especially pronounced during the early portion of the
mm-wave outburst, but tapered off during its later stages. It is possible that
a new superluminal knot was ejected during outburst $M2$, but could not be separated from 
$K3$ on the VLBA images. In this case, the multiple peaks of similar amplitude at 230~GHz, as well as the secondary 
optical maxima, can be explained by a knot passing through several standing shocks in the core 
region, as occurred during the passage of $K1$, $K2$, and $K3$ (Fig.~\ref{multi}). In the absence of a new ejection, the mm-wave outburst 
is probably not associated with the optical variability, rather it is caused by a change of the
viewing angle of $K3$ when the knot was $\sim$50~pc from the core (see \S 10.3).
The optical emission could originate in the core according to
polarization measurements, in a similar manner as proposed for the quasar 0420-014 \citep{FRANI07}.

The $\gamma$-ray light curve during event $M2$ (bottom panel of Fig.~\ref{OptHE})
is well correlated with the variations in optical flux, and the X-ray variations
correlate with both the optical and $\gamma$-ray light curves, albeit less strongly.
The $\gamma$-ray and optical variations were simultaneous
within a day (Table~\ref{CCFparm}), while the X-ray variations preceded the
optical variations by 3$\pm$2~days, although the X-ray measurements are sparse. 
The value of the X-ray spectral index during $M2$
is similar to $\alpha_{\rm X}$ during  $O_{\rm K1}$. We suggest that the X-ray emission 
during $M2$ was produced
via the SSC mechanism  by the scattering of a wide spectrum of synchrotron 
photons --- from far-IR to optical frequencies --- by relativistic
electrons with Lorentz factors $\gamma\sim$10$^{2-3}$.
This can explain the moderate correlation between the X-ray and optical variations and smoother X-ray
variability. If the delay of the optical with respect to the X-ray variations is 
real, it can be understood in the manner proposed for the quasar 3C~279 \citep{RITABAN08},
in which relativistic electrons are accelerated gradually. In this case, the acceleration mechanism should be different 
from that for $O_{\rm K2}$ and $O_{\rm K3}$, and probably involves turbulence in the 
core. This is consistent with a significantly lower degree of polarization in the
43~GHz relative to optical polarization if the two emission regions are co-spatial
but occupy different volumes (Table~\ref{O7evpa}). 

Our analysis of the $\gamma$-ray and optical light curves confirms a strong 
correlation between the two wavelengths found by \citet{BON09}. The correlation persists beyond $M2$, with 
$\gamma$-ray and optical variations having similar amplitudes of variability (see Fig.~\ref{OptHE}).  
\citet{BON09} have concluded that the $\gamma$-ray outburst during $M2$
was dominated by EC scattering of IR/optical photons from the accretion disk and/or broad-line
region by electrons with $\gamma\sim$10$^{3-4}$. Although this may seem a reasonable explanation 
for the observed strong correlation between $\gamma$-ray and optical variations,  polarization
observations (see Fig.~\ref{Poldat} and Table~\ref{O7evpa}) indicate that during $M2$
(RJD: 4550-4840) the position angle of the optical polarization tended to align with $\varphi_{\rm 43}$
in the core. This implies that the variable optical emission 
arose in the vicinity of the core, as discussed above.
For this reason, we favor the alternative scenario, wherein  the
$\gamma$-rays are produced by scattering of synchrotron seed photons from the sheath of the jet.

We can be more specific regarding the nature of the seed photons by considering the
steepening of the $\gamma$-ray
spectrum above 2~GeV reported by \citet{Abdo09b}, from a spectral index of 1.3 to 2.5,  
along with steepening of the spectral index of synchrotron emission 
from 1.4 at near-IR to 2.3 at soft UV wavelengths (Table \ref{SpecInd}).
We can explain the steepening of the $\gamma$-ray spectrum, as well as
the rapid fall-off in the synchrotron emission toward higher ultraviolet frequencies
by a steepening of the electron energy distribution
above an energy $\gamma \sim 10^4$ in rest-mass units, where we assume a magnetic field
strength $\lesssim 1$ G. Recent theoretical calculations by \citet{REY09}
show that, for an inhomogeneous source, more pronounced steepening than by 0.5 in the source's
integrated spectral index is possible due to a combination of synchrotron losses
and geometrical effects. If the 2~GeV $\gamma$-rays are produced by the EC process, 
the seed photons
should have observed frequencies near $10^{12}(\Gamma/25)^{-2}(\delta/25)^{-2}$ Hz. This favors
either the putative dust torus \citep{SMM08} or sheath of the jet as the source of the seed photons.
The similarity of the $\gamma$-ray and optical light curves, including very rapid variations (see \S 7), can then be explained as a consequence
of the variations being caused by changes in the number of electrons with
$\gamma \sim 3\times 10^3$-$10^4$ over a limited volume within a given disturbance.
 
\section {Conclusions}
The optical, X-ray, and radio light curves of blazars that have been
assembled and analyzed over the past several years \citep[e.g.,][]{VIL04,BACH06,RAI08b,RITABAN08,VIL09,MAR10}
strongly support the conclusion that the long-term variability at all wavelengths 
is governed by physical activity in the relativistic jet that we can image with the VLBA. 
Our multifrequency study of the quasar 3C~454.3 during the dramatic activity
of 2005-2008, including intense monitoring of the structure of the parsec-scale jet, reveals 
strong connections among the observed optical, X-ray, and $\gamma$-ray outbursts, 
its spectral and polarization behavior, and disturbances propagating down the jet.

We have found that the location of the most prominent X-ray and optical outburst (in 2005)
was within$\sim$10~pc from the BH. The event was the result of a disturbance propagating 
down the jet through a photon field that included synchrotron radiation from both the high-$\Gamma$
spine and slower sheath of the jet, and perhaps a hot dust torus \citep{SMM08} with X-ray 
emission produced  by both the SSC and EC mechanisms.
 
Our intensive monitoring with the VLBA at 43~GHz reveals that interaction between disturbances
in the jet and the mm-wave core results in X-ray and optical outbursts and
can cause optical intra-day variability. This implies that the mm-wave core
is a physical feature of the jet, e.g., a standing shock \citep{CAW06,J07,FRANI07},
that is different from the cm-wave core, the location of which is probably determined
by the jet opacity. 

We have observed three events when the passage of a superluminal knot
through the core coincided with a series of optical outbursts
and an increase in the X-ray emission. We infer that the structure of the 
optical outbursts reflects the fine-scale structure in the core, a scenario supported by super-resolved
VLBA images. The X-ray production in the core region can be dominated by either the
EC or SSC mechanism, depending on the properties of the disturbance, as
suggested by \citet{KG07}.

\citet{VER09} have reported a good correlation between $\gamma$-ray and optical
variations in autumn 2007, which coincides with a series of optical outbursts
that we relate to an event in the core. This suggests that $\gamma$-ray emission 
is also produced during interactions within the core. If the $\gamma$-rays originate via 
the EC mechanism, then the seed photons should be local to the 43~GHz core and
could come from the sheath of the jet. 

We have observed three episodes of rotation of the position angle of optical polarization
near the times of passage of superluminal components through the core.
We interpret this as evidence that the mm-wave core of
3C~454.3 is located near the end of the acceleration and collimation zone of the jet flow, where
the magnetic field has a toroidal structure \citep[e.g.,][]{VLAH06}.
Our findings agree with other studies reporting such rotations related to disturbances in
the jet \citep{MAR08,LAR08,MAR10,Abdo10}. This implies that rotation of the optical polarization 
position angle is a common occurrence during the early stages of flares in blazars. 
In the case of 3C~454.3, we tentatively infer an inverse
relationship between the Lorentz factor of the superluminal component and the rate of
rotation of the optical EVPA, as expected if higher values of $\Gamma$ are reached
farther from the BH where the jet has expanded to a larger cross-sectional radius. 

Our study of the $\gamma$-ray bright quasar 3C~454.3 demonstrates an obvious connection
between events at different wavelengths, the complexity of their relationships, and the primary
role of the parsec-scale jet in the generation of these observed events.
The ability of the Fermi Large Area Telescope to produce well-sampled $\gamma$-ray light curves
for the first time, combined with the ultra-high resolution imaging  of the VLBA and flux and polarization
curves at other wave bands,
provides an unprecedented opportunity to locate the events within the most compact
regions of the jets of blazars. This endeavor makes possible a deeper understanding of the dynamics
of relativistic jets and physics of the radiative processes. 

\acknowledgments
The authors thank the referee for very useful comments. 
The research at Boston University (BU) was funded in part by NASA Fermi Guest Investigator grants NNX08AV65G and NNX08AV61G, and through Astrophysical Data Analysis Program grant NNX08AJ64G, and by the National Science Foundation (NSF) through grant AST-0907893. The VLBA is an instrument of the National Radio Astronomy Observatory, a facility of the National Science Foundation operated under cooperative agreement by Associated Universities, Inc. The St. Petersburg State University 
team acknowledges support from RFBR grant 09-02-00092.
The research at the IAA-CSIC is supported in part by the  Spanish ``Ministerio de Ciencia e
Innovaci\'on'' through grant AYA2007-67626-C03-03. The effort at Steward Observatory
was funded in part by NASA through Fermi Guest Investigator grant NNX08AV65G. The Mets\"ahovi team
acknowledges support from the Academy of Finland. We are grateful to the IRAM Director for providing discretionary observing time at the 30\,m Telescope. The PRISM camera at
Lowell Observatory was developed by K.\ Janes et al. at BU and Lowell Observatory, with funding from
the NSF, BU, and Lowell Observatory. The Calar Alto
Observatory is jointly operated by the Max-Planck-Institut f\"ur Astronomie and the Instituto de 
Astrof\'{\i}sica de Andaluc\'{\i}a-CSIC. The Liverpool Telescope is operated on the island of La Palma by Liverpool John Moores University in the
Spanish Observatorio del Roque de los Muchachos of the Instituto de Astrofisica de Canarias, with funding
from the UK Science and Technology Facilities Council. 
Partly based on data taken and assembled by the WEBT collaboration and stored in the WEBT archive at the Osservatorio Astronomico di Torino - INAF (http://www.oato.inaf.it/blazars/webt/).
The Submillimeter Array is a joint project 
between the Smithsonian Astrophysical Observatory and the Academia Sinica Institute of Astronomy and Astrophysics, and is funded by the Smithsonian Institution and the Academia Sinica. 
The IRAM 30\,m telescope is supported by INSU/CNRS (France), MPG (Germany) and IGN (Spain).
We acknowledge the Swift team for providing the public archive of Swift-XRT data, and we acknowledge Fermi GI grant NNX09AU07G for supporting the public archive of processed data.

\clearpage

\begin{deluxetable}{lrrrrrrrrr}
\singlespace
\tablecolumns{10}
\tablecaption{\bf Parameters of Moving Components \label{Kparm}}
\tabletypesize{\footnotesize}
\tablehead{
\colhead{Knot}&\colhead{$T_\circ$}&\colhead{$T_\circ$}&\colhead{$S_{max}$}&\colhead{$T_{S_{max}}$}&\colhead{$\mu$}&\colhead{$\beta_{app}$}&\colhead{$\Gamma$}&\colhead{$\delta$}&\colhead{$\Theta_\circ$} \\
\colhead{}&\colhead{yr}&\colhead{RJD}&\colhead{Jy}&\colhead{RJD}&\colhead{mas~yr$^{-1}$}&\colhead{}&\colhead{}&\colhead{}&\colhead{$^\circ$}\\
\colhead{(1)}&\colhead{(2)}&\colhead{(3)}&\colhead{(4)}&\colhead{(5)}&\colhead{(6)}&\colhead{(7)}&\colhead{(8)}&\colhead{(9)}&\colhead{(10)}
}
\startdata
$K1$ &2005.50$\pm$0.08&3553$\pm$29&7.00$\pm$0.34&3772&0.10$\pm$0.02&4.9$\pm$0.9&12.3&23.4&1.0 \\
$K2$ &2007.49$\pm$0.07&4279$\pm$25&3.81$\pm$0.56&4343&0.18$\pm$0.05&8.3$\pm$2.3&10.7&17.6&2.5 \\
$K3$ &2007.93$\pm$0.10&4439$\pm$35&11.74$\pm$0.52&4720&0.09$\pm$0.03&4.1$\pm$1.4&24.7&49.1&0.2 \\ 
\enddata
\tablecomments{Columns: 1 - name of component; 2 - time of ejection from the core in years; 3 -  time of ejection from the core in RJD; 
4 - maximum flux of component, 5 - epoch of the maximum flux of component; 6 - proper motion;
7 - apparent speed; 8 - Lorentz factor; 9 - Doppler factor; 10 - viewing angle}
\end{deluxetable}

\begin{deluxetable}{lrrrrrr}
\singlespace
\tablecolumns{7}
\tablecaption{\bf Results of Correlation Analysis \label{CCFparm}}
\tabletypesize{\footnotesize}
\tablehead{
\colhead{}&\multicolumn{2}{c}{RJD:3502-3619}&\multicolumn{2}{c}{RJD:4292-4450}&\multicolumn{2}{c}{RJD:4613-4858} \\
\colhead{Waves}&\colhead{$f_{\rm max}$}&\colhead{$\tau$}&\colhead{$f_{\rm max}$}&\colhead{$\tau$}&\colhead{$f_{\rm max}$}&\colhead{$\tau$} \\
\colhead{}&\colhead{}&\colhead{days}&\colhead{}&\colhead{days}&\colhead{}&\colhead{days}\\
\colhead{(1)}&\colhead{(2)}&\colhead{(3)}&\colhead{(4)}&\colhead{(5)}&\colhead{(6)}&\colhead{(7)}
}
\startdata
$\gamma$-ray/X-ray&\nodata&\nodata&\nodata&\nodata&+0.75&+2$\pm$3 \\
$\gamma$-ray/Opt&\nodata&\nodata&\nodata&\nodata&+0.91&+0$\pm$1 \\
X-ray/Opt&+0.92&+1$\pm$1&\nodata&\nodata&+0.48&$-$3$\pm$2 \\
X-ray/230&+0.88&$-$42$\pm$3&\nodata&\nodata&+0.58&$-$16$\pm$3 \\
Opt/230  &+0.98&$-$48$\pm$5&+0.46&$-$21$\pm$2&+0.59&$-$13$\pm$4 \\
\enddata
\tablecomments{Columns: 1 - two bands used for correlation analysis; $f_{\rm max}$ -
maximum coefficient of linear correlation; $\tau$ - time delay between the two bands at
the maximum coefficient, negative delay corresponds to higher frequency variations leading}
\end{deluxetable}

\begin{deluxetable}{lrrrrrrrrr}
\singlespace
\tablecolumns{10}
\tablecaption{\bf Parameters of Total Flux Outbursts \label{Outparm}}
\tabletypesize{\footnotesize}
\tablehead{
\colhead{}&\colhead{}&\colhead{Optical}&\colhead{}&\colhead{}&\colhead{230 GHz}&\colhead{}&\colhead{}&\colhead{37 GHz}&\colhead{} \\
\colhead{Name}&\colhead{$T_{\rm opt}$}&\colhead{$E_{\rm opt}$}&\colhead{$w_{\rm opt}$}&\colhead{$T_{\rm 230}$}&\colhead{$E_{\rm 230}$}&\colhead{$w_{\rm 230}$}&\colhead{$T_{\rm 37}$}&\colhead{$E_{\rm 37}$}&\colhead{$w_{\rm 37}$} \\
\colhead{}&\colhead{RJD}&\colhead{mJy~day}&\colhead{days}&\colhead{RJD}&\colhead{Jy~day}&\colhead{days}&\colhead{RJD}&\colhead{Jy~day}&\colhead{days}\\
\colhead{(1)}&\colhead{(2)}&\colhead{(3)}&\colhead{(4)}&\colhead{(5)}&\colhead{(6)}&\colhead{(7)}&\colhead{(8)}&\colhead{(9)}&\colhead{(10)}
}
\startdata
$O_{\rm K1}/M_{\rm K1}$&3501$\pm$1&2194&47&3534$\pm$5&9343&104&3700$\pm$5&6200&180 \\
&3537$\pm$1&166&15&\nodata&\nodata&\nodata&\nodata&\nodata&\nodata \\
&3561$\pm$1&115&18&\nodata&\nodata&\nodata&\nodata&\nodata&\nodata  \\
&3673$\pm$1&96&19&\nodata&\nodata&\nodata&\nodata&\nodata&\nodata  \\
$M1$&\nodata&\nodata&\nodata  &3781$\pm$5&1140&38&3788$\pm$5&528&24  \\
$O_{\rm K2}/M_{\rm K2}$&4301$\pm$1&373&19&4330$\pm$5&1985&50&4331$\pm$5&1871&104 \\
&4334$\pm$1&170&12&\nodata&\nodata&\nodata&\nodata&\nodata&\nodata  \\
&4382$\pm$1&113&16&\nodata&\nodata&\nodata&\nodata&\nodata&\nodata  \\
$O_{\rm K3}/M_{\rm K3}$&4437$\pm$1&410&16&4454$\pm$5&1020&51&4450$\pm$5&2090&64 \\
&4491$\pm$1&253&27&\nodata&\nodata&\nodata&\nodata&\nodata&\nodata  \\
&4661$\pm$1&164&16&\nodata&\nodata&\nodata&\nodata&\nodata&\nodata  \\
$M2$&\nodata&\nodata&\nodata&4705$\pm$5&6810&134&4707$\pm$5&6663&155 \\
\enddata
\tablecomments{Columns: 1 - designation of outburst; 2, 3, 4  - epoch of the peak, power,
and time scale of optical outburst, respectively; 5, 6, 7  - epoch of the peak, power, 
and time scale of outburst at 230~GHz, respectively; 8, 9, 10  - epoch of the peak, power, 
and time scale of outburst at 37~GHz, respectively; uncertainties of the epochs of peaks
correspond to the smoothing time of light curves}
\end{deluxetable}

\begin{deluxetable}{lrrr}
\singlespace
\tablecolumns{4}
\tablecaption{\bf Connection between Ejection and Peak of Outbursts \label{Delparm}}
\tabletypesize{\footnotesize}
\tablehead{
\colhead{Knot}&\colhead{$\Delta T_{\rm opt}$}&\colhead{$\Delta T_{\rm 230}$}&\colhead{$\Delta T_{\rm 37}$} \\
\colhead{}&\colhead{days}&\colhead{days}&\colhead{days}\\
\colhead{(1)}&\colhead{(2)}&\colhead{(3)}&\colhead{(4)}
}
\startdata
K1&$-$52$\pm$30&$-$19$\pm$34&147$\pm$34 \\
K2&22$\pm$26&51$\pm$30&52$\pm$30 \\
K3&$-$2$\pm$36&15$\pm$40&11$\pm$40 \\
\enddata
\tablecomments{Columns: 1 - name of component; 2, 3, 4 - delay of the peak of optical, 230~GHz,
and 37~GHz outbust, respectively,  with respect to the ejection time of component, $\Delta T_\nu=T_\nu-T_\circ$}
\end{deluxetable}

\clearpage
\begin{deluxetable}{lrrrrrl}
\singlespace
\tablecolumns{7}
\tablecaption{\bf Optical, near-IR, UV, and X-ray Spectral Indices$^a$ of the Synchrotron Emission Components\label{SpecInd}}
\tabletypesize{\footnotesize}
\tablehead{
\colhead{State}&\colhead{$\alpha_{\rm IR}^{\rm syn}$}&\colhead{$\alpha_{\rm opt}^{\rm syn}$}&\colhead{$\alpha_{\rm UV1}^{\rm syn}$}&\colhead{$\alpha_{\rm UV2}^{\rm syn}$}&\colhead{$\alpha_{\rm x}$}&\colhead{Epoch$_{\rm x}$ (RJD)} \\
\colhead{(1)}&\colhead{(2)}&\colhead{(3)}&\colhead{(4)}&\colhead{(5)}&\colhead{(6)}&\colhead{(7)}
}
\startdata
$O_{\rm K1}^b$&1.88$\pm$0.03&1.67$\pm$0.11&\nodata&\nodata&0.72$\pm$0.04&3505.5814 \\
$O_{\rm K2}$+$O_{K3}$&1.31$\pm$0.02&1.68$\pm$0.01&2.31$\pm$0.14&0.81$\pm$0.17&\nodata&\nodata \\
$O_{\rm K2}$&\nodata&\nodata&\nodata&\nodata&0.41$\pm$0.13&4299.1418 \\
$O_{\rm K3}$&\nodata&\nodata&\nodata&\nodata&0.56$\pm$0.12&4426.7780 \\
$M2$&1.65$\pm$0.02&1.64$\pm$0.01&2.33$\pm$0.22&1.16$\pm$0.15&0.66$\pm$0.18&4706.2752 \\ 
$QS^c$&1.42$\pm$0.01&1.97$\pm$0.03&\nodata&\nodata&0.57$\pm$0.01&4087.5 \\ 
\enddata
\tablecomments{Columns: 1 - state of activity; 2 - IR spectral index of a synchrotron component; 3 - optical spectral index of a synchrotron component; 
4 - UV (2500-3500 \'A) spectral index of a synchrotron component; 
5 - UV (2000-2500 \'A) spectral index of a synchrotron component;
6 - X-ray (2.4-10 keV) spectral index; 7 - epoch of $\alpha_{\rm x}$ measurement}
\tablenotetext{a}{All spectral indices are defined as $S_\nu\propto\nu^{-\alpha}$}
\tablenotetext{b}{The values of $\alpha_{\rm opt}$ and $\alpha_{\rm ir}$ are from \citet{HT09}}
\tablenotetext{c}{The values of $\alpha_{\rm x}$ are from \citet{RAI07}}
\end{deluxetable}

\begin{deluxetable}{rrrrrrr}
\singlespace
\tablecolumns{7}
\tablecaption{\bf Quasi-Simultaneous Optical and 43~GHz Polarization Measurements \label{O7evpa}}
\tabletypesize{\footnotesize}
\tablehead{
\colhead{Epoch$_{\rm opt}^a$}&\colhead{${\rm p}_{\rm opt}$}&\colhead{$\varphi_{\rm opt}$}&\colhead{Epoch$_{43}$}&\colhead{${\rm p}_{43}$}&\colhead{$\varphi_{43}^b$}&\colhead{$\Theta_{\rm jet}$} \\
\colhead{RJD}&\colhead{\%}&\colhead{deg}&\colhead{RJD}&\colhead{\%}&\colhead{deg}&\colhead{deg}\\
\colhead{(1)}&\colhead{(2)}&\colhead{(3)}&\colhead{(4)}&\colhead{(5)}&\colhead{(6)}&\colhead{(7)}
}
\startdata
$^*$3563.4871&4.90$\pm$0.20&$-$134.4$\pm$1.1&3568&1.0$\pm$0.2&$-$127.0$\pm$4.4&$-$94.8$\pm$3.4 \\
3656.3266&10.72$\pm$1.45&$-$145.9$\pm$3.9&3650&1.7$\pm$0.3&$-$95$\pm$4.9&$-$98.4$\pm$4.4 \\
3684.2745&8.09$\pm$2.06&$-$21.3$\pm$7.3&3689&0.8$\pm$0.2&7.1$\pm$4.6&$-$89.0$\pm$5.3 \\
3697.3281&9.89$\pm$1.99&$-$26.1$\pm$5.8&3695&2.1$\pm$0.4&9.5$\pm$2.9&$-$101.2$\pm$7.5 \\
3729.2215&8.93$\pm$0.58&17.1$\pm$1.9&3726&2.2$\pm$0.2&$-$0.5$\pm$2.4&$-$86.6$\pm$3.3 \\
$^*$3906.4829&6.46$\pm$3.50&$-$114.6$\pm$15.6&3908&1.8$\pm$0.3&$-$136.7$\pm$5.5&$-$90.9$\pm$2.9 \\
3932.5166&12.10$\pm$6.53&$-$59.3$\pm$15.5&3935&1.6$\pm$0.5&$-$92.3$\pm$8.4&$-$90.6$\pm$2.0 \\
$^*$4014.3535&15.83$\pm$6.40&$-$30.5$\pm$11.6&4014&1.1$\pm$0.5&$-$22.5$\pm$14.4&$-$101.0$\pm$4.6 \\
$^*$4050.7222&2.00$\pm$0.70&$-$53.3$\pm$10.0&4057&1.4$\pm$0.5&$-$49.0$\pm$10.7&$-$100.4$\pm$5.3 \\
4086.1988&9.31$\pm$0.82&$-$161.9$\pm$2.5&4087&0.6$\pm$0.5&$-$67.9$\pm$22.8&$-$94.7$\pm$2.1 \\
$^*$4269.4898&2.66$\pm$1.09&$-$124.1$\pm$11.8&4266&0.7$\pm$0.2&$-$126.0$\pm$5.9&$-$86.7$\pm$6.5 \\
$^*$4300.4824&21.44$\pm$0.24&$-$100.8$\pm$0.6&4294&2.6$\pm$0.3&$-$97.0$\pm$3.4&$-$86.9$\pm$4.7 \\
4319.4380&5.76$\pm$0.61&$-$58.2$\pm$3.0&4320&3.2$\pm$0.4&$-$127.0$\pm$3.4&$-$103.6$\pm$3.3 \\
4343.4985&7.05$\pm$1.85&$-$61.7$\pm$7.5&4343&3.1$\pm$0.3&$-$111.8$\pm$3.0&$-$102.2$\pm$4.1 \\ 
$^*$4405.2629&6.48$\pm$0.53&$-$117.0$\pm$2.4&4406&1.2$\pm$0.3&$-$120.7$\pm$7.4&$-$103.6$\pm$2.8 \\
4627.4868&7.82$\pm$0.55&$-$12.9$\pm$2.0&4629&2.0$\pm$0.5&$-$129.7$\pm$6.5&$-$90.3$\pm$3.4 \\
4654.4839&8.61$\pm$0.52&$-$147.8$\pm$1.7&4653&1.4$\pm$0.4&$-$109.0$\pm$7.4&$-$104.4$\pm$2.2 \\
4794.6152&2.60$\pm$0.07&$-$143.1$\pm$0.7&4787&1.4$\pm$0.5&$-$155.3$\pm$9.5&$-$123.2$\pm$3.0 \\
$^*$4828.6533&0.79$\pm$0.16&$-$155.4$\pm$4.1&4822&2.5$\pm$0.3&$-$162.0$\pm$3.9&$-$112.6$\pm$2.5 \\
\enddata
\tablecomments{Columns: 1 - epoch of optical observation; 2 - optical degree of polarization;
3 - optical position angle of polarization; 4 - epoch of VLBA observation; 5 - degree of polarization
in the core region at 43~GHz; 6 - position angle of polarization in the core region at 43~GHz;
7 - direction of the inner jet in projection on the sky plane}
\tablenotetext{a}{symbol ($^*$) marks epoch at which the optical polarization position angle and EVPA in the 43~GHz core align within 1~$\sigma$ uncertainty}
\tablenotetext{b}{position angles of polarization at 43~GHz are corrected for $RM$ according
to \citet{J07}}
\end{deluxetable}

\begin{deluxetable}{rrrrrrr}
\singlespace
\tablecolumns{7}
\tablecaption{\bf Quasi-Simultaneous Optical and 86~GHz Polarization Measurements \label{O3evpa}}
\tabletypesize{\footnotesize}
\tablehead{
\colhead{Epoch$_{\rm opt}^a$}&\colhead{${\rm p}_{\rm opt}$}&\colhead{$\varphi_{\rm opt}$}&\colhead{Epoch$_{86}$}&\colhead{${\rm p}_{86}$}&\colhead{$\varphi_{86}$}&\colhead{$\Theta_{\rm jet}$} \\
\colhead{RJD}&\colhead{\%}&\colhead{deg.}&\colhead{RJD}&\colhead{\%}&\colhead{deg.}&\colhead{deg.}\\
\colhead{(1)}&\colhead{(2)}&\colhead{(3)}&\colhead{(4)}&\colhead{(5)}&\colhead{(6)}&\colhead{(7)}
}
\startdata
4304.5098&13.84$\pm$0.94&$-$82.1$\pm$1.8&4304&2.61$\pm$0.59&$-$110.4$\pm$6.5&$-$86.9$\pm$4.7 \\
4313.4829& 6.18$\pm$0.83&$-$78.0$\pm$3.9&4314&6.93$\pm$1.19&$-$112.2$\pm$4.9&$-$103.6$\pm$3.3 \\
$^*$4342.4062& 3.70$\pm$1.31&$-$110.9$\pm$10.1&4339&4.09$\pm$0.54&$-$103.4$\pm$3.8&$-$102.2$\pm$4.1 \\
4346.4395& 9.35$\pm$1.37&$-$28.4$\pm$4.2&4346&3.62$\pm$0.57&$-$105.9$\pm$4.5&$-$102.2$\pm$4.1 \\
4355.4492& 3.57$\pm$1.93&$-$3.5$\pm$9.7&4353&2.80$\pm$0.47&$-$96.8$\pm$4.8&$-$102.2$\pm$4.1 \\
4359.4316&12.91$\pm$3.87&$-$41.9$\pm$6.6&4357&2.11$\pm$0.57&$-$100.0$\pm$7.7&$-$103.5$\pm$2.0 \\
4371.3677&26.98$\pm$1.89&$-$51.6$\pm$2.0&4372&1.86$\pm$0.56&$-$91.5$\pm$8.6&$-$103.5$\pm$2.0 \\
4395.3691&23.48$\pm$4.11&$-$169.1$\pm$5.0&4393&1.45$\pm$0.46&$-$102.4$\pm$9.2&$-$103.6$\pm$2.8 \\
$^*$4410.3267&14.09$\pm$1.01&$-$99.5$\pm$2.1&4408&1.46$\pm$0.47&$-$87.3$\pm$9.2&$-$103.6$\pm$2.8 \\
4768.4941&8.26$\pm$1.72&$-$61.0$\pm$8.6&4768&1.39$\pm$1.02&$-$7.9$\pm$21.1&$-$132.2$\pm$3.5 \\
4769.5107&4.11$\pm$0.60&$-$58.0$\pm$6.0&4769&1.79$\pm$0.47&$-$169.6$\pm$7.5&$-$132.2$\pm$3.5 \\
4773.7139&5.20$\pm$1.42&$-$79.4$\pm$7.8&4773&1.46$\pm$0.47&$-$167.1$\pm$9.1&$-$132.2$\pm$3.5 \\
\enddata
\tablecomments{Columns: 1 - epoch of optical observation; 2 - optical degree of polarization;
3 - optical position angle of polarization; 4 - epoch of observation at 86~GHz; 5 - degree of polarization at 86~GHz; 6 - position angle of polarization at 86~GHz;
7 - direction of the inner jet in projection on the sky plane}
\tablenotetext{a}{symbol ($^*$) marks epoch at which the optical polarization position angle and EVPA at 86~GHz
align within 1~$\sigma$ uncertainty}
\end{deluxetable}

\begin{deluxetable}{lrrrr}
\singlespace
\tablecolumns{5}
\tablecaption{\bf Longitudinal Size of the VLBI Core \label{Zmax}}
\tabletypesize{\footnotesize}
\tablehead{
\colhead{Knot}&\colhead{$z_{\rm max}$}&\colhead{$a_{\rm core}^{\rm l,theor}$}&\colhead{$\Delta T_{\rm s}$}&\colhead{$a_{\rm core}^{\rm l,obs}$} \\
\colhead{}&\colhead{mas}&\colhead{mas}&\colhead{days}&\colhead{mas}\\
\colhead{(1)}&\colhead{(2)}&\colhead{(3)}&\colhead{(4)}&\colhead{(5)} 
}
\startdata
K1&0.50&0.026&50&0.013 \\
K2&0.56&0.041&50&0.025 \\
K3&1.01&0.011&65&0.018 \\
\enddata
\tablecomments{Columns: 1 - name of event; 2 - distance between constrictions; 3 - 
estimate of longitudinal size of the core from the theoretical model;
4 - estimate of longitudinal size of the core from observational consideration}
\end{deluxetable}

\clearpage

\clearpage

\clearpage

\clearpage 

\end{document}